\documentclass{aa}
\usepackage{graphicx}
\usepackage{xcolor}
\usepackage{caption}
\definecolor{links}{rgb}{0, 0, 255}
\usepackage[colorlinks=true,allcolors=links]{hyperref}
\usepackage[varg]{txfonts}

\newcommand{\msun}{M_{\sun}}
\newcommand{\msunh}{h^{-1}M_{\sun}}

\begin{document}

\title{The VariableTNG project: mass-dependent regulation of galaxy morphology by baryonic feedback}

\author{Yikai Liu\inst{\ref{inst1},\ref{inst2}}
\and Hong Guo\inst{\ref{inst1}}\fnmsep\thanks{Corresponding Author; guohong@shao.ac.cn}
\and Volker Springel\inst{\ref{inst3}}    
\and Kexin Liu\inst{\ref{inst1},\ref{inst2}}
\and Luis C. Ho\inst{\ref{inst4},\ref{inst5}}
\and BoCheng Zhu\inst{\ref{inst6},\ref{inst7}}
}

\institute{
Shanghai Astronomical Observatory, Chinese Academy of Sciences, Shanghai 200030, China.\label{inst1} 
\and University of Chinese Academy of Sciences, Beijing 100049, China.\label{inst2}
\and Max-Planck-Institut f\"{u}r Astrophysik, Karl-Schwarzschild-Stra\ss{}e 1, 85740 Garching bei M\"{u}nchen, Germany \label{inst3}
\and Kavli Institute for Astronomy and Astrophysics, Peking University, Beijing 100871, China \label{inst4}
\and Department of Astronomy, School of Physics, Peking University, Beijing 100871, China \label{inst5}
\and Institute of Astrophysics, School of Physics, Zhengzhou University, China \label{inst6}
\and Key Laboratory for Computational Astrophysics, National Astronomical Observatories, Chinese Academy of Sciences, Beijing 100101, China \label{inst7}
}

\abstract
{Galaxy morphology is shaped by both assembly history and baryonic processes, but their relative roles remain uncertain.
We use the VariableTNG (VTNG) simulation suite to quantify how systematic variations in baryonic feedback regulate the morphological diversity of galaxies, while keeping the initial conditions fixed. VTNG is a suite of cosmological magnetohydrodynamic simulations performed with the moving-mesh code {\sc AREPO}, in which eight parameters governing stellar and AGN feedback are varied. At $z=1$, we quantify galaxy morphology using the kinematic indicators $\kappa_{\rm co}$ and $v/\sigma$, together with the structural indicator $c/a$.
We find that baryonic feedback produces substantial morphological diversity even among galaxies with identical large-scale initial conditions. The dominant regulating mechanism depends strongly on stellar mass. For $M_\ast \lesssim 10^{11}\,{\rm M_\odot}$, morphology is primarily controlled by the supernova temperature $T_{\rm SN}$: larger $T_{\rm SN}$ delays early star formation, promotes the buildup of a denser and more rotationally supported gas reservoir, and favours the subsequent growth of stellar discs through in-situ star formation. At higher masses, morphology becomes increasingly sensitive to AGN feedback, especially the quasar-mode coupling efficiency $\epsilon_{\rm f,high}$. A higher $\epsilon_{\rm f,high}$ suppresses early black hole growth and thus weakens the later radio-mode feedback that is more directly associated with gas depletion and the destruction of rotational support. The morphology--mass relation is therefore non-monotonic, with maximum rotational support at intermediate stellar masses. Our results show that galaxy morphology responds strongly to mass-dependent feedback pathways under controlled initial conditions. In VTNG, stellar feedback primarily drives the morphological response in lower-mass galaxies, whereas AGN self-regulation and subsequent kinetic feedback become increasingly important at the massive end.}

\maketitle
\nolinenumbers

\section{Introduction}

Understanding the origin of galaxy morphology remains one of the central problems in galaxy formation theory. The Hubble sequence organizes galaxies into spirals, lenticulars, ellipticals, and irregulars \citep{1926ApJ....64..321H,1961hag..book.....S,1976ApJ...206..883V}, but morphology is more than a descriptive classification: it encodes how galaxies assembled their stellar mass, angular momentum, and gas reservoirs over cosmic time. Morphological type is tightly correlated with a wide range of galaxy properties, including stellar mass, colour, stellar age, and size \citep{2003MNRAS.341...54K,2003ApJS..149..289B,2014ARA&A..52..291C,2014ApJ...788...28V}. Explaining the physical origin of this diversity is therefore essential for any successful theory of galaxy evolution.

In the standard hierarchical picture, discs form as gas cools and settles within dark matter haloes, while mergers and dynamical heating transform part of the galaxy population into spheroid-dominated systems \citep{1969ApJ...155..393P,1980MNRAS.193..189F,1998MNRAS.295..319M,1978MNRAS.184..185W,1984ApJ...286...38W,1992ARA&A..30..705B,2014MNRAS.444.3357N,2017MNRAS.467.3083R}. This framework captures the broad connection between morphology and assembly history, but it is incomplete. In particular, purely merger-driven pictures struggle to explain the large diversity of galaxy structure across cosmic time, especially once the evolution of the gas reservoir and the regulation of star formation are taken into account.

Recent observations with the James Webb Space Telescope (JWST) have sharpened this tension by revealing a surprisingly large population of morphologically settled galaxies at $z>4$ \citep[e.g.,][]{2022ApJ...938L..15C,2023ApJ...948L..14C,2023ApJ...955...94F,2023MNRAS.518.6011D,2023MNRAS.520.4554D,2024ApJ...960..104S}. These results suggest that ordered disc-like and spheroidal structures can emerge much earlier than expected in simple merger-dominated scenarios \citep{2023ApJ...955...94F,2025A&A...699A.360C}. At the same time, several theoretical studies have shown that mergers do not inevitably produce long-lived spheroids, because sufficiently gas-rich systems may reform discs after the merger event \citep{2015MNRAS.452.4347K,2018MNRAS.480.2266M}. This points to a more general picture in which morphology depends not only on assembly history, but also on the baryonic processes that regulate gas supply, star formation, and disc regrowth.

Baryonic feedback from supernovae and active galactic nuclei (AGN) is now recognized as a fundamental ingredient of galaxy formation models \citep{2006MNRAS.370..645B,2018RMxAA..54..217S,2020MNRAS.492.3073L}. By heating, ejecting, and redistributing gas, feedback regulates star formation, modifies angular-momentum retention, and changes the conditions under which stellar discs can form or survive. A number of studies have shown that supernova-driven winds and AGN-driven outflows can influence galaxy structure \citep{2005ApJ...620L..79S,2008MNRAS.389.1137S,2009ApJS..182..216K,2014MNRAS.443.2092U,2013MNRAS.433.3297D,2016MNRAS.463.3948D}. However, the structural impact of baryonic feedback remains difficult to quantify. In conventional single-realization cosmological hydrodynamical simulations, morphological differences between galaxies generally reflect several effects at once, including differences in merger history, gas accretion history, environment, and feedback physics. As a result, when two galaxies in a single simulation show different morphologies, it is often difficult to determine how much of that difference is driven by their assembly history and how much is due to the adopted feedback model. In particular, it remains unclear how strongly feedback can alter galaxy morphology once the assembly history is broadly controlled, and through which physical pathways this influence operates.

Addressing this question requires controlled numerical experiments in which the feedback model is varied systematically while the initial conditions are kept fixed. In this paper, we use the VariableTNG (VTNG) simulation suite, which is based on the {\sc Arepo} code \citep{2010MNRAS.401..791S} and the IllustrisTNG framework \citep{2017MNRAS.465.3291W,2018MNRAS.473.4077P,2018MNRAS.475..676S,2019ComAC...6....2N,2020ApJS..248...32W}. VTNG varies eight parameters governing stellar and AGN feedback while adopting identical initial conditions across all realizations. This controlled design, together with the numerical resolution of the suite, makes VTNG well suited to studying internal galaxy morphology and stellar kinematics.

In recent years, large simulation suites such as the CAMELS project have demonstrated the power of broad parameter-space exploration for cosmological and astrophysical inference \citep{2021ApJ...915...71V,2023ApJ...959..136N}. However, CAMELS was primarily designed for statistical applications and machine-learning studies, rather than for detailed analyses of internal galaxy morphology and stellar kinematics. Its standard simulations adopt a box size of $25\,h^{-1}\,{\rm Mpc}$ with $256^3$ dark matter particles and $256^3$ gas resolution elements, which is not optimal for robust studies of galaxy internal structure. Moreover, its different simulation sets do not provide the same type of controlled framework as VTNG. The Latin-hypercube (LH) realizations vary physical parameters across a broad parameter space while also spanning different random initial conditions, whereas the fewer one-parameter (1P) runs keep the seed fixed but vary only one parameter at a time. By contrast, VTNG adopts identical initial conditions across all realizations while varying multiple feedback parameters simultaneously to sample the parameter space more efficiently with a limited number of simulation runs. 

VTNG is therefore complementary to CAMELS for a systematic analysis of how feedback regulates galaxy morphology. Rather than attempting to fully disentangle mergers and feedback as independent drivers of morphology, we use VTNG to quantify how strongly morphology responds to feedback variations under controlled initial conditions, and to identify the mass-dependent physical pathways through which stellar and AGN feedback promote, suppress, or rebuild rotationally supported stellar structures. We focus on simulation outputs down to $z\sim1$, where the diversity of galaxy structure is already well developed and the imprint of both stellar and AGN feedback is clearly visible. Section~\ref{secfeedback} describes the simulation setup and the variable feedback parameters. Section~\ref{sec:results} presents the morphology diagnostics and their dependence on stellar mass and feedback parameters. Our discussions and conclusions are presented in Section~\ref{sec:discuss} and~\ref{sec:conclusion}, respectively.

\section{\label{secfeedback}Feedback Model in VTNG}
The VTNG simulation suite is built on the moving-mesh magnetohydrodynamics code {\sc Arepo} \citep{2010MNRAS.401..791S}, using the same general framework as IllustrisTNG \citep[e.g.,][]{2018MNRAS.475..676S,2019ComAC...6....2N}. The model includes the key baryonic processes required for galaxy formation, including radiative cooling, star formation, stellar feedback, black hole growth, and AGN feedback. While the fiducial IllustrisTNG model successfully reproduces many low-redshift galaxy observables \citep{2018MNRAS.473.4077P}, its performance at high redshift is less satisfactory, suggesting that the adopted feedback prescriptions may not yet capture the full range of physically plausible galaxy-growth pathways. VTNG is designed to explore this uncertainty in a controlled way.

We therefore construct a suite of simulations in which eight parameters governing stellar and AGN feedback are varied systematically around their fiducial TNG values. The explored ranges are listed in Table~\ref{table-parameters}. We generate 100 parameter combinations using a Sobol sequence in order to sample the high-dimensional parameter space efficiently while minimizing clustering and degeneracy. All simulations start from identical initial conditions at $z\sim127$, so that differences in galaxy morphology can be attributed to the adopted feedback parameters rather than to variations in assembly history or large-scale environment. We adopt a Planck 2018 cosmology \citep{2020A&A...641A...6P} with $\{\Omega_m,\Omega_\Lambda,\Omega_b,\sigma_8,h\}=\{0.3158,0.6842,0.04939,0.8120,0.6732\}$. The simulation volume is a periodic box of side length $50\,h^{-1}\,{\rm Mpc}$, resolved with $1024^3$ dark matter particles and $1024^3$ gas cells, corresponding to mass resolutions of $1.28\times10^7\,{\rm M_\odot}$ for dark matter and $2.37\times10^6\,{\rm M_\odot}$ for baryons.

In the following subsections, we briefly summarize the sub-grid ingredients most relevant for the present analysis. We focus on the physical role of the varied parameters and defer more technical details of the implementation to \citet{2003MNRAS.339..289S}, \citet{2013MNRAS.436.3031V}, \citet{2017MNRAS.465.3291W}, and \citet{2018MNRAS.473.4077P}.

\begin{table*}[ht!]
	\caption{\label{table-parameters} Baryonic feedback parameters in the VTNG simulation suite.}
	\centering
	\begin{tabular*}{\textwidth}{@{\extracolsep{\fill}}lccc@{}}
		\hline\hline
		Parameter & Symbol & Fiducial (TNG) & Range (VTNG) \\
		\hline
        Seed black hole mass ($\msunh$) & $M_{\rm seed}$ & $8\times10^5$ & $(5\times10^5, 5\times10^6)$ \\ 
        Radio-mode feedback factor & $\epsilon_{\rm m}$ & $1.0$ & $(0.1, 10)$ \\ 
        Quasar-mode feedback factor & $\epsilon_{\rm f,high}$ & $0.1$ & $(0.01, 1)$ \\ 
        Quasar threshold & $\chi_0$ & $0.002$ & $(0.001, 0.004)$ \\ 
        Max star formation timescale (Gyr) & $t^{\star}_{0}$ & $2.27$ & $(1.0, 3.0)$ \\ 
        Supernova temperature (K) & $T_{\rm SN}$ & $5.73\times10^7$ & $(10^7, 10^8)$ \\ 
        Thermal wind fraction & $\tau_{w}$ & $0.1$ & $(0.01, 0.2)$ \\ 
        Wind velocity factor & $\kappa_w$ & $7.4$ & $(3, 10)$ \\ 		
	    \hline
	\end{tabular*}
	\tablefoot{The eight VTNG parameters varied in this work. Four regulate star formation and stellar feedback, and four govern black-hole growth and AGN feedback. Their physical roles are summarized in Section~\ref{secfeedback}.}
\end{table*}

\subsection{Star Formation Model and Stellar Feedback}

In {\sc Arepo}, star formation is modeled using a sub-grid multiphase prescription \citep{2003MNRAS.339..289S}. A gas cell becomes eligible for star formation once its density exceeds a critical threshold, $\rho_{\rm crit}$, which prevents spurious star formation in diffuse gas. Each star-forming gas cell is resolved as a single stellar population (SSP) characterized by a \citet{2003PASP..115..763C} initial mass function (IMF).

The star formation timescale is density-dependent and follows:
\begin{equation}
    t_{\star}(\rho) = t^{\star}_{0} \left( \frac{\rho}{\rho_{\rm th}} \right)^{-1/2},
\end{equation}
where $\rho$ is the gas density, $\rho_{\rm th}$ is the threshold density, and $t^{\star}_{0}$ is the star-formation timescale at the threshold. This scaling captures the empirical Kennicutt–Schmidt relation \citep{1998ApJ...498..541K}. The threshold density is determined by the balance between star formation, cloud evaporation, and supernova feedback. The total supernova energy per unit stellar mass is parameterized by the supernova temperature, $T_{\rm SN}$, which is a key variable in VTNG. Effectively, $T_{\rm SN}$ regulates $\rho_{\rm th}$: a higher $T_{\rm SN}$ shifts $\rho_{\rm th}$ toward higher values, suppressing star formation in lower-density environments.

The evolution of stellar and cold gas densities ($\rho_\star$ and $\rho_c$) is expressed as:
\begin{equation}
    \frac{{\rm d}\rho_{\star}}{{\rm d}t} = (1-\beta) \frac{\rho_{c}}{t_{\star}},
\end{equation}
where $\beta$ represents the mass fraction of short-lived massive stars ($M_\ast > 8\msun$) that promptly explode as supernovae, injecting thermal and kinetic energy as well as metals into the surrounding interstellar medium (ISM).

Stellar feedback triggers the generation of wind particles that transport mass, momentum, thermal energy, and metals into the circumgalactic medium (CGM). The wind velocity, $v_w$, is scaled by the local dark matter velocity dispersion, $\sigma_{\rm DM}$ \citep{2018MNRAS.473.4077P}:
\begin{equation}
    v_{w} = \max \left[ \kappa_{w} \sigma_{\rm DM} \left( \frac{H_0}{H(z)} \right)^{1/3}, v_{w,\min} \right],
\end{equation}
where $\kappa_{w}$ is a dimensionless wind velocity factor and $v_{w,\min}$ is the minimum wind velocity. The partition of feedback energy into thermal and kinetic channels is regulated by the thermal wind fraction, $\tau_{w}$. 
These prescriptions allow stellar feedback to redistribute metals and regulate gas cooling, thereby shaping the morphological evolution of galaxies \citep{2015MNRAS.454.2691M,2017MNRAS.472L.109A}.

The four key parameters related to star formation and stellar feedback in VTNG are the maximum star formation timescale $t^{\star}_{0}$, the supernova temperature $T_{\rm SN}$, the thermal wind fraction $\tau_{w}$, and the wind velocity factor $\kappa_{w}$. These parameters will be used to understand how different stellar feedback strengths alter the balance between star formation, gas recycling, and outflows, and consequently imprint themselves on galaxy morphology.

\subsection{AGN Feedback} \label{arepo:agn}
Centrally located supermassive black holes (SMBHs) exert a profound influence on galaxy evolution through AGN feedback. Since the detailed physics of black hole accretion occurs at scales well below our numerical resolution, we employ sub-grid prescriptions. In VTNG, SMBHs are seeded with an initial mass $M_{\rm seed}$ in halos exceeding $5 \times 10^{10}\msunh$. As detailed in \citet{2017MNRAS.465.3291W}, the feedback of SMBHs follows two primary modes—quasar mode and radio mode—determined by a dimensionless accretion threshold, $\chi$:
\begin{equation}
    \frac{\dot{M}_{\rm Bondi}}{\dot{M}_{\rm Edd}}\geq\chi
\end{equation}
where $\dot{M}_{\rm bondi}$ is the Bondi-Hoyle-Lyttleton accretion rate, $\dot{M}_{\rm Edd}$ is the Eddington accretion rate, and $\chi$ is set by
\begin{equation}
    \chi={\rm min}\left[\chi_{0}\left(\frac{M_{\rm BH}}{10^8\msun}\right)^{2},0.1\right],
\end{equation}
where $\chi_0$ is a variable parameter that determines the transition threshold. 

In the high-accretion (quasar) mode ($\dot{M}_{\rm Bondi}/\dot{M}_{\rm Edd} \geq \chi$), feedback energy is coupled to the surrounding gas as purely thermal energy:
\begin{equation}
    \Delta\dot{E}_{\rm high} = \epsilon_{\rm f,high} \epsilon_{\rm r} \dot{M}_{\rm BH} c^{2},
\end{equation}
where $\epsilon_{\rm r}$ is the radiative efficiency that regulates the amount of energy released by the accretion process and $\epsilon_{\rm f,high}$ is the quasar-mode feedback factor that determines the fraction of feedback energy coupled to the surrounding gas. 

In the low-accretion (radio) mode ($\dot{M}_{\rm Bondi}/\dot{M}_{\rm Edd} < \chi$), the AGN injects kinetic energy:
\begin{equation}
    \Delta\dot{E}_{\rm low} = \epsilon_{\rm m} \epsilon_{\rm r} \dot{M}_{\rm BH} c^{2},
\end{equation}
where $\epsilon_{\rm m}$ is the radio-mode feedback factor. In the radio-mode, kinetic energy is injected in a randomly oriented direction to mimic the impact of large-scale jets. This approach is necessary as current cosmological resolutions remain insufficient to explicitly resolve parsec-scale jet processes.

The varying parameters in VTNG related to the AGN feedback are $M_{\rm seed}$, $\epsilon_{\rm f,high}$, $\epsilon_{\rm m}$, and $\chi_0$, which aim to elucidate how AGN feedback dictates the structural and kinematic properties of galaxies, particularly in the high-mass regime.

\section{Results}\label{sec:results}
The full VTNG design consists of 100 feedback realizations. At the current stage, 33 of these simulations have been evolved to $z=1$, and the present analysis is based on this available subset. Although the results presented here are therefore limited to $z \geq 1$, this interval spans the peak of cosmic star formation and a critical epoch for the establishment of galaxy structure. It is thus well suited to identifying the early imprint of baryonic feedback on galaxy morphology. Extending the analysis to z=0 and to earlier epochs will be essential for connecting the VTNG trends to local morphology and JWST samples; this will be addressed when the full VTNG suite becomes available over a wider redshift range. Haloes and subhaloes are identified with the FoF and {\sc Subfind} algorithms. For the morphology analysis, we adopt $M_\ast>10^9\,{\rm M_\odot}$, corresponding to ~400 baryon particles at the VTNG baryonic mass resolution of $2.37*10^{6} {\rm M_\odot}$. This threshold is chosen to ensure that the kinematic and structural indicators are not dominated by particle noise, yielding approximately $3,000$ to $8,000$ galaxies per realization.

\begin{figure}
\centering
\includegraphics[width=0.95\linewidth]{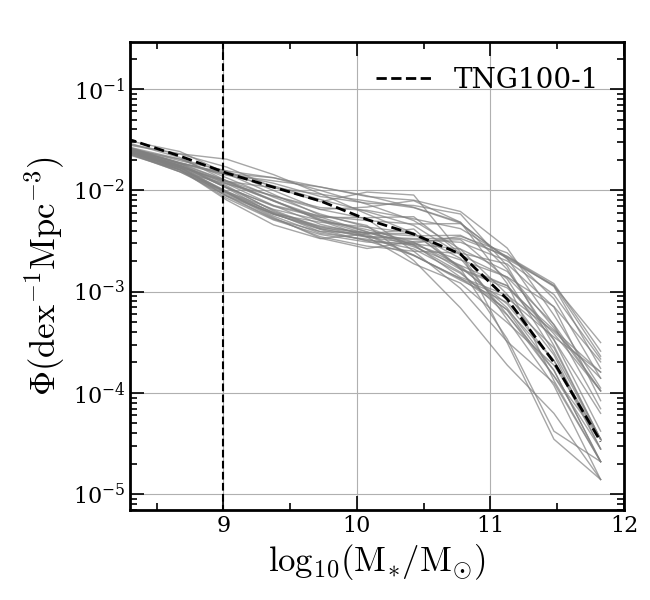}
        \caption{Stellar mass functions of the VTNG realizations at $z=1$. Each grey curve corresponds to one simulation with a different feedback parameter combination, while the fiducial TNG100-1 model is shown as the black dashed curve. Differences between realizations appear already at $M_\ast \sim 10^9\,{\rm M_\odot}$ and increase substantially toward the massive end, indicating that galaxy growth is highly sensitive to the adopted feedback model. Stellar mass below $10^9\,{\rm M_\odot}$ is shown only for context and is not used for morphology analysis}
\label{fig:stmassfunc}
\end{figure}

Figure~\ref{fig:stmassfunc} shows the stellar mass functions of the VTNG suite at $z=1$. For comparison, we also display the fiducial result from the TNG100-1 simulation, which has a slightly better resolution than VTNG. Differences between feedback realizations become visible already at $M_\ast\sim10^9\,{\rm M_\odot}$ and grow substantially toward the massive end. This confirms that the adopted feedback parameters have a strong impact on the overall efficiency of galaxy growth. In what follows, we examine how these differences propagate into galaxy structure and stellar kinematics.

\subsection{Morphology diagnostics and parameter sensitivity} \label{subsec:morpho} 
We quantify galaxy morphology using both kinematic and structural information measured within twice the stellar half-mass radius, $2r_{1/2}$. This choice allows us to characterize whether stellar systems are primarily supported by ordered rotation or by random motions, while also tracking the geometric thickness of the stellar component.

For the kinematic classification, we compute the fraction of stellar kinetic energy invested in ordered co-rotation, $\kappa_{\rm co}$, and the ratio of ordered to random motions, $v/\sigma$. The parameter $\kappa_{\rm co}$ is defined as \citep{2012MNRAS.423.1544S,2017MNRAS.472L..45C},
\begin{equation}
    \label{kappa_co}
    \kappa_{\rm co} = \frac{K_{\rm co}}{K} = \frac{1}{K} \sum_{r_{i}<2r_{1/2}} \frac{1}{2} m_{i} \left[ L_{z,i} / (m_{i} R_{i}) \right]^{2},
\end{equation}
where $K$ is the total stellar kinetic energy within $2r_{1/2}$, $m_i$ is the particle mass. $L_{z,i}$ is the positive angular momentum component aligned with the galaxy's total angular momentum vector $\vec{L}$, indicating the kinetic energy from the co-rotating star particles. $R_i$ is the projected distance from the rotation axis. 
For the parameter $v/\sigma$, $v$ is the mean rotational velocity measured about the direction of the total angular momentum and $\sigma$ is the velocity dispersion. Disc-dominated galaxies are expected to have high $\kappa_{\rm co}$ and large $v/\sigma$, whereas spheroid-dominated systems exhibit lower values. 

For the structural classification, we use the minor-to-major axis ratio $c/a$ derived from the stellar inertia tensor, with a small $c/a$ indicating flattened discs and a large $c/a$ indicating thicker or more spheroidal systems. The stellar inertia tensor, $I$, is defined as,
\begin{equation}
    \label{inertia_tensor}
    I_{ij} = \sum_{r_{k}<2r_{1/2}} m_{k} \left( \delta_{ij} |\mathbf{r}_{k}|^2 - r_{k,i} r_{k,j} \right),
\end{equation}
where $\mathbf{r}_{k}$ is the position vector relative to the galaxy centre. The eigenvectors of this tensor define the principal directions, and the corresponding principal axes $a$, $b$, and $c$ are inferred from its eigenvalues in the standard way for the rotational inertia tensor. We then use the minor-to-major axis ratio $c/a$ as the structural morphology indicator.

To evaluate how morphology responds to the VTNG feedback parameters, we combine two complementary statistical tools. First, we use a random-forest regressor to capture non-linear responses and parameter interactions, and quantify the relative importance of each parameter through permutation importance. Second, we use partial rank correlation coefficients (PRCCs) to assess the net monotonic influence of each parameter while controlling for the others. Although these statistical measures do not by themselves establish causality, the use of identical initial conditions across the VTNG suite ensures that the recovered parameter sensitivity is not driven by differences in large-scale environment or merger statistics.

\begin{figure*}
    \centering
    \includegraphics[width=0.32\linewidth]{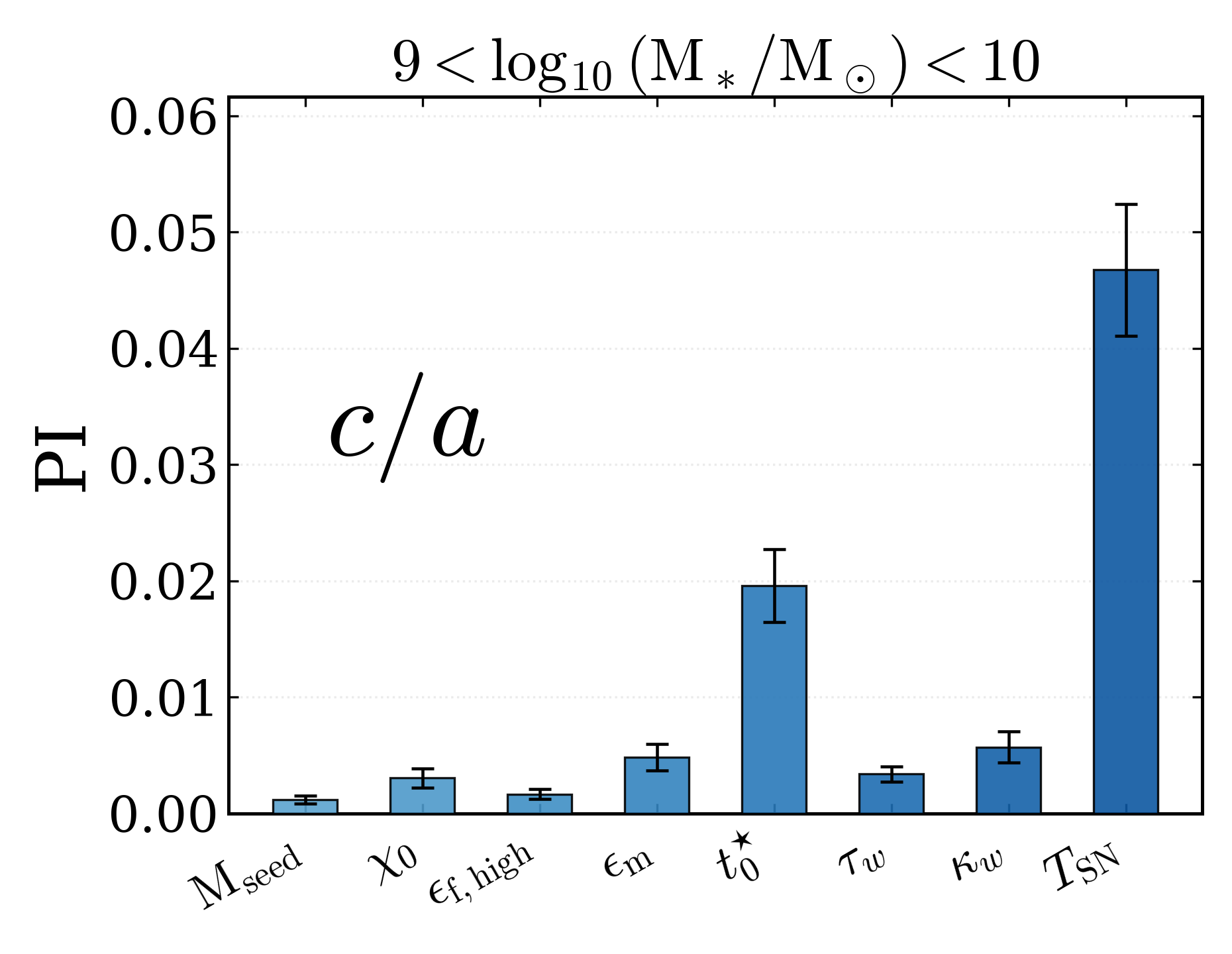}
    \hfill
    \includegraphics[width=0.32\linewidth]{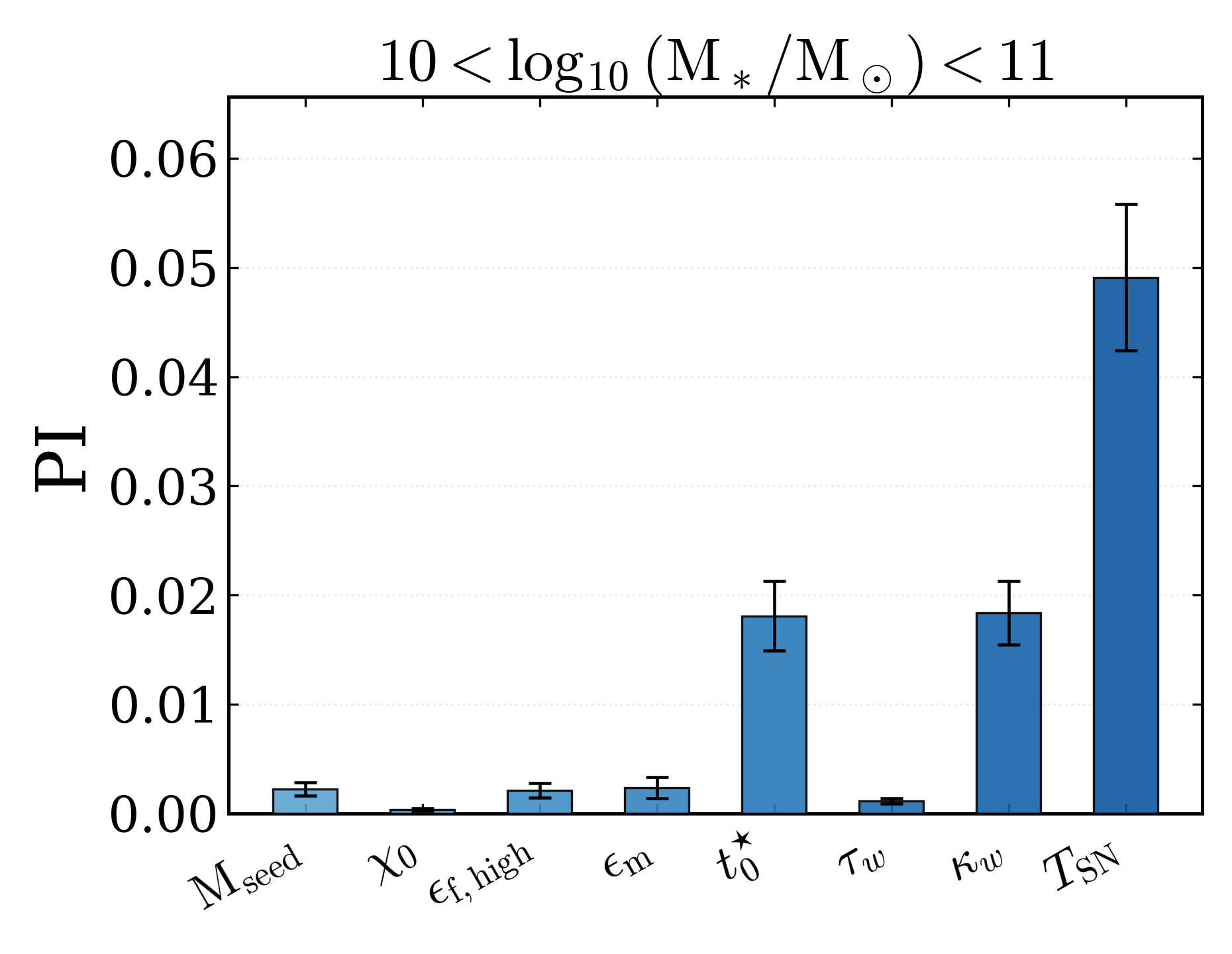}
    \hfill
    \includegraphics[width=0.32\linewidth]{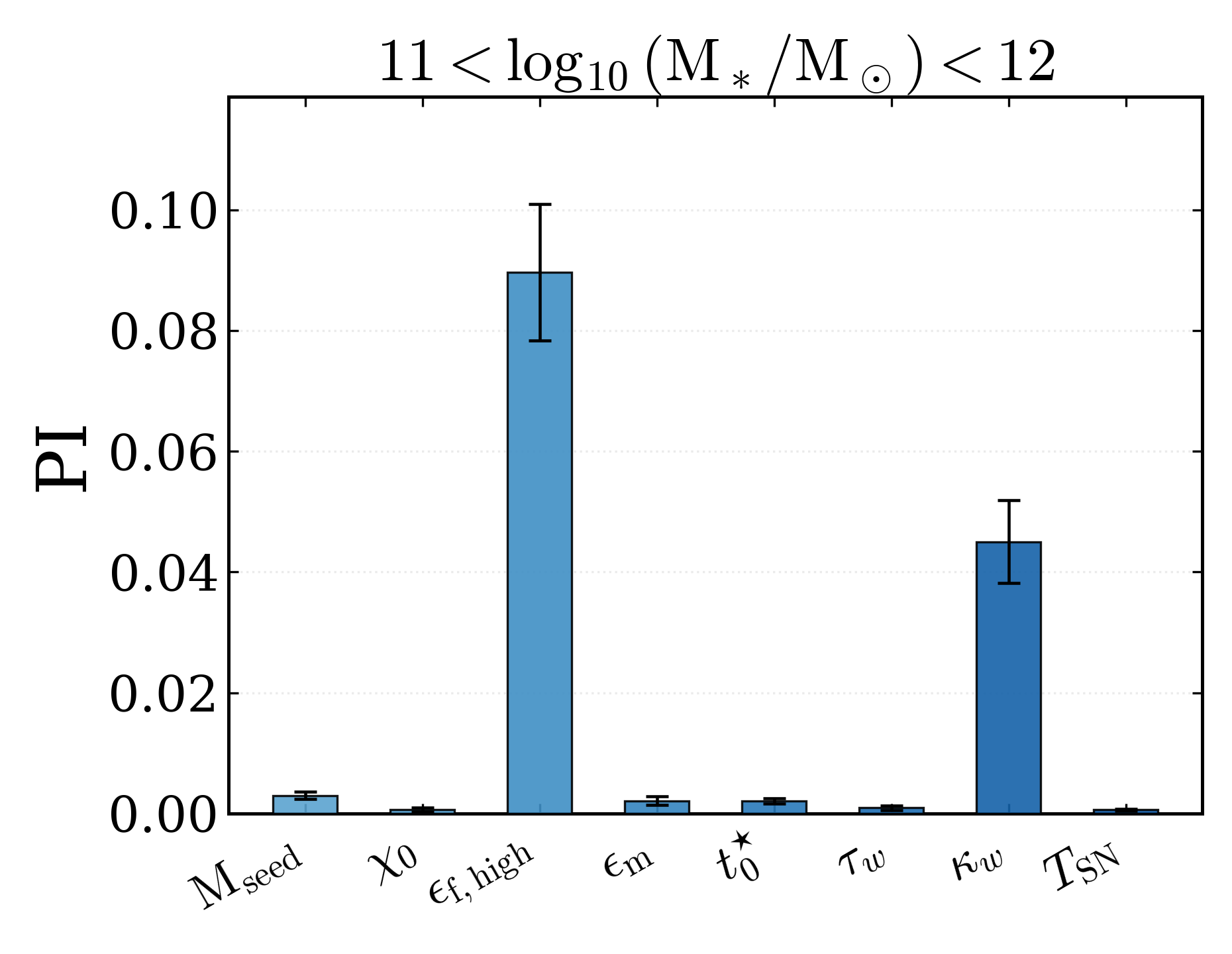}
    \includegraphics[width=0.32\linewidth]{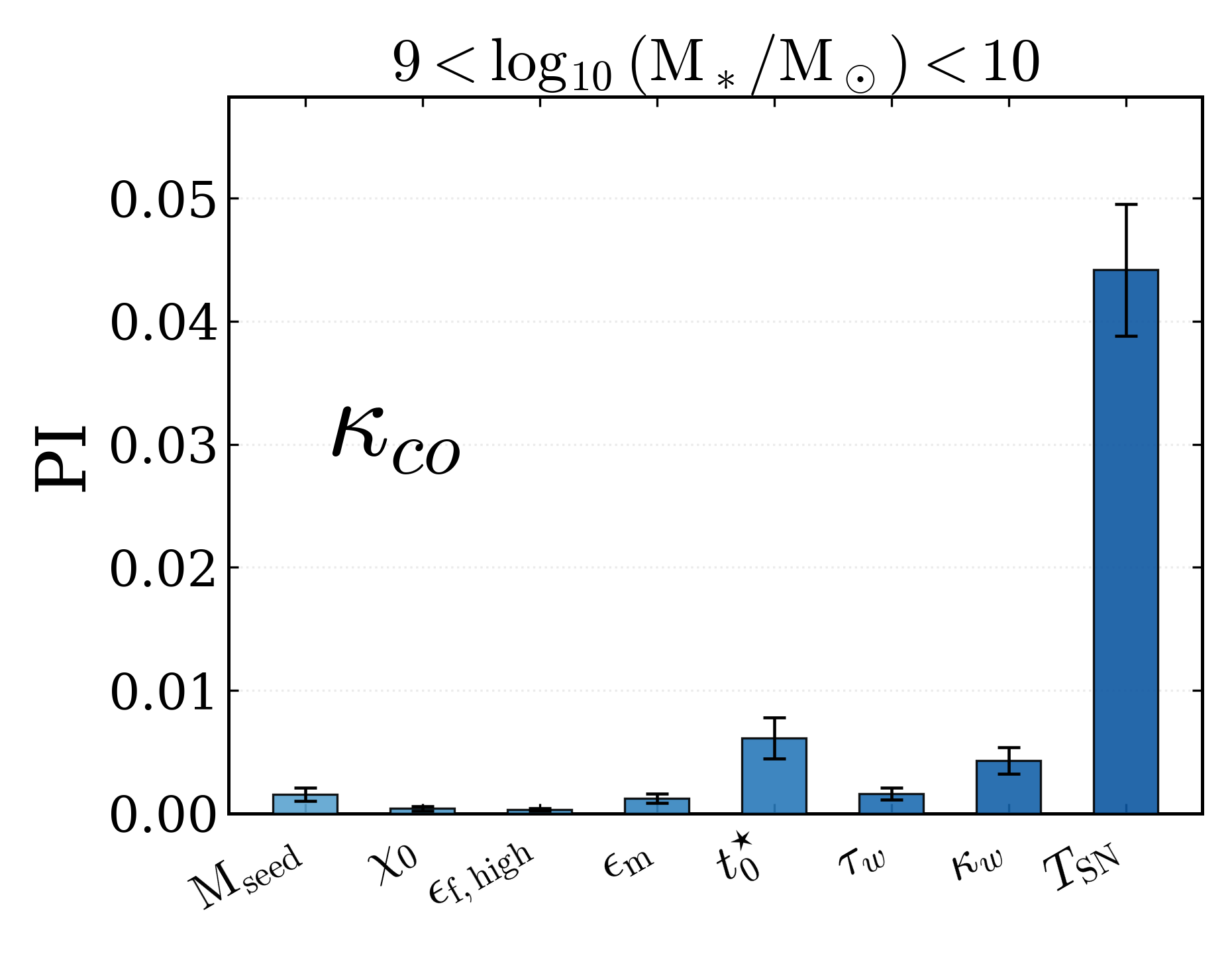}
    \hfill
    \includegraphics[width=0.32\linewidth]{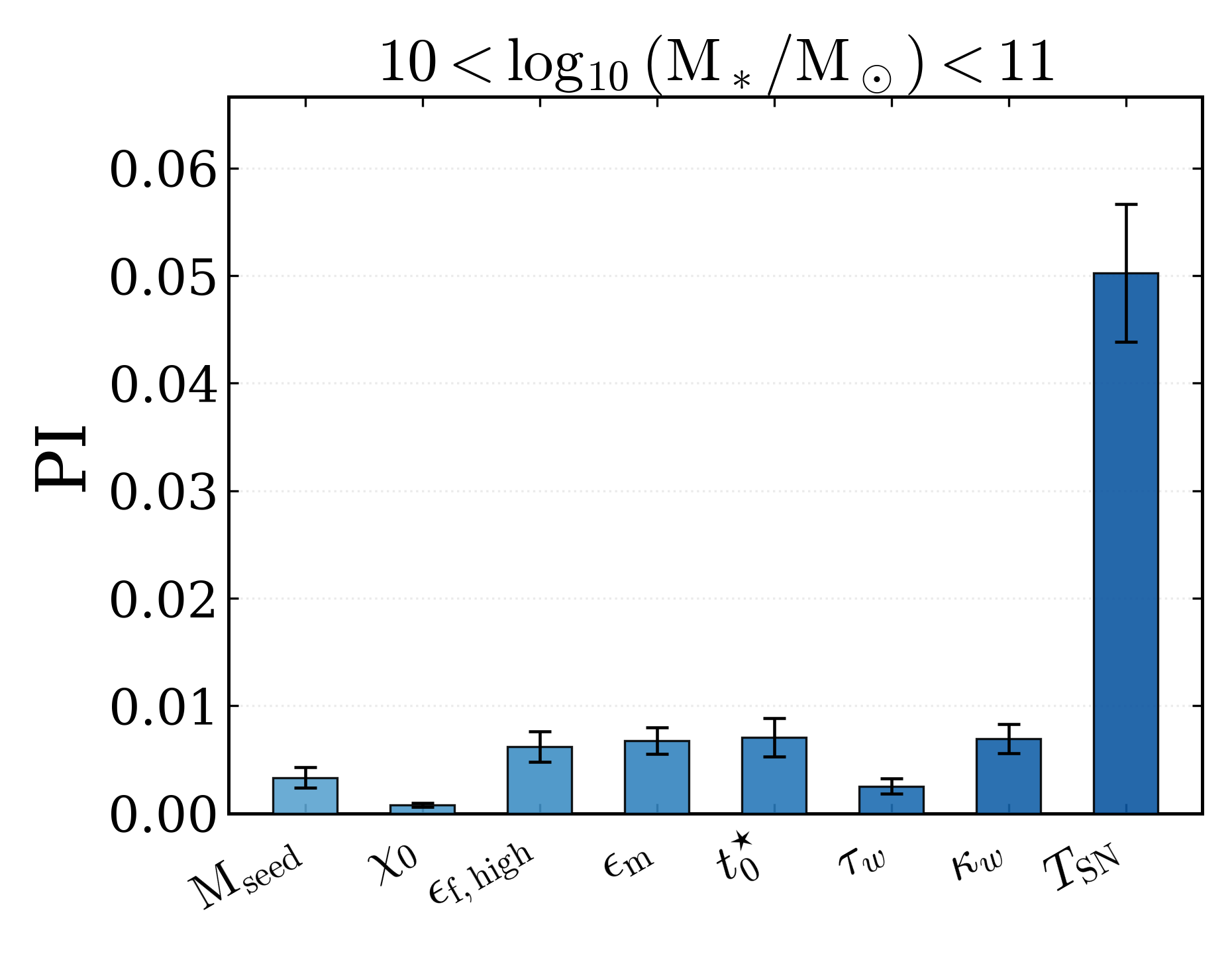}
    \hfill
    \includegraphics[width=0.32\linewidth]{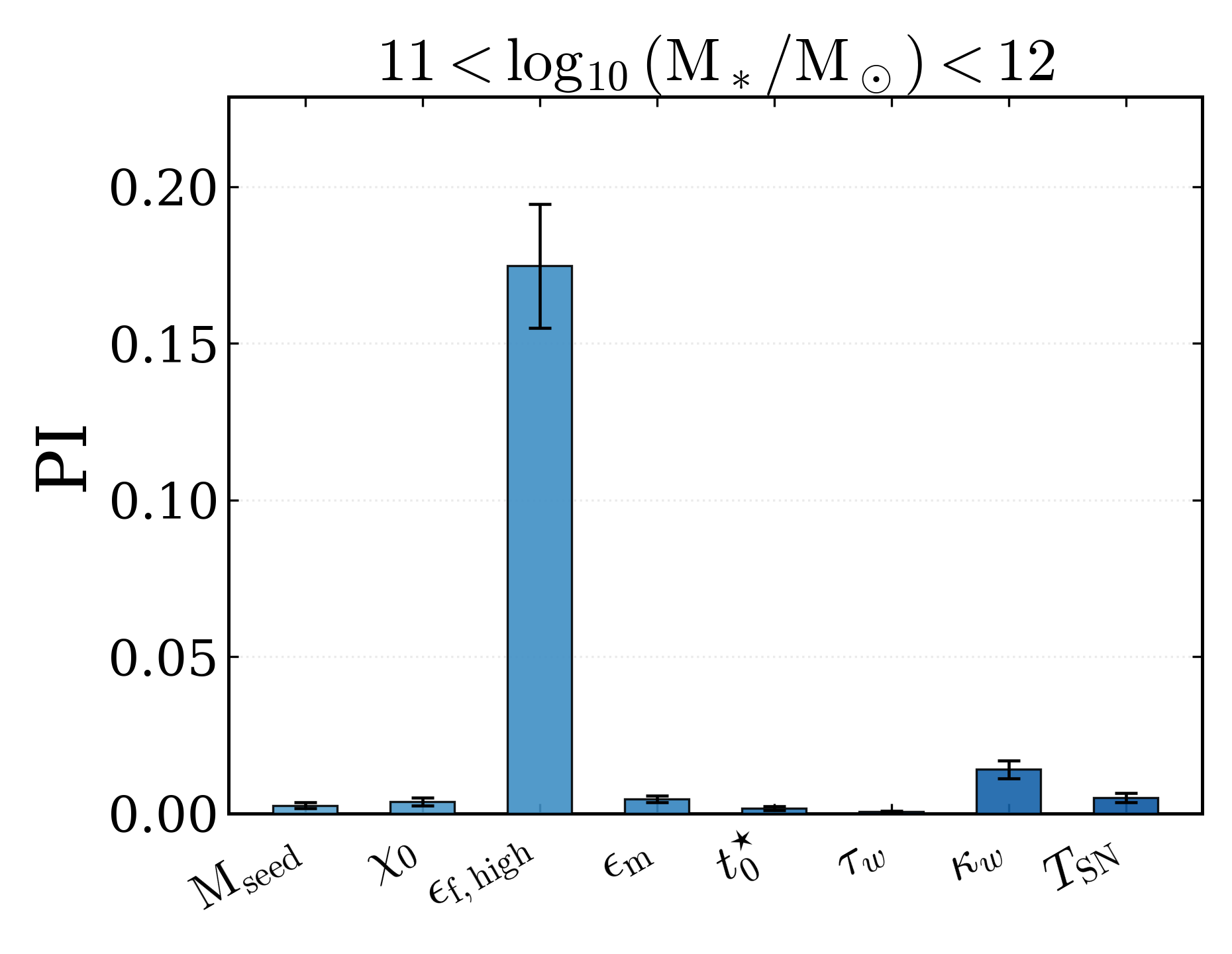}
    \includegraphics[width=0.32\linewidth]{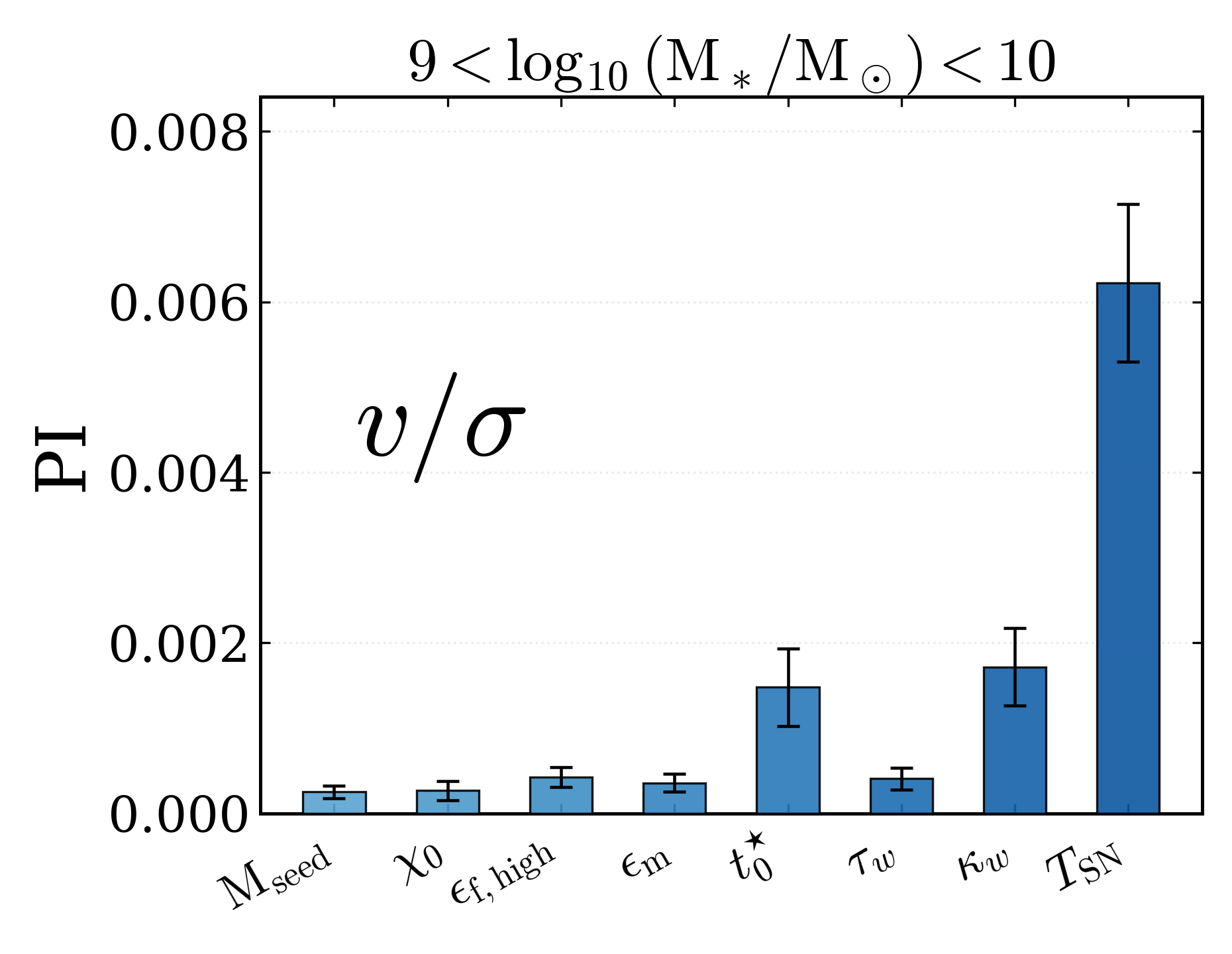}
    \hfill
    \includegraphics[width=0.32\linewidth]{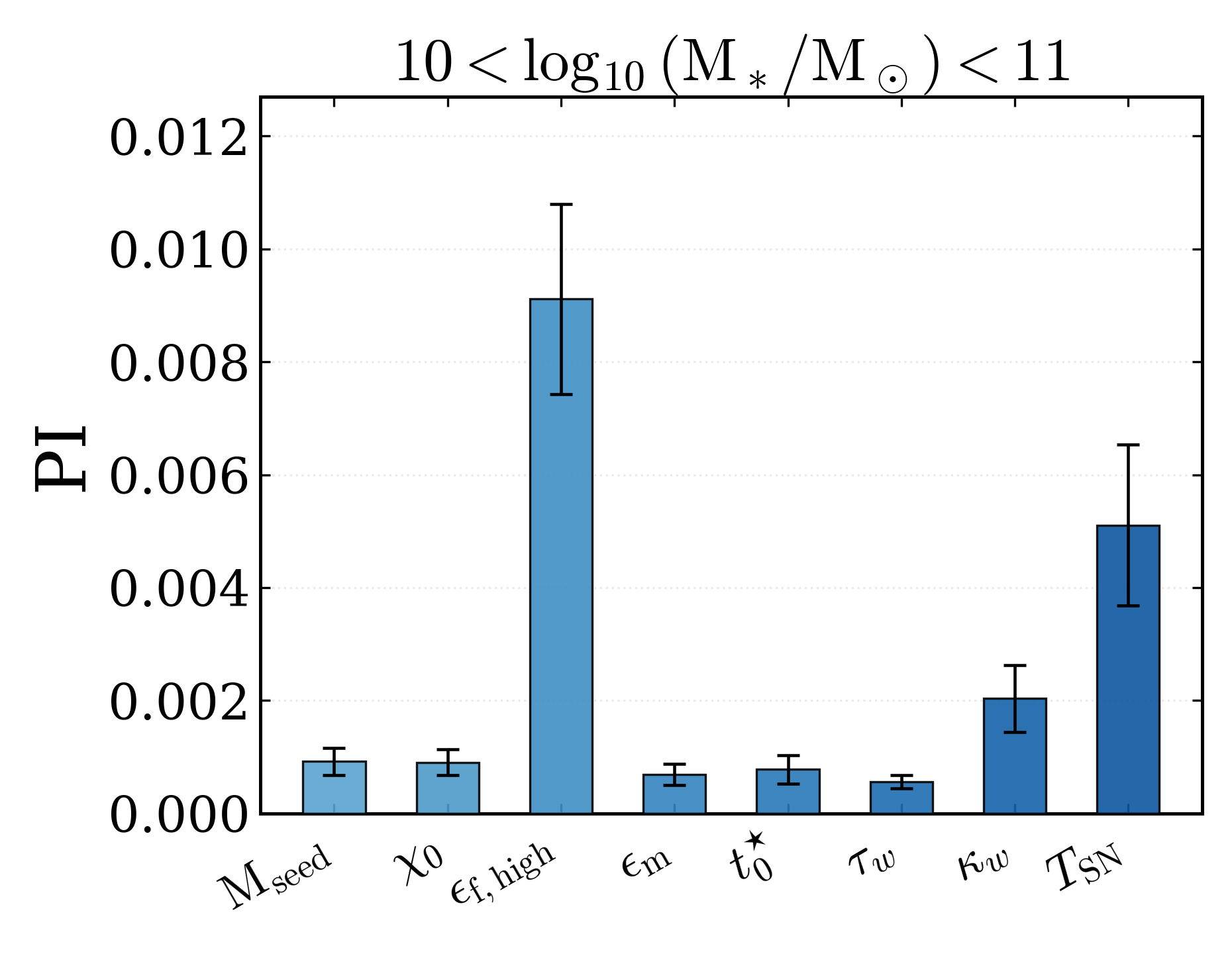}
    \hfill
    \includegraphics[width=0.32\linewidth]{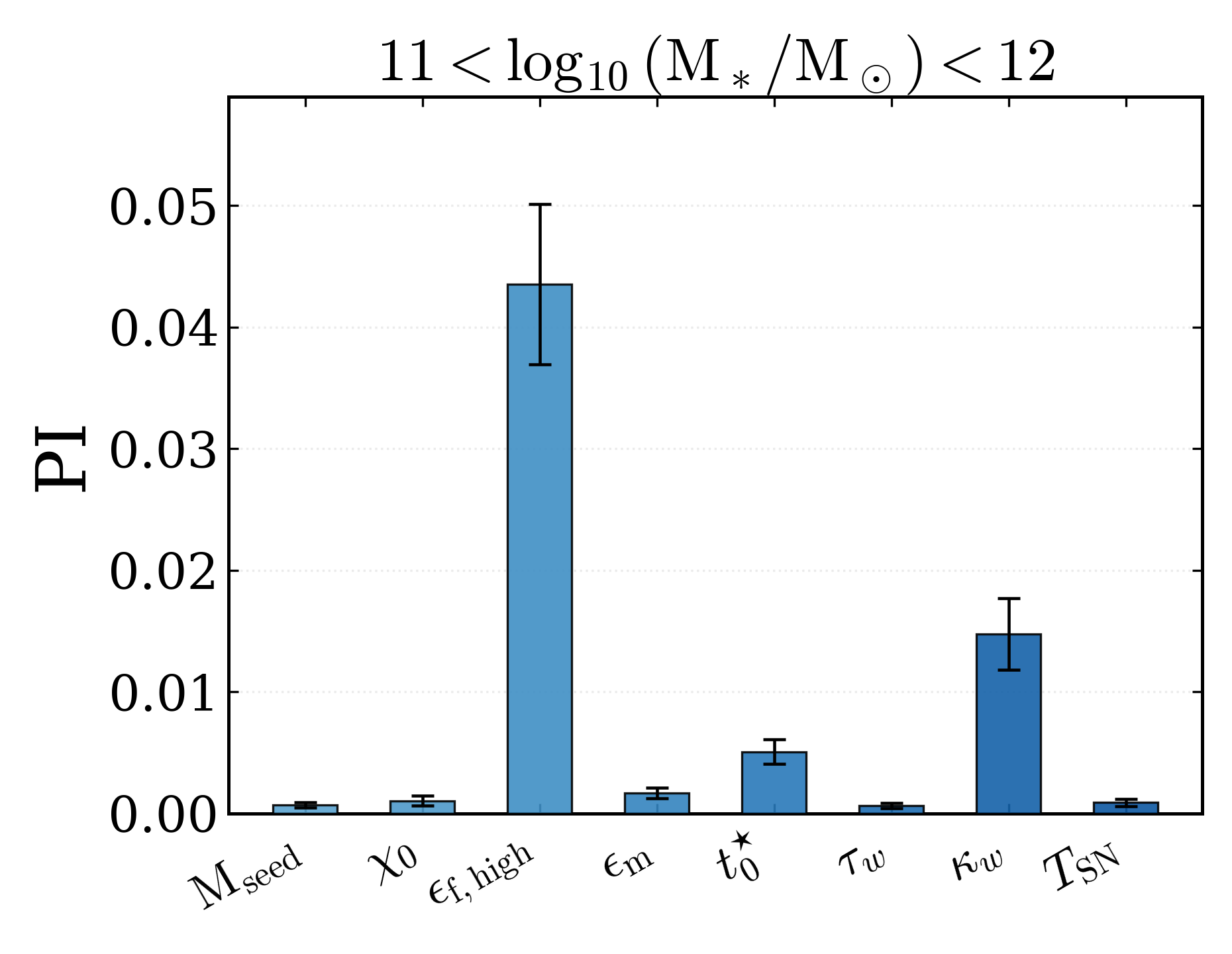}
    \caption{Permutation importance from the random-forest analysis for predicting the morphology indicators $c/a$, $\kappa_{\rm co}$, and $v/\sigma$ at $z=1$. Columns correspond to increasing stellar-mass bins. In low- and intermediate-mass galaxies, morphology is most strongly sensitive to the supernova temperature $T_{\rm SN}$, whereas in the most massive systems AGN-related parameters, especially $\epsilon_{\rm f,high}$, become dominant.}
    \label{fig:random_forest}
\end{figure*}

\begin{figure}
    \centering
    \includegraphics[width=0.95\linewidth]{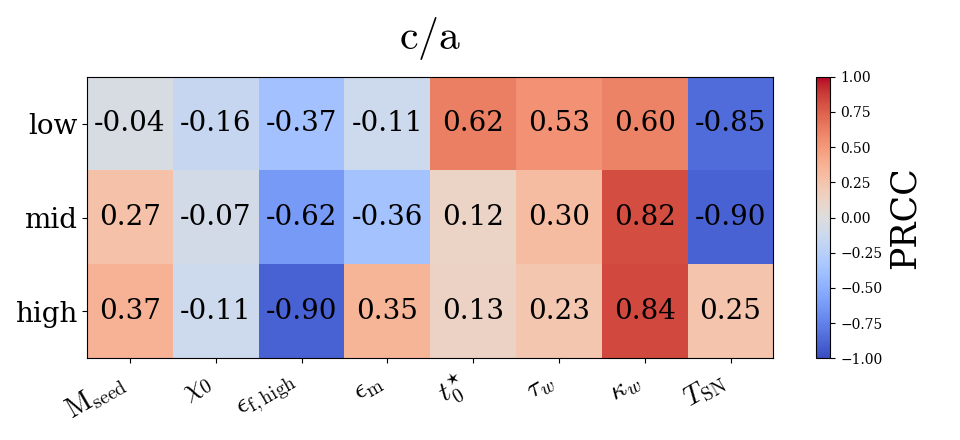}
    \includegraphics[width=0.95\linewidth]{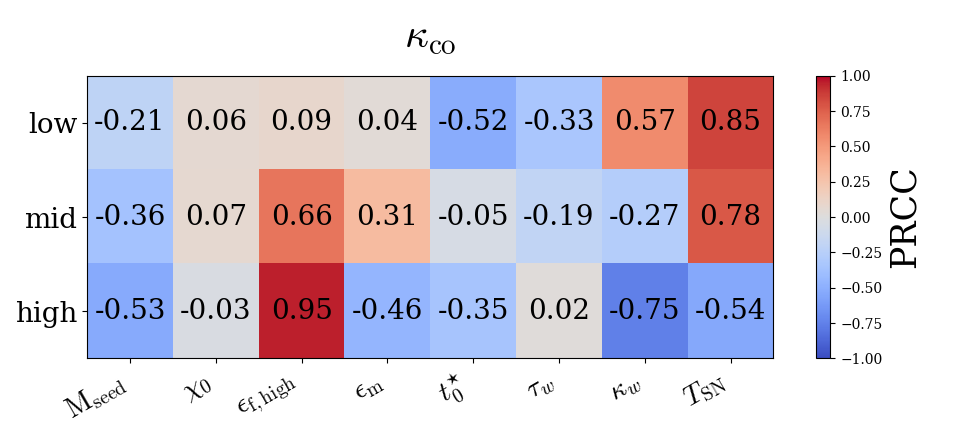}
    \includegraphics[width=0.95\linewidth]{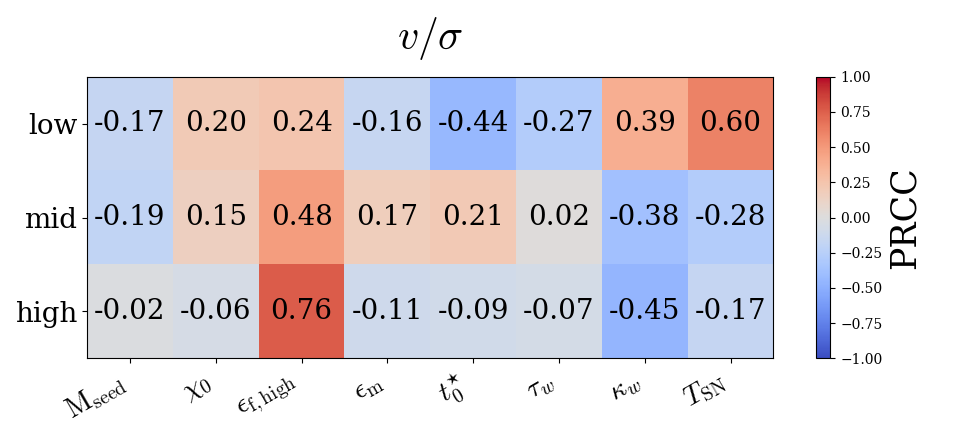}
    \caption{Partial rank correlation coefficients (PRCCs) between the VTNG feedback parameters and the morphology indicators in three stellar-mass bins. Red and blue indicate positive and negative monotonic correlations, respectively. The trends confirm the mass-dependent transition seen in Figure~\ref{fig:random_forest}: $T_{\rm SN}$ is the main driver of morphology at low masses, while AGN-related parameters dominate at high masses.}
    \label{fig:Spearman}
\end{figure}

Figures~\ref{fig:random_forest} and \ref{fig:Spearman} reveal a clear mass dependence in the physical origin of galaxy morphology. In the low- and intermediate-mass ranges, $10^9 < M_\ast/{\rm M_\odot} < 10^{11}$, morphology is primarily regulated by stellar feedback, with the supernova temperature $T_{\rm SN}$ emerging as the dominant parameter. Higher $T_{\rm SN}$ correlates with larger $\kappa_{\rm co}$ and $v/\sigma$, and with smaller $c/a$, indicating that stronger supernova heating favours thinner and more rotationally supported stellar structures. In the most massive systems, $M_\ast > 10^{11}\,{\rm M_\odot}$, the dominant sensitivity shifts to AGN-related parameters, especially the quasar-mode coupling efficiency $\epsilon_{\rm f,high}$.

An important and initially counterintuitive result is that larger $\epsilon_{\rm f,high}$ tends to be associated with more disc-like massive galaxies. As we show below, this does not imply that stronger AGN feedback directly promotes disc formation. Instead, it reflects the fact that strong early quasar-mode feedback suppresses black hole growth and thereby reduces the later radio-mode feedback that more efficiently depletes central gas reservoirs and suppresses rotational support. Because all three morphology diagnostics respond in a qualitatively similar way, we focus primarily on $\kappa_{\rm co}$ in the remainder of the paper.

This mass dependence is physically well motivated. In low-mass haloes, the gas binding energy is comparable to the energy injected by supernovae, so stellar feedback can efficiently regulate gas accretion, angular-momentum retention, and star formation \citep{2015MNRAS.454.2691M,2016ApJ...824...57C}. In more massive haloes, by contrast, the gravitational potential is too deep for stellar feedback alone to dominate the baryon cycle, and AGN feedback becomes increasingly important for heating or ejecting halo gas and suppressing renewed disc growth \citep{2006MNRAS.365...11C,2006MNRAS.370..645B,2015ARA&A..53...51S,2018MNRAS.475..648P}. VTNG therefore shows in a controlled way that the dominant feedback sensitivity shifts from stellar feedback to AGN feedback with increasing stellar mass.

\subsection{Mass-dependent morphology at $z=1$} \label{subsec:massdepend}
\begin{figure*}[htbp]
    \centering
    \includegraphics[width=0.30\linewidth]{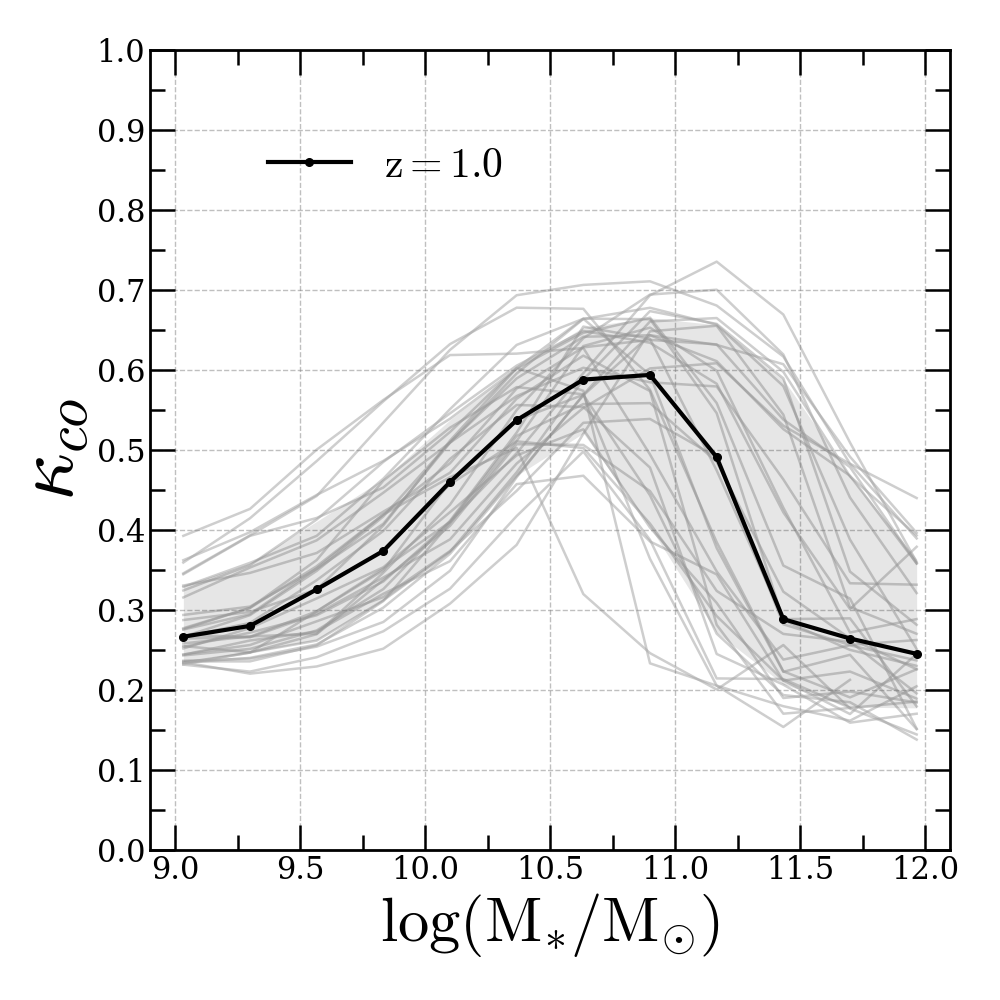}
    \hspace{10pt}
    \includegraphics[width=0.30\linewidth]{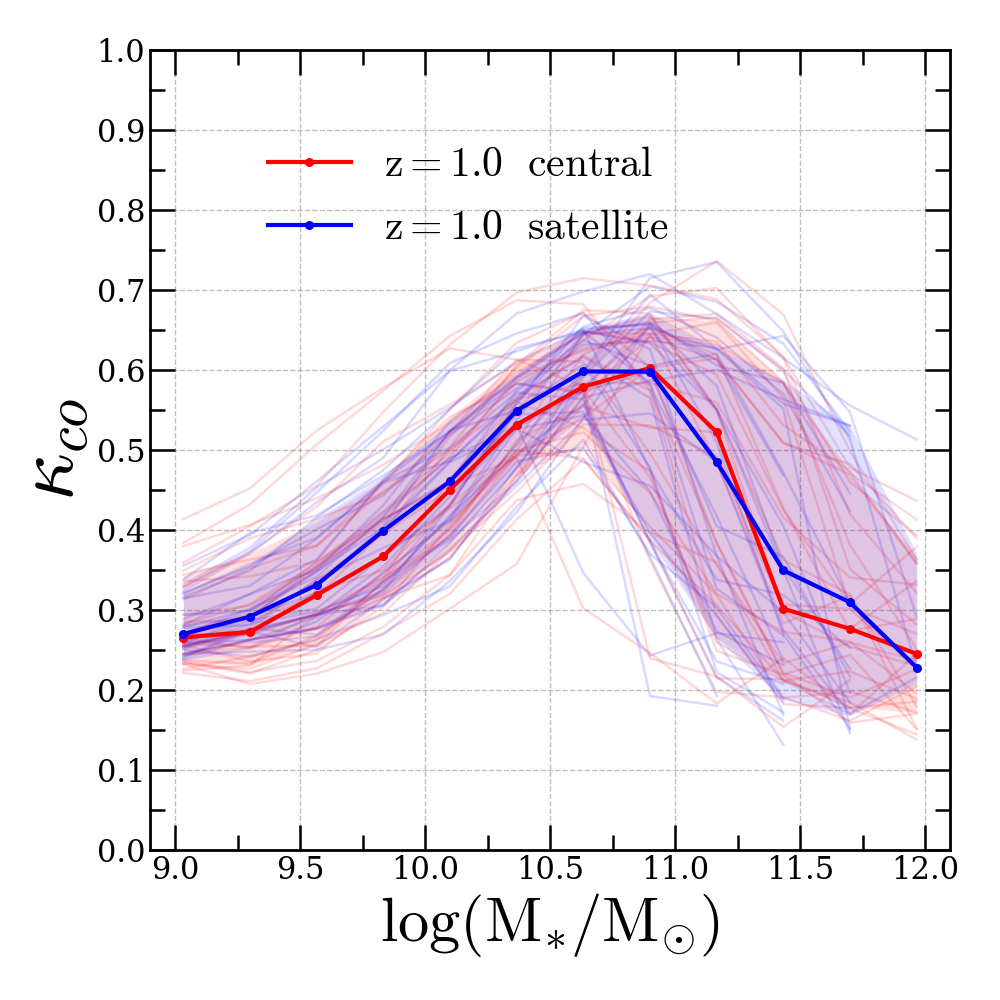}
    \hspace{10pt}
    \includegraphics[width=0.30\linewidth]{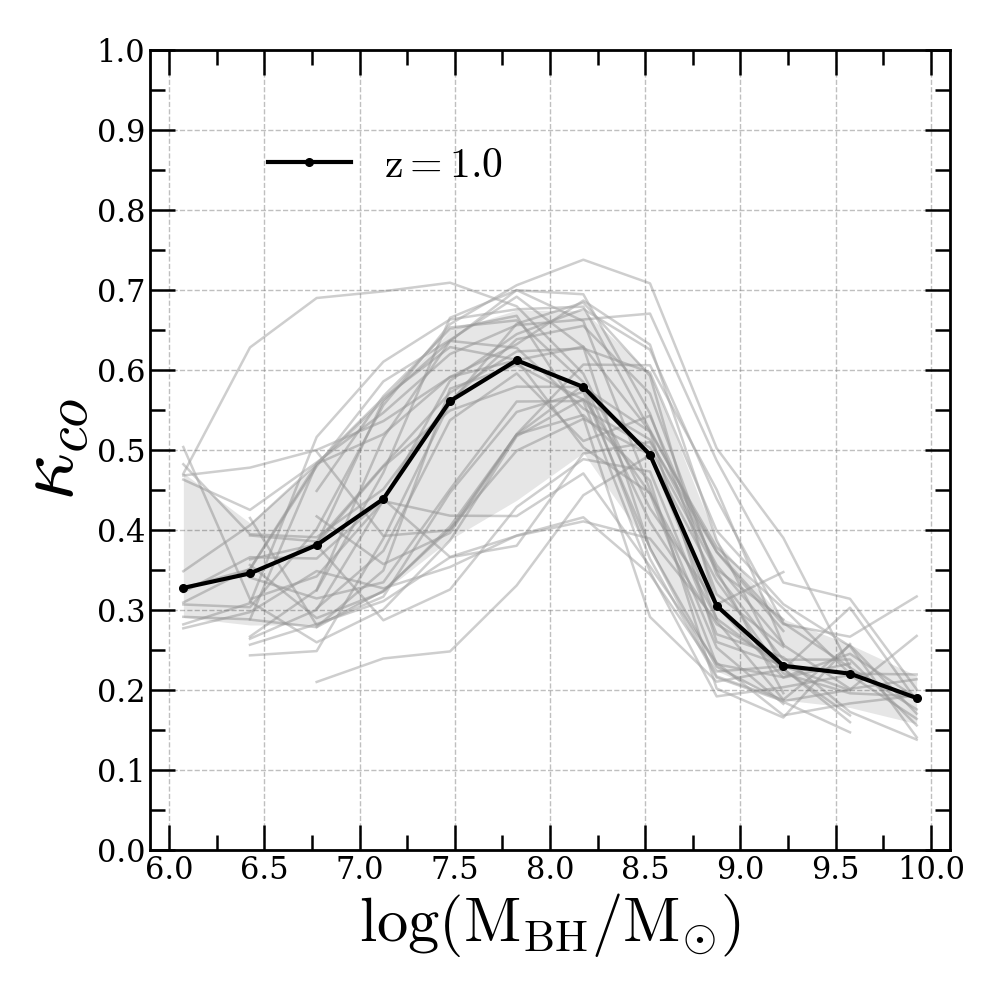}
    \caption{Median relation between the morphology indicator $\kappa_{\rm co}$ and stellar mass (left and middle) or black-hole mass (right) at z=1. Grey curves show the median relations for individual VTNG realizations, while thick curves show the median over the full suite. The middle panel separates the galaxy population into centrals (red) and satellites (blue). In all cases, the relation is non-monotonic, with the highest rotational support occurring at intermediate stellar and black-hole masses. The overall trends are similar for central and satellite galaxies.}
    \label{fig:mass-dependence}
\end{figure*}
Figure~\ref{fig:mass-dependence} shows the dependence of $\kappa_{\rm co}$ on stellar mass and black hole mass at $z=1$. In both cases, the relation is clearly non-monotonic. As stellar mass increases, $\kappa_{\rm co}$ rises gradually and reaches a maximum at $M_\ast \sim 10^{10.8}\,{\rm M_\odot}$, after which it declines toward the massive end. Galaxies therefore achieve their highest level of rotational support at intermediate masses, while both low- and high-mass systems are, on average, more dispersion-dominated. We additionally separate the sample into central and satellite galaxies in the middle panel and find that the overall $\kappa_{\rm co}-M_\ast$ relation remains very similar for the two populations, suggesting that the main trends reported here are not driven by the mixture of centrals and satellites. This weak central--satellite dependence is consistent with previous studies showing that the angular-momentum content or structural properties of galaxies are primarily controlled by stellar mass, and that residual central--satellite differences become small once stellar mass is matched \citep{2018ApJ...852...36G,2020ApJ...889...37W}. We therefore use the full galaxy sample in the following analysis, since separating centrals and satellites does not qualitatively change the feedback-driven morphology trends.

At the low-mass end, $M_\ast \lesssim 10^{10.5}\,{\rm M_\odot}$, galaxies have relatively low $\kappa_{\rm co}$. Their shallow potential wells make them more susceptible to supernova-driven outflows, dynamical perturbations, and non-circular gas motions, all of which hinder the formation of dynamically cold stellar discs \citep{2016MNRAS.463.3948D,2018MNRAS.473.1930E,2019MNRAS.487.5416T,2024MNRAS.532.2558Z}. In the intermediate-mass range, $M_\ast \sim 10^{10.5}$--$10^{11}\,{\rm M_\odot}$, deeper potentials stabilize the gas against disruption while continued gas accretion sustains in-situ star formation, producing the highest degree of rotational support. This is also the approximate mass scale at which the stellar-to-halo mass relation peaks and disc fractions are typically largest \citep{2006MNRAS.373.1389C,2019MNRAS.488.3143B}.

Above $M_\ast \gtrsim 10^{11}\,{\rm M_\odot}$, $\kappa_{\rm co}$ declines again. In this regime, stellar dynamical heating driven by mergers, virial shock heating, and AGN feedback jointly reduce the cold-gas supply and suppress the continued growth of stellar discs \citep{2006MNRAS.368....2D}. Massive galaxies at $z=1$ are therefore predominantly dispersion-supported systems. A similar non-monotonic trend appears as a function of black hole mass, with the maximum rotational support occurring near $M_{\rm BH}\sim10^{7.8}\,{\rm M_\odot}$. This correspondence further supports the view that black hole growth, feedback history, and galaxy structure are tightly linked.

The broad morphology--mass relation in VTNG can thus be understood as the outcome of processes whose relative importance varies with mass: stellar feedback in shallow potential wells, sustained gas accretion and in-situ star formation at intermediate masses, and AGN feedback together with dynamical heating at the massive end. The relation itself is generic across the suite, but the scatter among realizations demonstrates that baryonic feedback strongly modulates where individual systems lie around this median trend.

\subsection{Physical origin of the feedback-driven morphological diversity} \label{sec:baryon}
The statistical trends presented above show that galaxy morphology in VTNG depends sensitively on the adopted feedback parameters. The key physical question is therefore not whether feedback matters, but through which physical pathways it shapes galaxy structure under controlled initial conditions. Because all VTNG realizations share identical initial conditions, corresponding galaxies can be traced across runs and compared directly. This makes it possible to identify how different feedback prescriptions generate divergent structural evolution in matched systems.

\begin{table}[ht!]
	\caption{Key parameter values for the simulations used in the detailed case studies.}                
	\label{table-choosensimu}    
	\centering                        
	\begin{tabular}{c c c c}      
		\hline\hline               
		Simulation& ${\rm T}_{\rm SN} (K)$  & Simulation & $\epsilon_{\rm f,high}$ \\         
		\hline                      
HighT & $7.09\times 10^{7}$ & High$\epsilon_{\rm f,high}$ & 0.192 \\
LowT & $1.08\times 10^{7}$ & Low$\epsilon_{\rm f,high}$ & 0.0218 \\
		\hline                                  
	\end{tabular}
\end{table}

\subsubsection{Supernova feedback} \label{subsec:sn}
\begin{figure}[htbp]
    \centering
    
    \begin{minipage}[c]{0.46\textwidth}
        \centering
        \includegraphics[width=\linewidth]{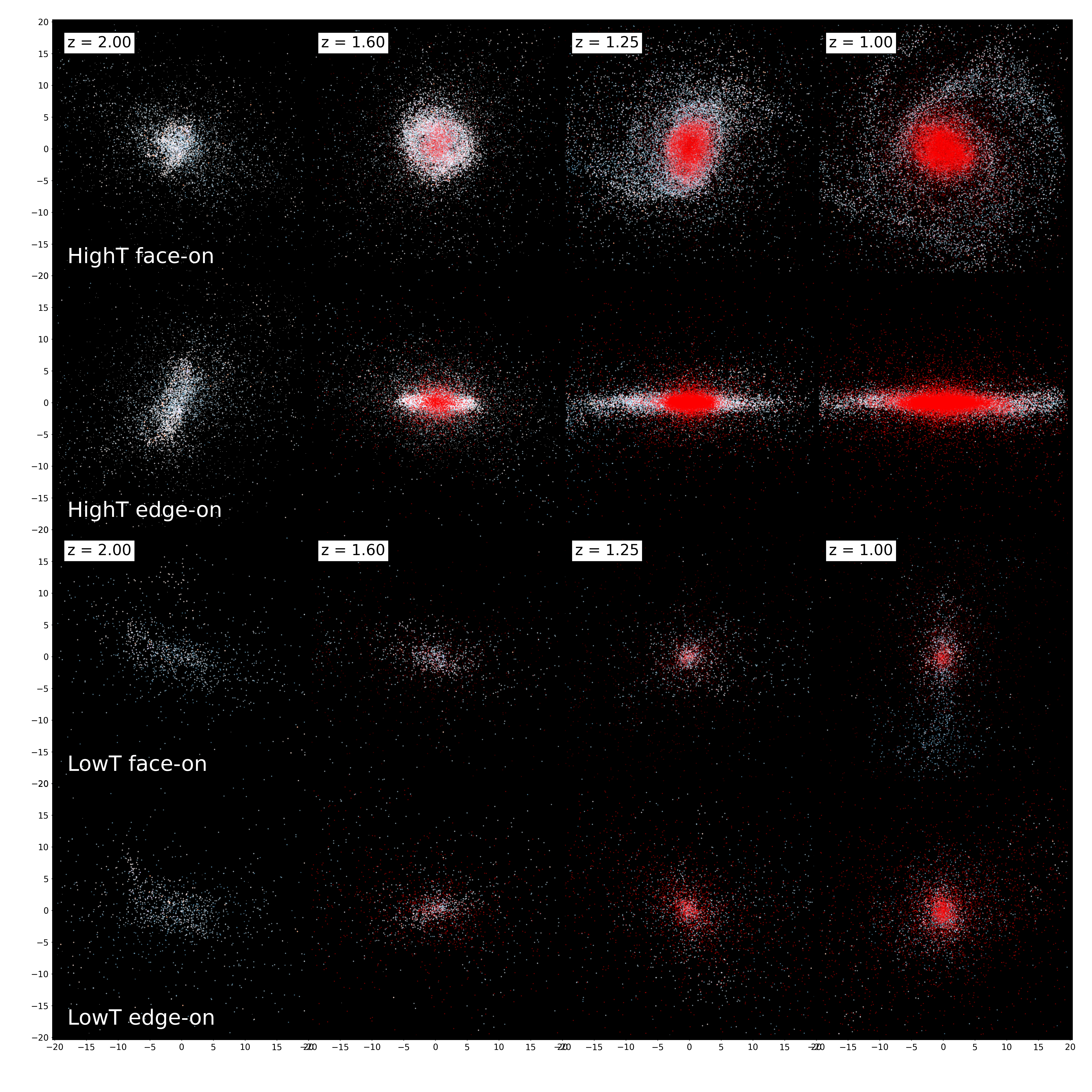}
    \end{minipage}\hfill

    \begin{minipage}[t]{0.48\textwidth}
        \centering
        \includegraphics[width=\linewidth]{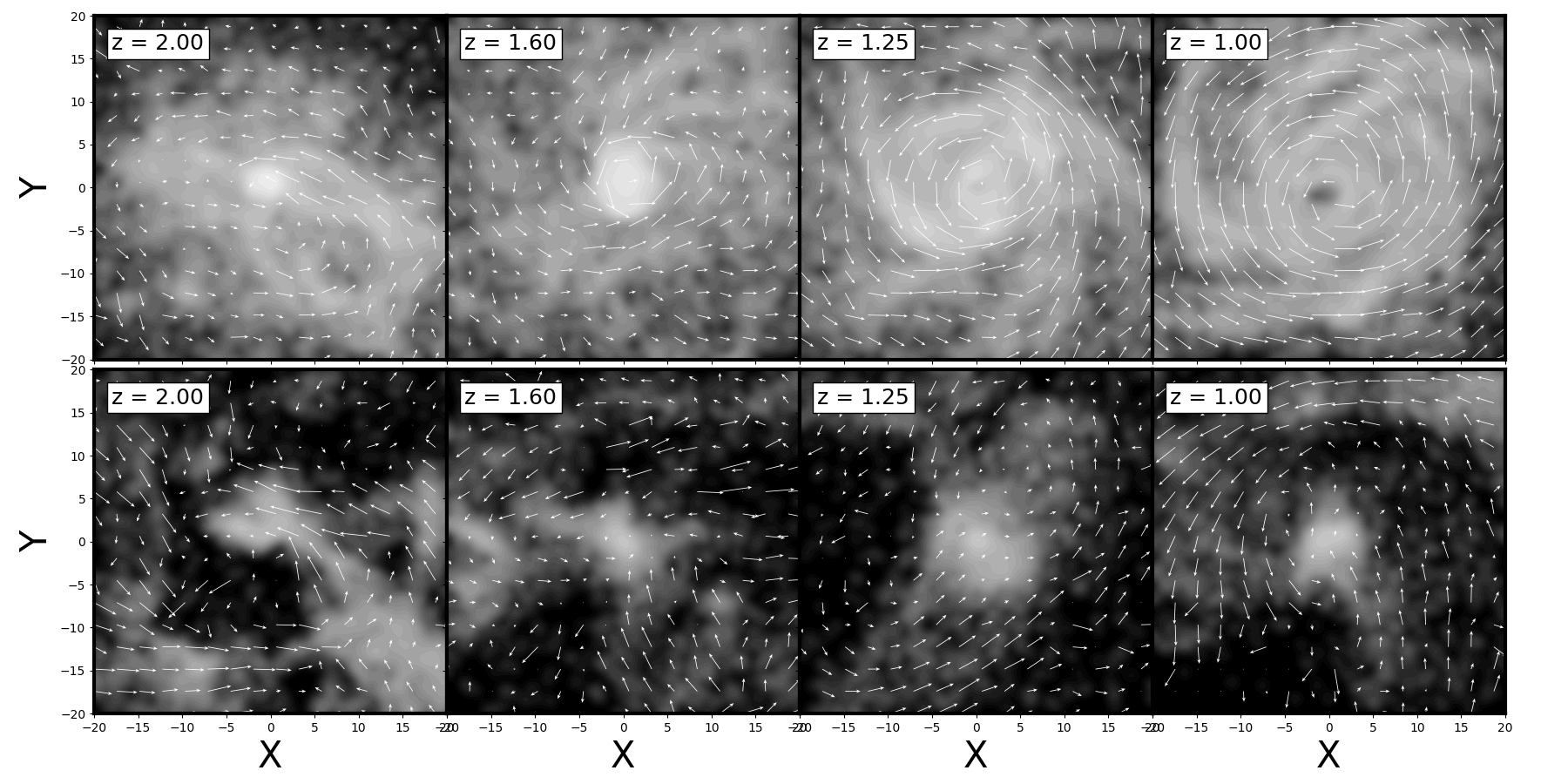}
    \end{minipage}
    \caption{Evolution of a matched galaxy in the HighT and LowT simulations from $z\sim2$ to $z\sim1$. Upper: stellar distributions in face-on and edge-on projections. Red points mark stars already present in the previous snapshot, and white points mark newly formed stars. Lower: gas surface-density maps with velocity vectors, shown in the same face-on orientation as the stellar maps. The HighT run develops a denser and more coherent rotating gas disc, which supports continued in-situ disc growth, whereas the LowT run remains more disturbed and less rotationally supported.}
    \label{fig:sn_gal}
\end{figure}

\begin{figure}[htbp]
    \centering
    
    \begin{minipage}[c]{0.42\textwidth}
        \centering
        \includegraphics[width=\linewidth]{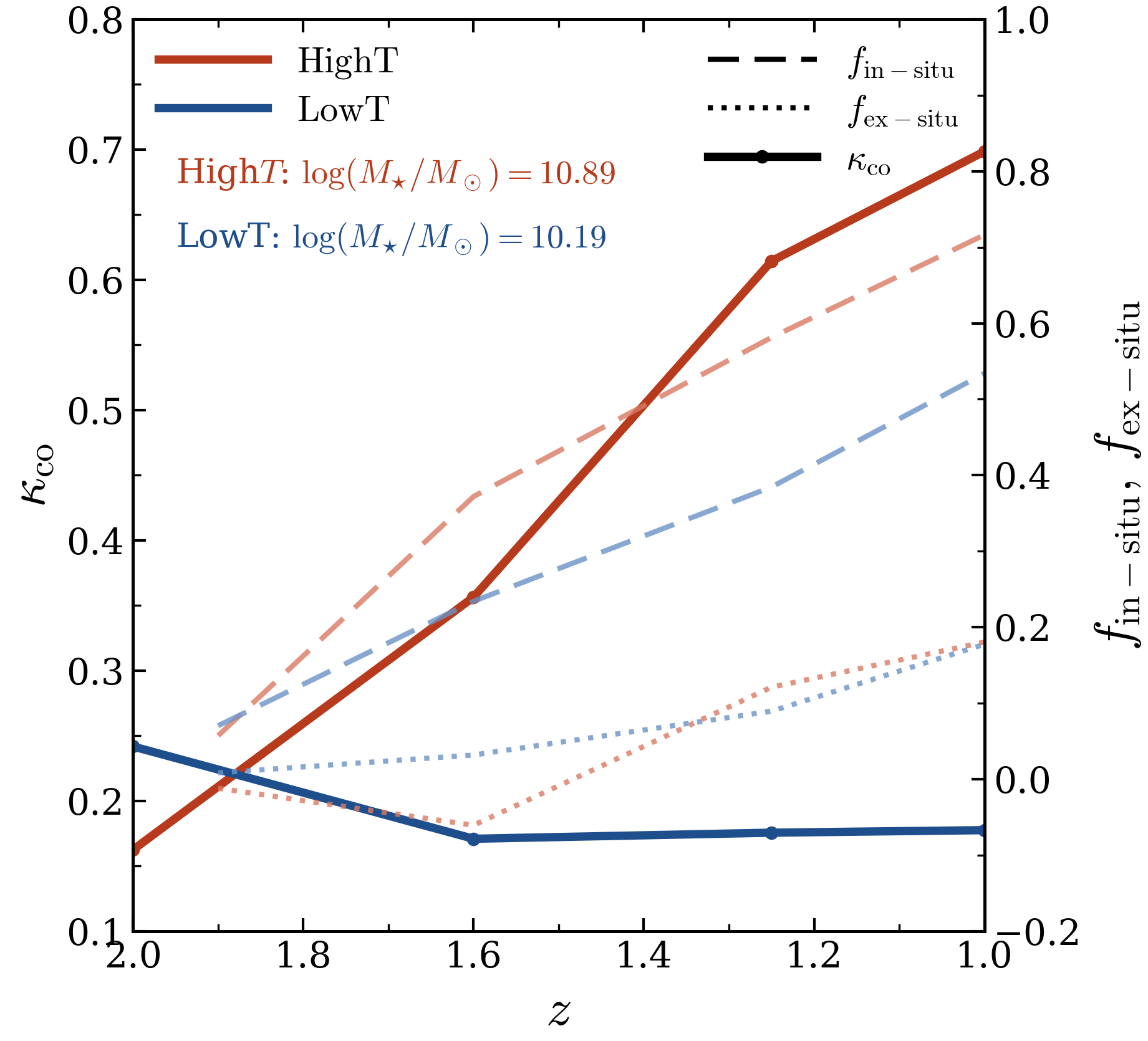}
    \end{minipage}
    \caption{Evolution of $\kappa_{\rm co}$ (solid), together with the in-situ (dashed) and ex-situ (dotted) stellar mass fractions for the galaxies in Fig~\ref{fig:sn_gal}.}

    \label{fig:sn_f}
\end{figure}

Figures~\ref{fig:random_forest} and \ref{fig:Spearman} show that, in low-mass galaxies, morphology is most sensitive to the stellar-feedback parameters, especially the supernova temperature $T_{\rm SN}$. At first sight, this trend appears counterintuitive: one might expect a larger $T_{\rm SN}$ to correspond to stronger feedback, weaker star formation, and therefore a larger spheroidal fraction. Instead, VTNG shows the opposite behaviour. Higher $T_{\rm SN}$ is associated with higher $\kappa_{\rm co}$ and a larger disc fraction. This indicates that the dominant role of $T_{\rm SN}$ is not the instantaneous suppression of star formation, but the regulation of how gas accumulates, cools, and settles before forming stars.

To illustrate this mechanism, we compare a matched galaxy in two realizations with high and low $T_{\rm SN}$, denoted HighT and LowT in Figure~\ref{fig:sn_gal}. Because the simulations share identical initial conditions, the corresponding systems can be identified by matching the dark matter particles in their host subhaloes. At $z=2$, both galaxies are spheroid-dominated. By $z=1$, however, the HighT system has evolved into a disc-dominated galaxy, whereas the LowT system shows only limited structural change. The increase in $\kappa_{\rm co}$ in the HighT run is accompanied by a substantial contribution from newly formed stars distributed in a flattened, extended component. In the LowT run, newly formed stars remain less ordered and contribute less efficiently to disc growth.

We further divide the stellar mass increase between two snapshots into the contributions from the in-situ star formation and from ex-situ stars of accretion or mergers. In the upper right panel of Fig.~\ref{fig:sn_f}, the dashed lines denote the in-situ stellar mass fraction, while the dotted lines represent the ex-situ fraction. The in-situ/ex-situ decomposition indicates that disc growth in the HighT case is driven primarily by in-situ star formation rather than by a simple rearrangement of the pre-existing stellar component. The HighT galaxy maintains a systematically larger in-situ contribution over $1<z<2$, implying that a larger fraction of its stellar mass is assembled from gas that remains within the main galaxy and forms stars in an ordered configuration. This growth mode is naturally more favourable for the buildup of a rotationally supported stellar disc.

This behaviour can be understood in the context of the {\sc Arepo} multiphase star-formation model \citep{2003MNRAS.339..289S}. 
In this framework, $T_{\rm SN}$ enters through the effective supernova energy scale and regulates the star-formation threshold density. A larger $T_{\rm SN}$ raises the threshold density, delaying the onset of star formation. Gas can then accumulate for a longer time before being converted into stars, allowing a denser and more coherent gas disc to form. Once that threshold is reached, star formation proceeds in a gas reservoir that is already more centrally concentrated and more rotationally supported. We further verify this picture by measuring the gas cooling rate within the inner $10\,{\rm kpc}$. From $z\sim2$ to $z\sim1$, the high-$T_{\rm SN}$ simulations exhibit systematically higher cooling rates than the low-$T_{\rm SN}$ runs. This is consistent with the AREPO cooling model, in which the net cooling rate depends on gas density and metallicity. The higher cooling efficiency in the high-$T_{\rm SN}$ runs is therefore driven by both the larger central gas density and the stronger metal enrichment associated with subsequent star formation, which enhances metal-line cooling. By contrast, in low-$T_{\rm SN}$ runs, star formation is triggered earlier, producing outflows and turbulence before the gas has settled into a stable disc.

The gas maps in the lower panels of Figure~\ref{fig:sn_gal} support this interpretation. In the LowT case, the gas is more diffuse and the velocity field is more disturbed, with little sign of coherent rotation. In the HighT case, the gas is more centrally concentrated and exhibits a much more ordered rotational pattern within the inner $\sim 20$ kpc. The stellar component therefore inherits the kinematic state of the star-forming gas: when the gas settles into a rotationally supported configuration, the newly formed stars build a disc; when the gas remains turbulent or misaligned, disc growth is much less efficient.

Overall, VTNG shows that stronger supernova heating does not simply suppress disc formation. Instead, by postponing early star formation and reshaping the gas cycle, a higher $T_{\rm SN}$ can promote the buildup of a denser and more rotationally supported gas reservoir, from which a dynamically colder stellar disc subsequently forms.

\begin{figure}
\centering
    \begin{minipage}[c]{0.48\textwidth}
        \centering
        \includegraphics[width=\linewidth]{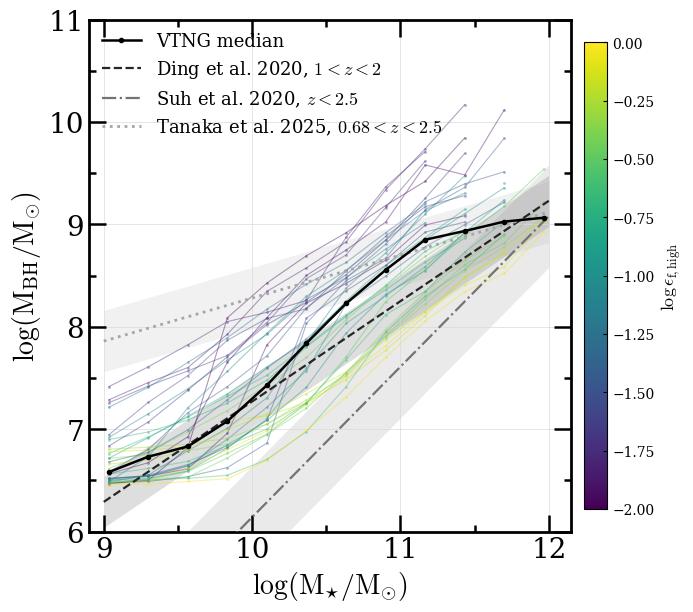}
    \end{minipage}
    \hfill
    \begin{minipage}[c]{0.48\textwidth}
        \centering
        \includegraphics[width=\linewidth]{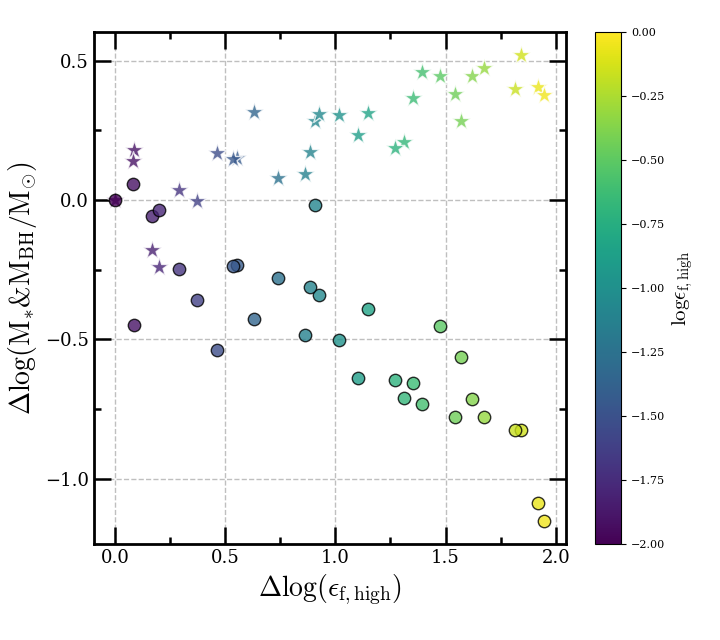}
    \end{minipage}
\caption{Upper panel: Relation between stellar mass and black-hole mass in the VTNG simulations for different values of the quasar-mode coupling efficiency $\epsilon_{\rm f,high}$. Each curve corresponds to one realization, colour-coded by $\log \epsilon_{\rm f,high}$. At fixed stellar mass, larger $\epsilon_{\rm f,high}$ produces systematically smaller black-hole masses, showing that strong early quasar-mode feedback suppresses subsequent black-hole growth. For comparison, we also show the observation results from \cite{2020ApJ...888...37D,2020ApJ...889...32S,2025ApJ...979..215T}.  Lower panel: Relative variations of the median stellar mass and median black-hole mass as a function of $\epsilon_{\rm f,high}$. Taking the simulation with the lowest $\epsilon_{\rm f,high}$ as the reference model, we select galaxies with $\rm M_{*}>10^{11}\,{\rm M_{\odot}}$ and compare their matched counterparts across the simulation suite. The plotted quantities show the changes in the median stellar mass(star) and median black-hole mass(circle) relative to the reference simulation.}\label{fig:metal_bhmass}
\end{figure}

\begin{figure}
\centering
\includegraphics[width=0.85\linewidth]{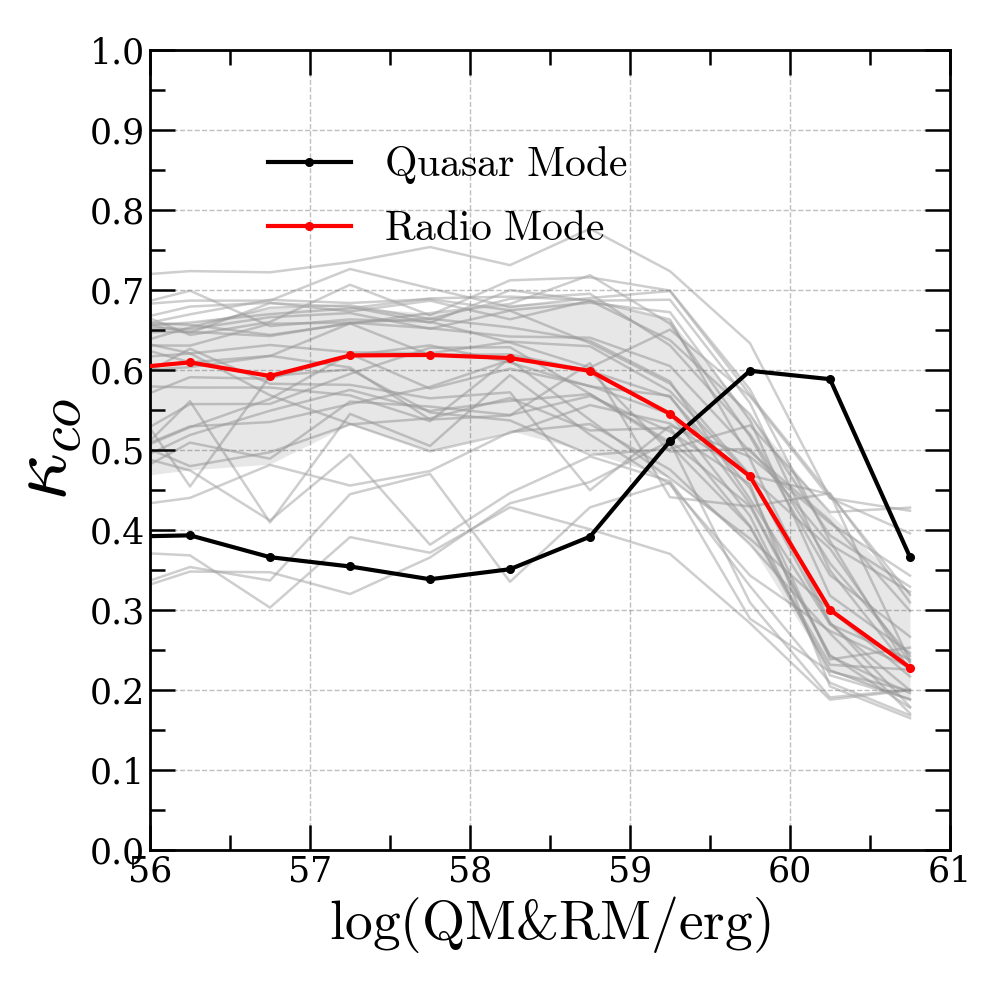}
\caption{Dependence of $\kappa_{\rm co}$ on the cumulative AGN energy released in the quasar and radio modes. The black and red curves show the median trends for the quasar-mode and radio-mode energies, respectively. The shaded region indicates the 25th--75th percentile range for the radio-mode relation, and the faint grey curves show the radio-mode trends in individual realizations. The radio-mode energy correlates much more directly with morphology, whereas the quasar-mode relation is weaker and non-monotonic.}\label{fig:QRM}
\end{figure}

\begin{figure}
\centering
\includegraphics[width=0.95\linewidth]{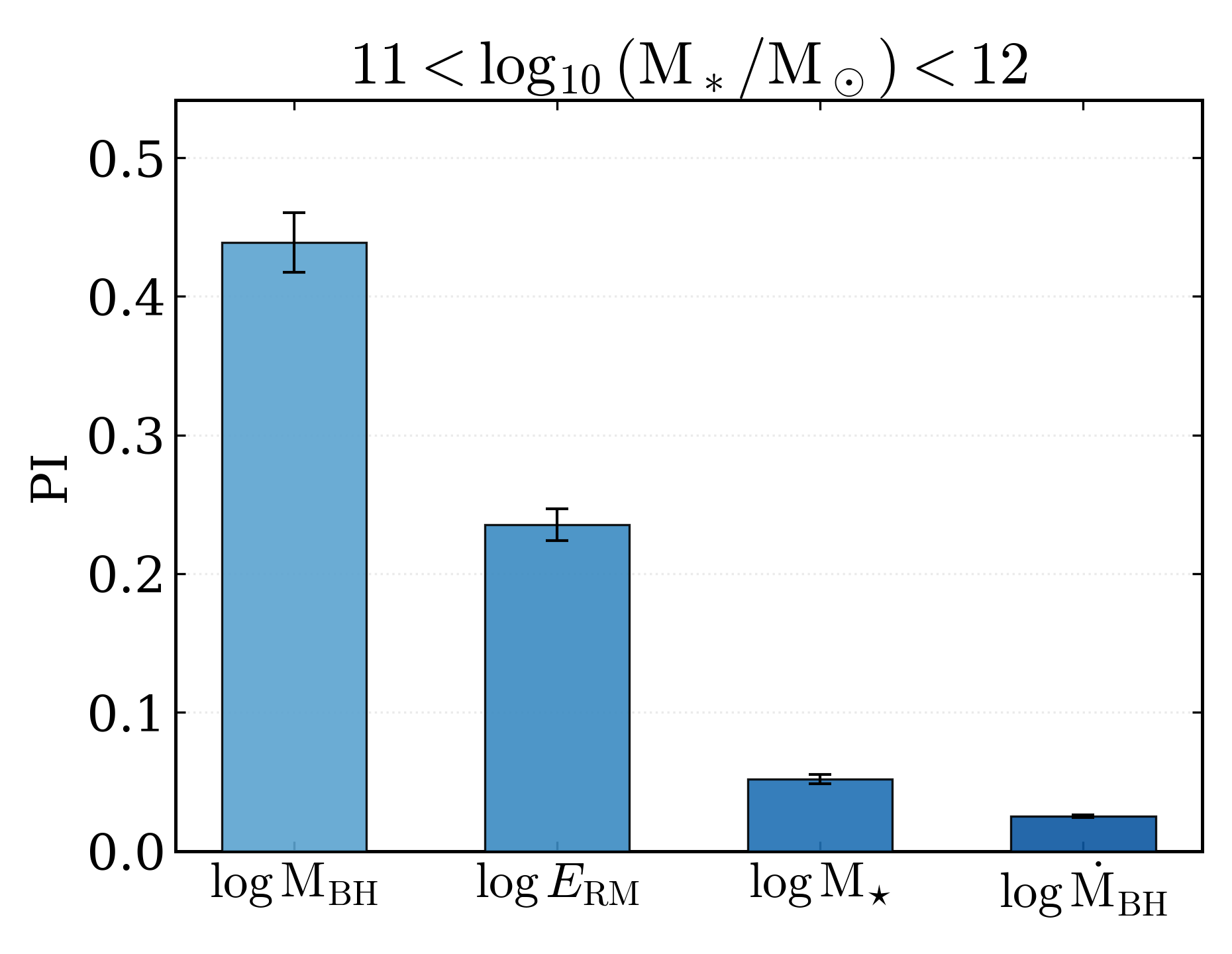}
\caption{Permutation importance from the random-forest analysis for predicting $\kappa_{\rm co}$ in galaxies with $11 < \log_{10}(M_\ast/{\rm M_\odot}) < 12$. The input variables are $\log M_{\rm BH}$, the cumulative radio-mode AGN energy $\log E_{\rm RM}$, stellar mass $\log M_\ast$, and the instantaneous black-hole accretion rate $\log \dot{M}_{\rm BH}$. Error bars show the scatter across the validation folds. Black-hole mass is the most informative predictor, followed by the cumulative radio-mode energy.}\label{fig:bh_rf}
\end{figure}

\begin{figure}[htbp]
    \centering
    
    \begin{minipage}[c]{0.46\textwidth}
        \centering
        \includegraphics[width=\linewidth]{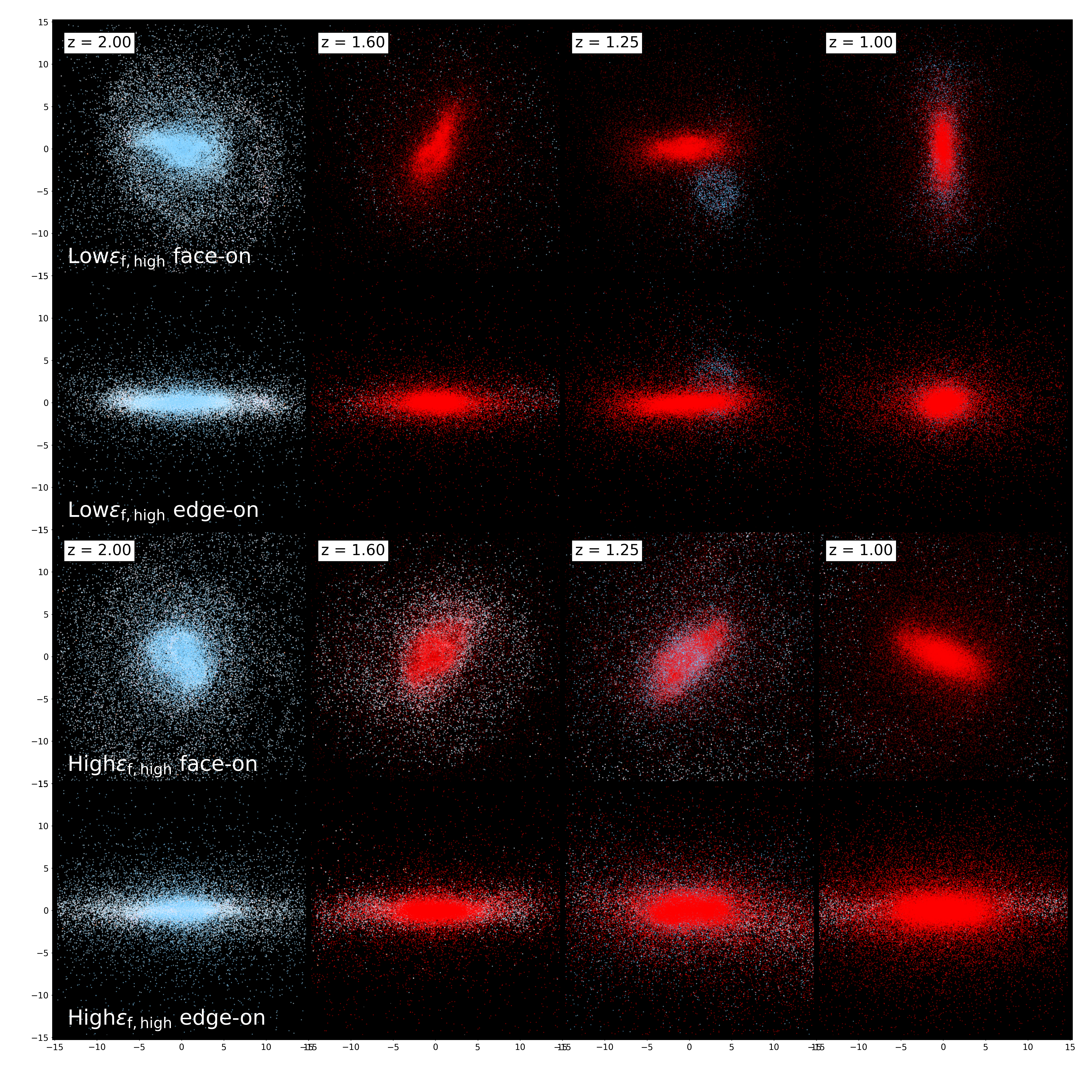}
    \end{minipage}\hfill

    \begin{minipage}[t]{0.48\textwidth}
        \centering
        \includegraphics[width=\linewidth]{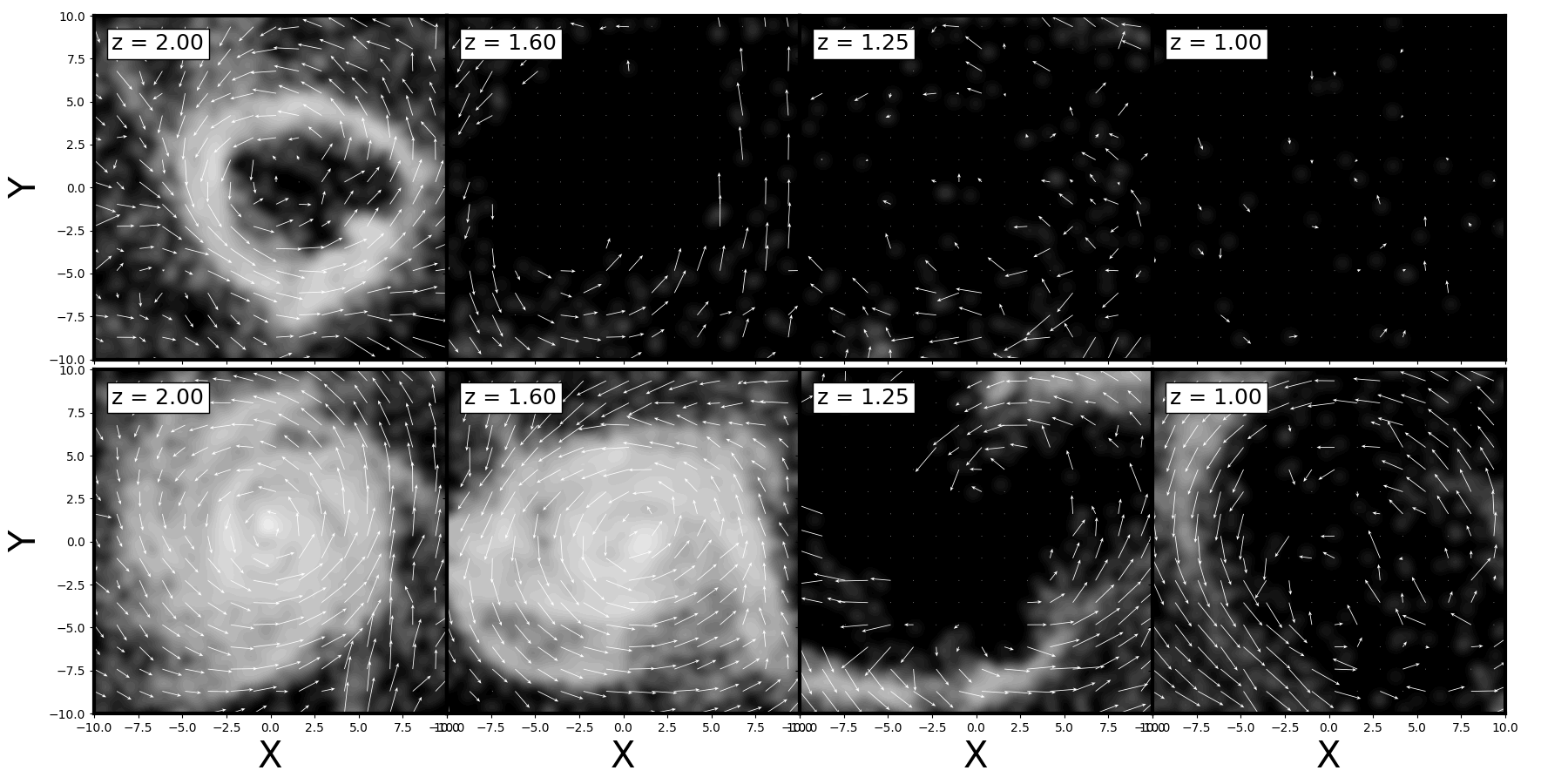}
    \end{minipage}
    \caption{Evolution of a matched galaxy in the High$\epsilon_{\rm f,high}$ and Low$\epsilon_{\rm f,high}$ simulations from $z\sim2$ to $z\sim1$. Upper: stellar distributions in face-on and edge-on projections. Red points mark stars already present in the previous snapshot, and blue points mark newly formed stars.  Lower: gas surface-density maps with velocity vectors, shown in the same face-on orientation as the stellar images. The High$\epsilon_{\rm f,high}$ run retains a more coherent gas disc and sustains later in-situ star formation, whereas the Low$\epsilon_{\rm f,high}$ run becomes more gas-poor and more dispersion-dominated.}
    \label{fig:agn_gal}
\end{figure}

\begin{figure}[htbp]
    \centering

    \begin{minipage}[c]{0.42\textwidth}
        \centering
        \includegraphics[width=\linewidth]{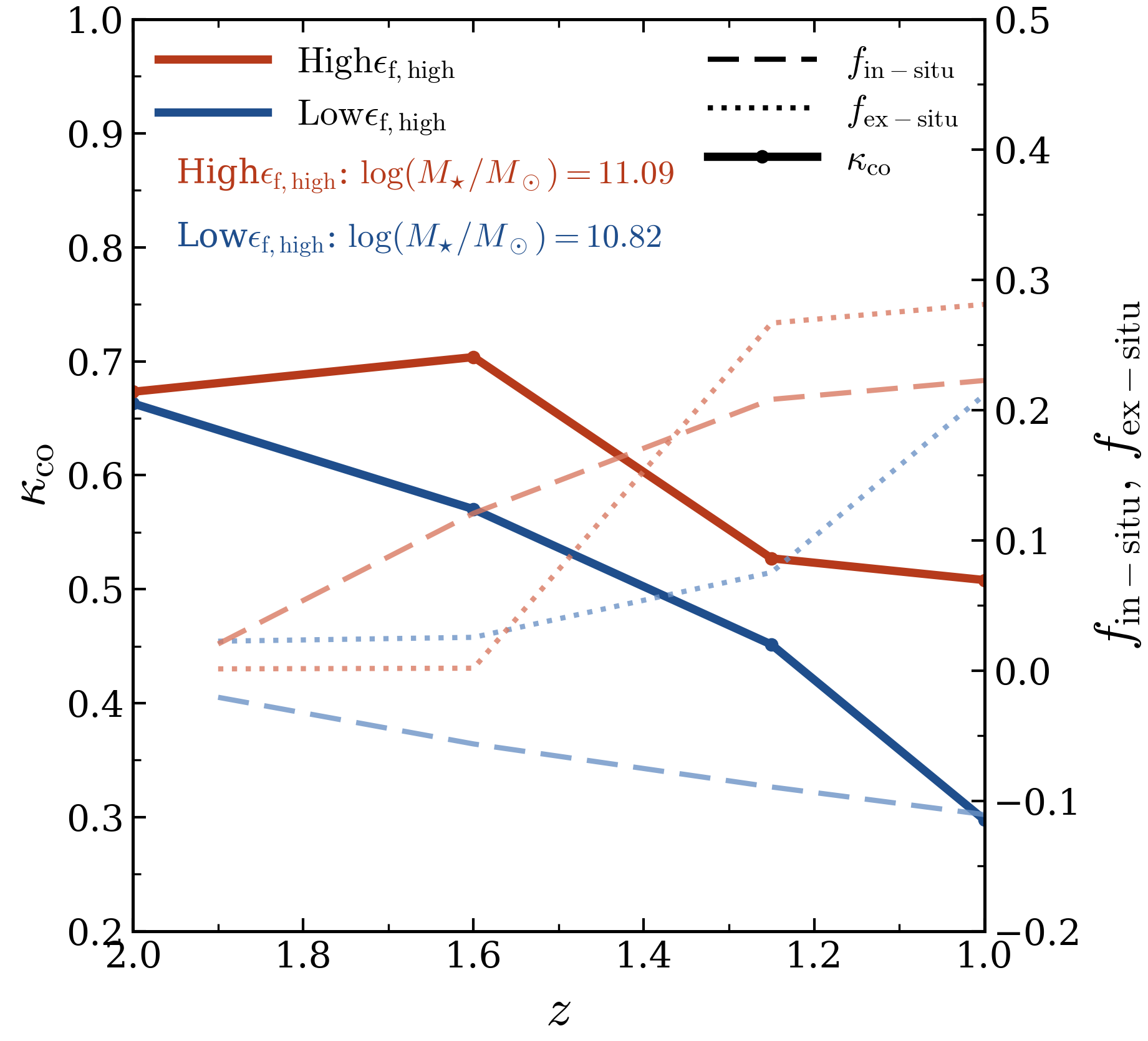}
    \end{minipage}

    \caption{Evolution of $\kappa_{\rm co}$ (solid), together with the in-situ (dashed) and ex-situ (dotted) stellar mass fractions for the galaxies in Fig~\ref{fig:agn_gal}.}
    \label{fig:agn_f}
\end{figure}

\subsubsection{AGN feedback}

Among the AGN-related parameters varied in VTNG, the quasar-mode coupling efficiency $\epsilon_{\rm f,high}$ shows the strongest connection to galaxy morphology. This parameter is most relevant when black holes are still relatively small but accreting rapidly. A larger $\epsilon_{\rm f,high}$ couples more thermal energy to the surrounding gas during this early phase, which suppresses subsequent black hole growth by reducing the amount of gas available for later accretion. Its influence on morphology is therefore indirect, acting through the later black hole growth history and the feedback channels that become important at later times.

Figure~\ref{fig:metal_bhmass} demonstrates this effect clearly. As shown in the upper panel, at fixed stellar mass, galaxies in realizations with larger $\epsilon_{\rm f,high}$ systematically host less massive black holes. Stronger early quasar-mode feedback heats and expels gas more efficiently, thereby limiting the subsequent growth of the central black hole. Conversely, weaker quasar-mode coupling allows more prolonged gas accretion and produces larger black hole masses at fixed stellar mass. We further compare the simulated $M_{\rm BH}-M_{\star}$ relation with recent observational constraints at $z\sim1$--2 \citep{2020ApJ...888...37D,2020ApJ...889...32S,2025ApJ...979..215T}. The shaded regions indicate the adopted $1\sigma$ scatter around the corresponding observational relations in the $\log M_{\rm BH}$ direction. At the massive end, the median VTNG relation is broadly consistent with these observations, suggesting that the simulations reproduce a reasonable normalization of black-hole growth by $z\sim1$. Since the observational samples are mainly active broad-line AGN, while our relation is measured from the simulated galaxy population, this agreement should be regarded as an approximate consistency check. One may further ask whether variations in $\epsilon_{\rm f,high}$ primarily shift the black-hole mass, the stellar mass, or both. We select galaxies with $\rm M_{*}>10^{11}\,{\rm M_{\odot}}$ in the simulation with the lowest $\epsilon_{\rm f,high}$ and compare their matched counterparts across the simulation suite. The lower panel of Figure~\ref{fig:metal_bhmass} shows that changing $\epsilon_{\rm f,high}$ affects both $\rm M_{*}$ and $\rm M_{\rm BH}$. However, the response in $M_{\rm BH}$ is substantially stronger: the median change in black-hole mass is approximately twice that in stellar mass. This is expected because $\epsilon_{\rm f,high}$ directly regulates the coupling efficiency of quasar-mode feedback and therefore has a more immediate impact on black-hole accretion and growth. A similar trend has been reported in CAMELS, where stronger high-accretion thermal feedback suppresses later AGN activity by limiting black hole growth \citep{2023ApJ...959..136N}.

The next question is which AGN channel is more directly linked to morphology. Figure~\ref{fig:QRM} shows that $\kappa_{\rm co}$ correlates much more clearly with the cumulative radio-mode energy than with the cumulative quasar-mode energy. Once the cumulative radio-mode energy exceeds $\sim10^{59}\,{\rm erg}$, morphology becomes strongly dependent on the integrated strength of radio-mode feedback. By contrast, the quasar-mode relation mainly reflects the underlying black hole mass scale at which radio-mode feedback begins to dominate. Taken together with Figure~\ref{fig:metal_bhmass}, this suggests a two-step process: stronger early quasar-mode feedback suppresses black hole growth, which in turn weakens the later radio-mode feedback that more efficiently removes central gas and suppresses rotational support.

We further quantify this connection in Figure~\ref{fig:bh_rf}, where we use a random-forest analysis to predict $\kappa_{\rm co}$ for galaxies with $10^{11}<M_\ast/{\rm M_\odot}<10^{12}$. Among the tested quantities, black hole mass is the most important predictor, followed by the cumulative radio-mode energy. This indicates that morphology in the massive regime is more tightly linked to the long-term black hole growth history and the integrated effect of AGN feedback than to the instantaneous accretion rate. Morphology therefore responds primarily to the time-integrated regulation of the gas reservoir rather than to short-timescale AGN variability.

A representative example is shown in Figures~\ref{fig:agn_gal} and~\ref{fig:agn_f}, where we compare the same galaxy in two simulations with high and low $\epsilon_{\rm f,high}$. At $z\sim2$, both realizations have similar morphology. From $z\sim1.6$ to $z\sim1$, however, the two systems diverge. In the high-$\epsilon_{\rm f,high}$ case, the galaxy retains enough gas to sustain continued in-situ star formation and to preserve an extended stellar disc. In the low-$\epsilon_{\rm f,high}$ case, the gas reservoir is more strongly depleted, so the galaxy cannot counteract the disc-heating effect of mergers and evolves toward a more dispersion-dominated state. The in-situ and ex-situ components therefore act in opposite directions: external accretion tends to weaken rotational support, while continued in-situ star formation rebuilds it.

These results indicate that the two AGN modes play distinct but coupled roles in shaping morphology. Radio-mode feedback is more directly associated with gas removal, kinematic disturbance, and the destruction of disc structure, whereas quasar-mode feedback primarily acts by regulating the later black hole mass and therefore the subsequent radio-mode feedback budget. We therefore conclude that $\epsilon_{\rm f,high}$ is a key control parameter for the morphology of massive galaxies in VTNG, because it sets the long-term pathway of black hole growth, gas depletion, and disc survival.

\begin{figure*}[htbp]
    \centering
    \includegraphics[width=0.42\linewidth]{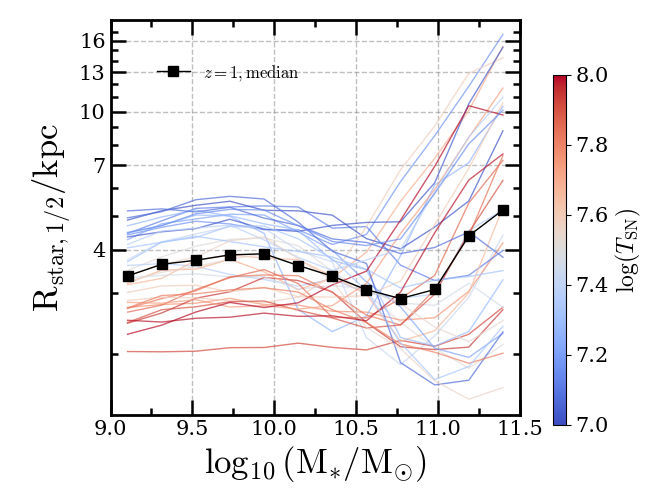}
    \hspace{20pt}
    \includegraphics[width=0.42\linewidth]{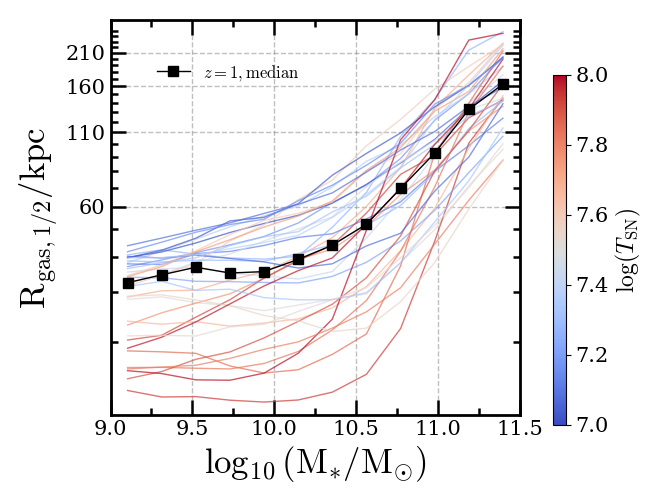}
    \caption{Half-mass radius as a function of stellar mass for the stellar component (left) and gas component (right) at $z=1$. Each curve corresponds to one VTNG realization. For the gas component, we use the total gas associated with each subhalo, including both cold gas and hot/diffuse CGM gas.} In low- and intermediate-mass galaxies, the size distribution is ordered primarily by the supernova temperature $T_{\rm SN}$, whereas the scatter becomes much larger above $M_\ast \gtrsim 10^{10.5}\,{\rm M_\odot}$, indicating an increasing role for AGN feedback in shaping galaxy sizes.
    \label{fig:mass-size}
\end{figure*}

\begin{figure*}[htbp]
    \centering
    \includegraphics[width=1.0\linewidth]{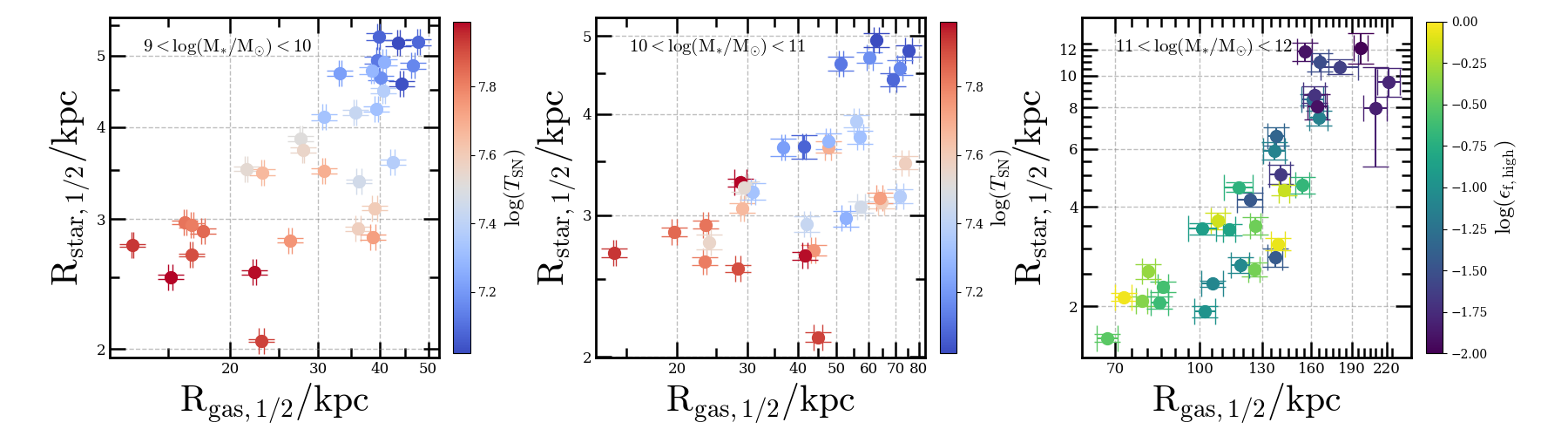}
    \caption{Relation between the stellar and gas half-mass radii in three stellar-mass bins: $9 < \log_{10}(M_\ast/{\rm M_\odot}) < 10$ (left), $10 < \log_{10}(M_\ast/{\rm M_\odot}) < 11$ (middle), and $11 < \log_{10}(M_\ast/{\rm M_\odot}) < 12$ (right). For each simulation, the points show the median stellar and gas sizes, with bootstrap uncertainties indicated by the error bars. The lower-mass panels are colour-coded by $T_{\rm SN}$, while the highest-mass panel is colour-coded by $\epsilon_{\rm f,high}$. Simulations with more extended gas discs also tend to host larger stellar components, linking the final stellar size to the structure of the star-forming gas reservoir.}
    \label{fig:gas-starsize}
\end{figure*}

\subsubsection{Mass-size relation} \label{subsec:masssize}

The mass--size relation provides an additional structural diagnostic of how feedback reshapes galaxies. Observationally, both early- and late-type galaxies follow relatively tight but distinct mass--size relations, with disc galaxies being systematically larger than spheroids at fixed stellar mass \citep{2003MNRAS.343..978S,2014ApJ...788...28V}. Because galaxy size records the cumulative outcome of gas accretion, star formation, feedback, and mergers, it offers an important complementary constraint on the feedback pathways identified above \citep{2015ARA&A..53...51S}. In this paper, we do not attempt a detailed observational calibration of the stellar mass--size relation here; the comparison is intended only as a structural diagnostic of feedback sensitivity.

Figure~\ref{fig:mass-size} shows the stellar and gas half-mass radii as functions of stellar mass for the VTNG realizations. Here the gas component includes all gas associated with the subhalo, without separating the cold star-forming disk from the diffuse halo gas. Therefore, the gas half-mass radius should be interpreted as the characteristic extent of the total gaseous reservoir, including both cold gas and hot/diffuse CGM gas, rather than as the size of the cold gas disk. This definition naturally allows the gas sizes to be much larger than the stellar sizes, especially in massive systems where feedback can redistribute gas to large radii. The scatter is substantial across the full mass range, demonstrating that baryonic feedback strongly affects structural growth. However, this scatter is not random. In the low- and intermediate-mass regimes, the clear dependence on $T_{\rm SN}$ indicates that stellar feedback is an important driver of size variation. In the massive regime, the scatter broadens rapidly and the ordering becomes increasingly sensitive to AGN-related parameters, indicating that size evolution there depends on AGN feedback in addition to merger history.

For $M_\ast \lesssim 10^{10.5}\,{\rm M_\odot}$, the mass--size relation is relatively flat in most realizations, but higher $T_{\rm SN}$ tends to produce larger gas and stellar sizes. This behaviour is consistent with the picture developed in Section~\ref{subsec:sn}: by delaying early star formation, a larger $T_{\rm SN}$ allows gas to settle into a more extended and coherent disc before stars form, which in turn favours the buildup of a larger stellar component. Stellar feedback may also contribute indirectly by driving radial migration through gas inflow and outflow cycles, especially for older stellar populations \citep{2016ApJ...820..131E,2024ApJ...974..247C}.

At higher masses, $M_\ast \gtrsim 10^{10.5}\,{\rm M_\odot}$, the scatter in galaxy size increases dramatically from one feedback realization to another. Some models continue to grow in size, whereas others remain much more compact. This divergence indicates that AGN feedback becomes a major source of size variation in massive galaxies, within an evolutionary context that also includes merger-driven growth. The gas--stellar size relation shown in Figure~\ref{fig:gas-starsize} provides the physical link: realizations with larger gas discs also tend to host larger stellar components, showing that changes in the extent of the star-forming gas reservoir are closely connected to the final stellar size.

Overall, the mass--size relation supports the same picture inferred from the morphology diagnostics. At low masses, structural diversity is governed mainly by stellar feedback through its regulation of gas settling and disc growth. At high masses, the larger scatter and stronger dependence on AGN parameters indicate that AGN feedback becomes the dominant source of structural variation.

\section{Discussion}\label{sec:discuss}

The VTNG results show that galaxy morphology responds strongly to changes in the baryonic feedback model, even under controlled initial conditions. Across the simulation suite, variations in the adopted feedback prescriptions generate substantial diversity in stellar kinematics and structure. This indicates that feedback is an important source of morphological variation within a common cosmological setting.

Our overall picture is broadly consistent with earlier theoretical studies showing that galaxy morphology is shaped not only by mergers, but also by the regulation of gas accretion, gas retention, and subsequent in-situ star formation. In the standard hierarchical framework, gas accreted by dark matter halos cools and settles into rotationally supported discs \citep{1969ApJ...155..393P,1980MNRAS.193..189F,1998MNRAS.295..319M}, while major mergers can disrupt these structures and drive the formation of spheroids \citep{1978MNRAS.184..185W,1992ARA&A..30..705B,2014MNRAS.444.3357N,2017MNRAS.467.3083R}. However, this picture alone is insufficient to explain observed galaxy populations, indicating that baryonic processes regulating star formation must also play a central role. Previous work has shown that mergers do not necessarily produce permanently spheroid-dominated systems, because discs can reform if sufficient gas is retained or reaccreted \citep{2009ApJ...691.1168H,2015MNRAS.452.4347K,2018MNRAS.480.2266M}. Cosmological simulations have also demonstrated that AGN feedback is essential for suppressing renewed disc growth in massive galaxies and for producing the massive spheroid-dominated population \citep{2005ApJ...620L..79S,2016MNRAS.463.3948D}. Our VTNG analysis supports this broader framework, but adds a more controlled perspective by showing how strongly morphology can respond to systematic variations in the feedback model when the initial conditions are fixed.

In the low-mass regime, the VTNG results suggest that the main role of stellar feedback is not simply to suppress star formation, but to regulate the conditions under which stars form. This regulation is supported by some observation results \citep{2019MNRAS.487L..61Z}. The positive correlation between $T_{\rm SN}$ and disc-like morphology indicates that stronger supernova heating can delay early star formation and allow gas to settle into a denser and more rotationally supported configuration before it is converted into stars. In this sense, the resulting discs are built not because feedback is weak, but because feedback reshapes the timing and thermodynamic state of the gas reservoir. This interpretation is broadly consistent with earlier work suggesting that stellar feedback can remove low-angular-momentum gas at early times and thereby facilitate the later growth of disc-like systems \citep{2008MNRAS.389.1137S,2014MNRAS.443.2092U,2015ApJ...804L..40G,2016ApJ...824...79A}. In VTNG, however, the clearest signature appears through the regulation of the star-forming gas from which the stellar disc subsequently grows. Recent cloud-scale observations suggest that stellar feedback can disperse dense star-forming gas on very short timescales after the onset of star formation, implying that feedback can efficiently regulate the central cold-gas structure and subsequent in-situ star formation.

At high masses, the relevant pathway is different. Here the structural state of galaxies is more closely linked to black-hole growth and to the cumulative impact of AGN feedback than to the instantaneous accretion state. Our results indicate that the quasar-mode coupling efficiency $\epsilon_{\rm f,high}$ affects morphology primarily through its long-term influence on black-hole mass and thus on the later strength of radio-mode feedback. This supports a time-integrated view of AGN-driven structural evolution, in which the final morphology depends on the cumulative depletion or preservation of the central gas reservoir. Such a picture is consistent with previous work showing that AGN feedback can suppress star formation, reduce central gas densities, and hinder the survival or regrowth of stellar discs in massive systems \citep{2005ApJ...620L..79S,2016MNRAS.463.3948D,2019MNRAS.489.2702F}. The VTNG suite extends these ideas by showing that different AGN-feedback prescriptions can lead to markedly different structural outcomes even within the same cosmological realization. Some other studies, such as \cite{2022MNRAS.513.3768I}, showed that AGN feedback can also strongly affect Milky-Way-mass galaxies by suppressing central star formation and weakening bar growth, which infer that AGN feedback may shape galaxy morphology at $M_\ast < 10^{11}\,{\rm M_\odot}$.

The structural imprint of feedback is also evident in galaxy sizes. Our results show that the mass-size relation is sensitive to feedback across the full mass range, although the dominant channels differ with stellar mass. At low masses, the size--mass relation is ordered primarily by the supernova temperature, with lower $T_{\rm SN}$ generally producing more extended gas and stellar components. At high masses, the much larger scatter in the VTNG mass-size relation suggests that size evolution reflects the coupled action of AGN feedback and assembly history. In our simulations, galaxies with more extended gas discs also tend to host larger stellar components, indicating that feedback can influence not only whether discs survive, but also how far the stellar distribution grows. AGN-driven outflows may evacuate central gas reservoirs and quench in-situ star formation, producing larger half-mass radii \citep{2018ApJ...866...91C,2023MNRAS.523.2409C}, while major mergers can both destroy discs and deposit stars at large radii, further increasing galaxy sizes \citep{2009ApJ...699L.178N}. In this respect, the VTNG results suggest that feedback affects morphology and size simultaneously, but not always in a one-to-one manner.

Minor mergers are likely to play an important role in the subsequent evolution of massive galaxies, especially below $z\sim1$, where repeated accretion of low-mass companions can increase galaxy sizes, heat stellar discs, and contribute to the build-up of spheroidal components. The impact of such mergers depends not only on the merger mass ratio, but also on the gas content, stellar structure, and orbital properties of the accreted satellites. The most obvious example is in the transformation of red nuggets at z = 2 to giant ellipticals (e.g., M87) at z = 0 \citep{2013ApJ...768L..28H,2013ApJ...766...47H,2016ApJ...821..114H}. These effects are not explicitly isolated in the present analysis. Our results should therefore be interpreted as describing how feedback variations shape galaxy morphology and baryon distributions at $z\sim1$, rather than as a complete model for the later transformation of massive galaxies into their $z\sim0$ descendants. A full assessment of minor-merger-driven size growth and morphological evolution will require following the VTNG galaxies to lower redshift and analysing their merger histories and satellite properties directly.

Although the present work does not attempt a direct comparison with observational morphology measurements, the VTNG trends have potentially important implications for the early emergence of ordered galaxy structures. It is reported that the number of high-redshift massive galaxies is greater than what was expected based on theoretical models \citep{2022ApJ...938L..15C,2023ApJ...948L..14C,2024ApJ...972..143C,2022ApJ...940L..14N,2023MNRAS.518.6011D}. There are added disk and elliptical structures of galaxies, suggesting that the Hubble sequence is present not later than $z\sim6$, which were expected to be more peculiar/irregular at that early epoch\citep{2023ApJ...955...94F,2025A&A...699A.360C}. In this context, our results support a physical picture in which the emergence and survival of disc-like systems depend sensitively on how feedback regulates the gas reservoir available for in-situ growth and disc regrowth. 

At last, the scope of the present paper should be stated clearly. We do not attempt to fully disentangle mergers and feedback as independent drivers of morphology, nor do we provide a formal decomposition of the relative contributions of assembly history and sub-grid physics. Even under identical initial conditions, these effects remain coupled, because feedback alters gas fractions, star-formation histories, and black-hole growth, which in turn modify how galaxies respond to external events. The strength of the VTNG framework is therefore different: it allows us to quantify how strongly morphology responds to feedback variations under controlled initial conditions, and to identify the mass-dependent physical pathways through which this response operates.

\section{Conclusions}\label{sec:conclusion}

We have presented the morphology-focused analysis of the VariableTNG (VTNG) simulation suite, a set of cosmological magnetohydrodynamic simulations based on {\sc Arepo} in which key stellar- and AGN-feedback parameters are systematically varied while the initial conditions are kept fixed. Using the outputs available down to $z\sim1$, we examined the stellar kinematics and structural properties of galaxies with $M_\ast > 10^9\,{\rm M_\odot}$ and traced the physical origin of the resulting morphological diversity. Our main results are as follows.

\begin{enumerate}
    \item Galaxy morphology in VTNG is strongly regulated by baryonic feedback, and the dominant feedback channel depends on stellar mass. Random-forest analysis and partial rank correlation coefficients show that stellar feedback is the primary source of morphological variation below $M_\ast \sim 10^{11}\,{\rm M_\odot}$, with the supernova temperature $T_{\rm SN}$ emerging as the most influential parameter. Above this mass scale, AGN-related parameters become dominant, in particular the quasar-mode coupling efficiency $\epsilon_{\rm f,high}$.

    \item The morphology--mass relation at $z=1$ is non-monotonic. Rotational support, as measured by $\kappa_{\rm co}$, increases from low masses to a maximum near $M_\ast \sim 10^{10.8}\,{\rm M_\odot}$ and declines again toward the massive end. A similar non-monotonic trend is found as a function of black hole mass, indicating a close connection between black hole growth, feedback history, and the structural state of galaxies.

    \item In low- and intermediate-mass galaxies, the structural impact of stellar feedback is mediated primarily through the gas cycle. A higher $T_{\rm SN}$ raises the star-formation threshold density, delays early star formation, and allows gas to accumulate into a denser and more rotationally supported reservoir before stars form. This promotes continued in-situ star formation in a disc-like configuration and leads to higher $\kappa_{\rm co}$ and lower $c/a$.

    \item In massive galaxies, morphology is shaped mainly through the coupling between black hole growth and AGN feedback. A higher $\epsilon_{\rm f,high}$ suppresses early black hole growth, thereby reducing the later radio-mode feedback that is more directly responsible for depleting central gas reservoirs and suppressing rotational support. The morphology of massive galaxies is therefore governed by the cumulative balance between early black hole self-regulation, later kinetic feedback, and the availability of cold gas for renewed disc growth.

    \item The mass--size relation is also strongly affected by feedback. At low masses, galaxy size correlates mainly with stellar-feedback parameters, consistent with the role of $T_{\rm SN}$ in regulating the extent of the gas disc from which stars form. At high masses, the much larger scatter and the stronger dependence on AGN-related parameters indicate that AGN feedback and assembly history dominate the structural evolution.
\end{enumerate}

Taken together, these results show that galaxy morphology in VTNG is regulated by mass-dependent feedback pathways in addition to merger-driven evolution. Stellar feedback primarily governs how discs form and survive in lower-mass galaxies, whereas AGN self-regulation and subsequent radio-mode feedback dominate the structural evolution of massive systems.

\begin{acknowledgements}
This work is supported by the National Natural Science Foundation of China (12595313), the National SKA Program of China (2025SKA0150100), and the CAS Project for Young Scientists in Basic Research (No. YSBR-092). LCH was supported by the National Science Foundation of China (12233001) and the China Manned Space Program (CMS-CSST-2025-A09). We acknowledge the use of the High Performance Computing Resource in the Core Facility for Advanced Research Computing at the Shanghai Astronomical Observatory.
\end{acknowledgements}

\bibliographystyle{aa}
\bibliography{ref}

@ARTICLE{2016ApJ...824...79A,
       author = {{Agertz}, Oscar and {Kravtsov}, Andrey V.},
        title = "{The Impact of Stellar Feedback on the Structure, Size, and Morphology of Galaxies in Milky-Way-sized Dark Matter Halos}",
      journal = {\apj},
         year = 2016,
        month = jun,
       volume = {824},
       number = {2},
          eid = {79},
        pages = {79},
          doi = {10.3847/0004-637X/824/2/79},
archivePrefix = {arXiv},
       eprint = {1509.00853},
 primaryClass = {astro-ph.GA},
       adsurl = {https://ui.adsabs.harvard.edu/abs/2016ApJ...824...79A}
}

@ARTICLE{1992ARA&A..30..705B,
       author = {{Barnes}, Joshua E. and {Hernquist}, Lars},
        title = "{Dynamics of interacting galaxies.}",
      journal = {\araa},
         year = 1992,
        month = jan,
       volume = {30},
        pages = {705-742},
          doi = {10.1146/annurev.aa.30.090192.003421},
       adsurl = {https://ui.adsabs.harvard.edu/abs/1992ARA&A..30..705B}
}

@ARTICLE{2006MNRAS.370..645B,
       author = {{Bower}, R.~G. and {Benson}, A.~J. and {Malbon}, R. and {Helly}, J.~C. and {Frenk}, C.~S. and {Baugh}, C.~M. and {Cole}, S. and {Lacey}, C.~G.},
        title = "{Breaking the hierarchy of galaxy formation}",
      journal = {\mnras},
         year = 2006,
        month = aug,
       volume = {370},
       number = {2},
        pages = {645-655},
          doi = {10.1111/j.1365-2966.2006.10519.x},
archivePrefix = {arXiv},
       eprint = {astro-ph/0511338},
 primaryClass = {astro-ph},
       adsurl = {https://ui.adsabs.harvard.edu/abs/2006MNRAS.370..645B}
}

@ARTICLE{2024ApJ...974..247C,
       author = {{Chamba}, Nushkia and {Marcum}, Pamela M. and {Saintonge}, Am{\'e}lie and {Borlaff}, Alejandro S. and {Hayes}, Matthew J. and {Le Gouellec}, Valentin J.~M. and {Chojnowski}, S. Drew and {Fanelli}, Michael N.},
        title = "{A Break in the Size{\textendash}Stellar Mass Relation: Evidence for Quenching and Feedback in Dwarf Galaxies}",
      journal = {\apj},
         year = 2024,
        month = oct,
       volume = {974},
       number = {2},
          eid = {247},
        pages = {247},
          doi = {10.3847/1538-4357/ad7377},
archivePrefix = {arXiv},
       eprint = {2408.13311},
 primaryClass = {astro-ph.GA},
       adsurl = {https://ui.adsabs.harvard.edu/abs/2024ApJ...974..247C}
}

@ARTICLE{2023MNRAS.523.2409C,
       author = {{Cochrane}, R.~K. and {Angl{\'e}s-Alc{\'a}zar}, D. and {Mercedes-Feliz}, J. and {Hayward}, C.~C. and {Faucher-Gigu{\`e}re}, C. -A. and {Wellons}, S. and {Terrazas}, B.~A. and {Wetzel}, A. and {Hopkins}, P.~F. and {Moreno}, J. and {Su}, K. -Y. and {Somerville}, R.~S.},
        title = "{The impact of AGN-driven winds on physical and observable galaxy sizes}",
      journal = {\mnras},
         year = 2023,
        month = aug,
       volume = {523},
       number = {2},
        pages = {2409-2421},
          doi = {10.1093/mnras/stad1528},
archivePrefix = {arXiv},
       eprint = {2303.12858},
 primaryClass = {astro-ph.GA},
       adsurl = {https://ui.adsabs.harvard.edu/abs/2023MNRAS.523.2409C}
}

@ARTICLE{2017MNRAS.472L..45C,
       author = {{Correa}, Camila A. and {Schaye}, Joop and {Clauwens}, Bart and {Bower}, Richard G. and {Crain}, Robert A. and {Schaller}, Matthieu and {Theuns}, Tom and {Thob}, Adrien C.~R.},
        title = "{The relation between galaxy morphology and colour in the EAGLE simulation}",
      journal = {\mnras},
         year = 2017,
        month = nov,
       volume = {472},
       number = {1},
        pages = {L45-L49},
          doi = {10.1093/mnrasl/slx133},
archivePrefix = {arXiv},
       eprint = {1704.06283},
 primaryClass = {astro-ph.GA},
       adsurl = {https://ui.adsabs.harvard.edu/abs/2017MNRAS.472L..45C}
}

@ARTICLE{2013MNRAS.433.3297D,
       author = {{Dubois}, Yohan and {Gavazzi}, Rapha{\"e}l and {Peirani}, S{\'e}bastien and {Silk}, Joseph},
        title = "{AGN-driven quenching of star formation: morphological and dynamical implications for early-type galaxies}",
      journal = {\mnras},
         year = 2013,
        month = aug,
       volume = {433},
       number = {4},
        pages = {3297-3313},
          doi = {10.1093/mnras/stt997},
archivePrefix = {arXiv},
       eprint = {1301.3092},
 primaryClass = {astro-ph.CO},
       adsurl = {https://ui.adsabs.harvard.edu/abs/2013MNRAS.433.3297D}
}

@ARTICLE{2016MNRAS.463.3948D,
       author = {{Dubois}, Yohan and {Peirani}, S{\'e}bastien and {Pichon}, Christophe and {Devriendt}, Julien and {Gavazzi}, Rapha{\"e}l and {Welker}, Charlotte and {Volonteri}, Marta},
        title = "{The HORIZON-AGN simulation: morphological diversity of galaxies promoted by AGN feedback}",
      journal = {\mnras},
         year = 2016,
        month = dec,
       volume = {463},
       number = {4},
        pages = {3948-3964},
          doi = {10.1093/mnras/stw2265},
archivePrefix = {arXiv},
       eprint = {1606.03086},
 primaryClass = {astro-ph.GA},
       adsurl = {https://ui.adsabs.harvard.edu/abs/2016MNRAS.463.3948D}
}

@ARTICLE{2016ApJ...820..131E,
       author = {{El-Badry}, Kareem and {Wetzel}, Andrew and {Geha}, Marla and {Hopkins}, Philip F. and {Kere{\v{s}}}, Dusan and {Chan}, T.~K. and {Faucher-Gigu{\`e}re}, Claude-Andr{\'e}},
        title = "{Breathing FIRE: How Stellar Feedback Drives Radial Migration, Rapid Size Fluctuations, and Population Gradients in Low-mass Galaxies}",
      journal = {\apj},
         year = 2016,
        month = apr,
       volume = {820},
       number = {2},
          eid = {131},
        pages = {131},
          doi = {10.3847/0004-637X/820/2/131},
archivePrefix = {arXiv},
       eprint = {1512.01235},
 primaryClass = {astro-ph.GA},
       adsurl = {https://ui.adsabs.harvard.edu/abs/2016ApJ...820..131E}
}

@ARTICLE{2018MNRAS.473.1930E,
       author = {{El-Badry}, Kareem and {Quataert}, Eliot and {Wetzel}, Andrew and {Hopkins}, Philip F. and {Weisz}, Daniel R. and {Chan}, T.~K. and {Fitts}, Alex and {Boylan-Kolchin}, Michael and {Kere{\v{s}}}, Du{\v{s}}an and {Faucher-Gigu{\`e}re}, Claude-Andr{\'e} and {Garrison-Kimmel}, Shea},
        title = "{Gas kinematics, morphology and angular momentum in the FIRE simulations}",
      journal = {\mnras},
         year = 2018,
        month = jan,
       volume = {473},
       number = {2},
        pages = {1930-1955},
          doi = {10.1093/mnras/stx2482},
archivePrefix = {arXiv},
       eprint = {1705.10321},
 primaryClass = {astro-ph.GA},
       adsurl = {https://ui.adsabs.harvard.edu/abs/2018MNRAS.473.1930E}
}

@ARTICLE{1980MNRAS.193..189F,
       author = {{Fall}, S.~M. and {Efstathiou}, G.},
        title = "{Formation and rotation of disc galaxies with haloes.}",
      journal = {\mnras},
         year = 1980,
        month = oct,
       volume = {193},
        pages = {189-206},
          doi = {10.1093/mnras/193.2.189},
       adsurl = {https://ui.adsabs.harvard.edu/abs/1980MNRAS.193..189F}
}

@ARTICLE{2015ApJ...804L..40G,
       author = {{Genel}, Shy and {Fall}, S. Michael and {Hernquist}, Lars and {Vogelsberger}, Mark and {Snyder}, Gregory F. and {Rodriguez-Gomez}, Vicente and {Sijacki}, Debora and {Springel}, Volker},
        title = "{Galactic Angular Momentum in the Illustris Simulation: Feedback and the Hubble Sequence}",
      journal = {\apjl},
         year = 2015,
        month = may,
       volume = {804},
       number = {2},
          eid = {L40},
        pages = {L40},
          doi = {10.1088/2041-8205/804/2/L40},
archivePrefix = {arXiv},
       eprint = {1503.01117},
 primaryClass = {astro-ph.GA},
       adsurl = {https://ui.adsabs.harvard.edu/abs/2015ApJ...804L..40G}
}

@ARTICLE{2009ApJS..182..216K,
       author = {{Kormendy}, John and {Fisher}, David B. and {Cornell}, Mark E. and {Bender}, Ralf},
        title = "{Structure and Formation of Elliptical and Spheroidal Galaxies}",
      journal = {\apjs},
         year = 2009,
        month = may,
       volume = {182},
       number = {1},
        pages = {216-309},
          doi = {10.1088/0067-0049/182/1/216},
archivePrefix = {arXiv},
       eprint = {0810.1681},
 primaryClass = {astro-ph},
       adsurl = {https://ui.adsabs.harvard.edu/abs/2009ApJS..182..216K}
}

@ARTICLE{2018ApJ...866...91C,
       author = {{Choi}, Ena and {Somerville}, Rachel S. and {Ostriker}, Jeremiah P. and {Naab}, Thorsten and {Hirschmann}, Michaela},
        title = "{The Role of Black Hole Feedback on Size and Structural Evolution in Massive Galaxies}",
      journal = {\apj},
         year = 2018,
        month = oct,
       volume = {866},
       number = {2},
          eid = {91},
        pages = {91},
          doi = {10.3847/1538-4357/aae076},
archivePrefix = {arXiv},
       eprint = {1809.02143},
 primaryClass = {astro-ph.GA},
       adsurl = {https://ui.adsabs.harvard.edu/abs/2018ApJ...866...91C}
}

@ARTICLE{2018MNRAS.480.2266M,
       author = {{Martin}, G. and {Kaviraj}, S. and {Devriendt}, J.~E.~G. and {Dubois}, Y. and {Pichon}, C.},
        title = "{The role of mergers in driving morphological transformation over cosmic time}",
      journal = {\mnras},
         year = 2018,
        month = oct,
       volume = {480},
       number = {2},
        pages = {2266-2283},
          doi = {10.1093/mnras/sty1936},
archivePrefix = {arXiv},
       eprint = {1807.08761},
 primaryClass = {astro-ph.GA},
       adsurl = {https://ui.adsabs.harvard.edu/abs/2018MNRAS.480.2266M}
}

@ARTICLE{1998MNRAS.295..319M,
       author = {{Mo}, H.~J. and {Mao}, Shude and {White}, Simon D.~M.},
        title = "{The formation of galactic discs}",
      journal = {\mnras},
         year = 1998,
        month = apr,
       volume = {295},
       number = {2},
        pages = {319-336},
          doi = {10.1046/j.1365-8711.1998.01227.x},
archivePrefix = {arXiv},
       eprint = {astro-ph/9707093},
 primaryClass = {astro-ph},
       adsurl = {https://ui.adsabs.harvard.edu/abs/1998MNRAS.295..319M}
}

@ARTICLE{2014MNRAS.444.3357N,
       author = {{Naab}, Thorsten and {Oser}, L. and {Emsellem}, E. and {Cappellari}, Michele and {Krajnovi{\'c}}, D. and {McDermid}, R.~M. and {Alatalo}, K. and {Bayet}, E. and {Blitz}, L. and {Bois}, M. and {Bournaud}, F. and {Bureau}, M. and {Crocker}, A. and {Davies}, R.~L. and {Davis}, T.~A. and {de Zeeuw}, P.~T. and {Duc}, P. -A. and {Hirschmann}, M. and {Johansson}, P.~H. and {Khochfar}, S. and {Kuntschner}, H. and {Morganti}, R. and {Oosterloo}, T. and {Sarzi}, M. and {Scott}, N. and {Serra}, P. and {van de Ven}, G. and {Weijmans}, A. and {Young}, L.~M.},
        title = "{The ATLAS$^{3D}$ project - XXV. Two-dimensional kinematic analysis of simulated galaxies and the cosmological origin of fast and slow rotators}",
      journal = {\mnras},
         year = 2014,
        month = nov,
       volume = {444},
       number = {4},
        pages = {3357-3387},
          doi = {10.1093/mnras/stt1919},
archivePrefix = {arXiv},
       eprint = {1311.0284},
 primaryClass = {astro-ph.CO},
       adsurl = {https://ui.adsabs.harvard.edu/abs/2014MNRAS.444.3357N}
}

@ARTICLE{1969ApJ...155..393P,
       author = {{Peebles}, P.~J.~E.},
        title = "{Origin of the Angular Momentum of Galaxies}",
      journal = {\apj},
         year = 1969,
        month = feb,
       volume = {155},
        pages = {393},
          doi = {10.1086/149876},
       adsurl = {https://ui.adsabs.harvard.edu/abs/1969ApJ...155..393P}
}

@ARTICLE{2017MNRAS.467.3083R,
       author = {{Rodriguez-Gomez}, Vicente and {Sales}, Laura V. and {Genel}, Shy and {Pillepich}, Annalisa and {Zjupa}, Jolanta and {Nelson}, Dylan and {Griffen}, Brendan and {Torrey}, Paul and {Snyder}, Gregory F. and {Vogelsberger}, Mark and {Springel}, Volker and {Ma}, Chung-Pei and {Hernquist}, Lars},
        title = "{The role of mergers and halo spin in shaping galaxy morphology}",
      journal = {\mnras},
         year = 2017,
        month = may,
       volume = {467},
       number = {3},
        pages = {3083-3098},
          doi = {10.1093/mnras/stx305},
archivePrefix = {arXiv},
       eprint = {1609.09498},
 primaryClass = {astro-ph.GA},
       adsurl = {https://ui.adsabs.harvard.edu/abs/2017MNRAS.467.3083R}
}

@ARTICLE{2012MNRAS.423.1544S,
       author = {{Sales}, Laura V. and {Navarro}, Julio F. and {Theuns}, Tom and {Schaye}, Joop and {White}, Simon D.~M. and {Frenk}, Carlos S. and {Crain}, Robert A. and {Dalla Vecchia}, Claudio},
        title = "{The origin of discs and spheroids in simulated galaxies}",
      journal = {\mnras},
         year = 2012,
        month = jun,
       volume = {423},
       number = {2},
        pages = {1544-1555},
          doi = {10.1111/j.1365-2966.2012.20975.x},
archivePrefix = {arXiv},
       eprint = {1112.2220},
 primaryClass = {astro-ph.CO},
       adsurl = {https://ui.adsabs.harvard.edu/abs/2012MNRAS.423.1544S}
}

@ARTICLE{2008MNRAS.389.1137S,
       author = {{Scannapieco}, Cecilia and {Tissera}, Patricia B. and {White}, Simon D.~M. and {Springel}, Volker},
        title = "{Effects of supernova feedback on the formation of galaxy discs}",
      journal = {\mnras},
         year = 2008,
        month = sep,
       volume = {389},
       number = {3},
        pages = {1137-1149},
          doi = {10.1111/j.1365-2966.2008.13678.x},
archivePrefix = {arXiv},
       eprint = {0804.3795},
 primaryClass = {astro-ph},
       adsurl = {https://ui.adsabs.harvard.edu/abs/2008MNRAS.389.1137S}
}

@ARTICLE{2005ApJ...620L..79S,
       author = {{Springel}, Volker and {Di Matteo}, Tiziana and {Hernquist}, Lars},
        title = "{Black Holes in Galaxy Mergers: The Formation of Red Elliptical Galaxies}",
      journal = {\apjl},
         year = 2005,
        month = feb,
       volume = {620},
       number = {2},
        pages = {L79-L82},
          doi = {10.1086/428772},
archivePrefix = {arXiv},
       eprint = {astro-ph/0409436},
 primaryClass = {astro-ph},
       adsurl = {https://ui.adsabs.harvard.edu/abs/2005ApJ...620L..79S}
}

@ARTICLE{2018RMxAA..54..217S,
       author = {{S{\'a}nchez}, S.~F. and {Avila-Reese}, V. and {Hernandez-Toledo}, H. and {Cortes-Su{\'a}rez}, E. and {Rodr{\'\i}guez-Puebla}, A. and {Ibarra-Medel}, H. and {Cano-D{\'\i}az}, M. and {Barrera-Ballesteros}, J.~K. and {Negrete}, C.~A. and {Calette}, A.~R. and {de Lorenzo-C{\'a}ceres}, A. and {Ortega-Minakata}, R.~A. and {Aquino}, E. and {Valenzuela}, O. and {Clemente}, J.~C. and {Storchi-Bergmann}, T. and {Riffel}, R. and {Schimoia}, J. and {Riffel}, R.~A. and {Rembold}, S.~B. and {Brownstein}, J.~R. and {Pan}, K. and {Yates}, R. and {Mallmann}, N. and {Bitsakis}, T.},
        title = "{SDSS IV MaNGA - Properties of AGN Host Galaxies}",
      journal = {\rmxaa},
         year = 2018,
        month = apr,
       volume = {54},
        pages = {217-260},
          doi = {10.48550/arXiv.1709.05438},
archivePrefix = {arXiv},
       eprint = {1709.05438},
 primaryClass = {astro-ph.GA},
       adsurl = {https://ui.adsabs.harvard.edu/abs/2018RMxAA..54..217S}
}

@ARTICLE{2019MNRAS.487.5416T,
       author = {{Tacchella}, Sandro and {Diemer}, Benedikt and {Hernquist}, Lars and {Genel}, Shy and {Marinacci}, Federico and {Nelson}, Dylan and {Pillepich}, Annalisa and {Rodriguez-Gomez}, Vicente and {Sales}, Laura V. and {Springel}, Volker and {Vogelsberger}, Mark},
        title = "{Morphology and star formation in IllustrisTNG: the build-up of spheroids and discs}",
      journal = {\mnras},
         year = 2019,
        month = aug,
       volume = {487},
       number = {4},
        pages = {5416-5440},
          doi = {10.1093/mnras/stz1657},
archivePrefix = {arXiv},
       eprint = {1904.12860},
 primaryClass = {astro-ph.GA},
       adsurl = {https://ui.adsabs.harvard.edu/abs/2019MNRAS.487.5416T}
}

@ARTICLE{2017MNRAS.465.3291W,
       author = {{Weinberger}, Rainer and {Springel}, Volker and {Hernquist}, Lars and {Pillepich}, Annalisa and {Marinacci}, Federico and {Pakmor}, R{\"u}diger and {Nelson}, Dylan and {Genel}, Shy and {Vogelsberger}, Mark and {Naiman}, Jill and {Torrey}, Paul},
        title = "{Simulating galaxy formation with black hole driven thermal and kinetic feedback}",
      journal = {\mnras},
         year = 2017,
        month = mar,
       volume = {465},
       number = {3},
        pages = {3291-3308},
          doi = {10.1093/mnras/stw2944},
archivePrefix = {arXiv},
       eprint = {1607.03486},
 primaryClass = {astro-ph.GA},
       adsurl = {https://ui.adsabs.harvard.edu/abs/2017MNRAS.465.3291W}
}

@ARTICLE{1978MNRAS.184..185W,
       author = {{White}, S.~D.~M.},
        title = "{Simulations of merging galaxies.}",
      journal = {\mnras},
         year = 1978,
        month = jul,
       volume = {184},
        pages = {185-203},
          doi = {10.1093/mnras/184.2.185},
       adsurl = {https://ui.adsabs.harvard.edu/abs/1978MNRAS.184..185W}
}

@ARTICLE{1984ApJ...286...38W,
       author = {{White}, S.~D.~M.},
        title = "{Angular momentum growth in protogalaxies}",
      journal = {\apj},
         year = 1984,
        month = nov,
       volume = {286},
        pages = {38-41},
          doi = {10.1086/162573},
       adsurl = {https://ui.adsabs.harvard.edu/abs/1984ApJ...286...38W}
}

@ARTICLE{2024MNRAS.532.2558Z,
       author = {{Zeng}, Guangquan and {Wang}, Lan and {Gao}, Liang and {Yang}, Hang},
        title = "{Kinematic morphology of low-mass galaxies in IllustrisTNG}",
      journal = {\mnras},
         year = 2024,
        month = aug,
       volume = {532},
       number = {2},
        pages = {2558-2569},
          doi = {10.1093/mnras/stae1651},
archivePrefix = {arXiv},
       eprint = {2404.14184},
 primaryClass = {astro-ph.GA},
       adsurl = {https://ui.adsabs.harvard.edu/abs/2024MNRAS.532.2558Z}
}

@ARTICLE{2014MNRAS.443.2092U,
       author = {{{\"U}bler}, Hannah and {Naab}, Thorsten and {Oser}, Ludwig and {Aumer}, Michael and {Sales}, Laura V. and {White}, Simon D.~M.},
        title = "{Why stellar feedback promotes disc formation in simulated galaxies}",
      journal = {\mnras},
         year = 2014,
        month = sep,
       volume = {443},
       number = {3},
        pages = {2092-2111},
          doi = {10.1093/mnras/stu1275},
archivePrefix = {arXiv},
       eprint = {1403.6124},
 primaryClass = {astro-ph.GA},
       adsurl = {https://ui.adsabs.harvard.edu/abs/2014MNRAS.443.2092U}
}

@ARTICLE{2019ComAC...6....2N,
       author = {{Nelson}, Dylan and {Springel}, Volker and {Pillepich}, Annalisa and {Rodriguez-Gomez}, Vicente and {Torrey}, Paul and {Genel}, Shy and {Vogelsberger}, Mark and {Pakmor}, Ruediger and {Marinacci}, Federico and {Weinberger}, Rainer and {Kelley}, Luke and {Lovell}, Mark and {Diemer}, Benedikt and {Hernquist}, Lars},
        title = "{The IllustrisTNG simulations: public data release}",
      journal = {Computational Astrophysics and Cosmology},
         year = 2019,
        month = may,
       volume = {6},
       number = {1},
          eid = {2},
        pages = {2},
          doi = {10.1186/s40668-019-0028-x},
archivePrefix = {arXiv},
       eprint = {1812.05609},
 primaryClass = {astro-ph.GA},
       adsurl = {https://ui.adsabs.harvard.edu/abs/2019ComAC...6....2N}
}

@ARTICLE{2018MNRAS.475..676S,
       author = {{Springel}, Volker and {Pakmor}, R{\"u}diger and {Pillepich}, Annalisa and {Weinberger}, Rainer and {Nelson}, Dylan and {Hernquist}, Lars and {Vogelsberger}, Mark and {Genel}, Shy and {Torrey}, Paul and {Marinacci}, Federico and {Naiman}, Jill},
        title = "{First results from the IllustrisTNG simulations: matter and galaxy clustering}",
      journal = {\mnras},
         year = 2018,
        month = mar,
       volume = {475},
       number = {1},
        pages = {676-698},
          doi = {10.1093/mnras/stx3304},
archivePrefix = {arXiv},
       eprint = {1707.03397},
 primaryClass = {astro-ph.GA},
       adsurl = {https://ui.adsabs.harvard.edu/abs/2018MNRAS.475..676S}
}

@ARTICLE{2018MNRAS.475..648P,
       author = {{Pillepich}, Annalisa and {Nelson}, Dylan and {Hernquist}, Lars and {Springel}, Volker and {Pakmor}, R{\"u}diger and {Torrey}, Paul and {Weinberger}, Rainer and {Genel}, Shy and {Naiman}, Jill P. and {Marinacci}, Federico and {Vogelsberger}, Mark},
        title = "{First results from the IllustrisTNG simulations: the stellar mass content of groups and clusters of galaxies}",
      journal = {\mnras},
         year = 2018,
        month = mar,
       volume = {475},
       number = {1},
        pages = {648-675},
          doi = {10.1093/mnras/stx3112},
archivePrefix = {arXiv},
       eprint = {1707.03406},
 primaryClass = {astro-ph.GA},
       adsurl = {https://ui.adsabs.harvard.edu/abs/2018MNRAS.475..648P}
}

@ARTICLE{2013MNRAS.436.3031V,
       author = {{Vogelsberger}, Mark and {Genel}, Shy and {Sijacki}, Debora and {Torrey}, Paul and {Springel}, Volker and {Hernquist}, Lars},
        title = "{A model for cosmological simulations of galaxy formation physics}",
      journal = {\mnras},
         year = 2013,
        month = dec,
       volume = {436},
       number = {4},
        pages = {3031-3067},
          doi = {10.1093/mnras/stt1789},
archivePrefix = {arXiv},
       eprint = {1305.2913},
 primaryClass = {astro-ph.CO},
       adsurl = {https://ui.adsabs.harvard.edu/abs/2013MNRAS.436.3031V}
}

@ARTICLE{2003MNRAS.339..289S,
       author = {{Springel}, Volker and {Hernquist}, Lars},
        title = "{Cosmological smoothed particle hydrodynamics simulations: a hybrid multiphase model for star formation}",
      journal = {\mnras},
         year = 2003,
        month = feb,
       volume = {339},
       number = {2},
        pages = {289-311},
          doi = {10.1046/j.1365-8711.2003.06206.x},
archivePrefix = {arXiv},
       eprint = {astro-ph/0206393},
 primaryClass = {astro-ph},
       adsurl = {https://ui.adsabs.harvard.edu/abs/2003MNRAS.339..289S}
}

@ARTICLE{2023ApJ...959..136N,
       author = {{Ni}, Yueying and {Genel}, Shy and {Angl{\'e}s-Alc{\'a}zar}, Daniel and {Villaescusa-Navarro}, Francisco and {Jo}, Yongseok and {Bird}, Simeon and {Di Matteo}, Tiziana and {Croft}, Rupert and {Chen}, Nianyi and {de Santi}, Natal{\'\i} S.~M. and {Gebhardt}, Matthew and {Shao}, Helen and {Pandey}, Shivam and {Hernquist}, Lars and {Dave}, Romeel},
        title = "{The CAMELS Project: Expanding the Galaxy Formation Model Space with New ASTRID and 28-parameter TNG and SIMBA Suites}",
      journal = {\apj},
         year = 2023,
        month = dec,
       volume = {959},
       number = {2},
          eid = {136},
        pages = {136},
          doi = {10.3847/1538-4357/ad022a},
archivePrefix = {arXiv},
       eprint = {2304.02096},
 primaryClass = {astro-ph.CO},
       adsurl = {https://ui.adsabs.harvard.edu/abs/2023ApJ...959..136N}
}

@ARTICLE{2018MNRAS.473.4077P,
       author = {{Pillepich}, Annalisa and {Springel}, Volker and {Nelson}, Dylan and {Genel}, Shy and {Naiman}, Jill and {Pakmor}, R{\"u}diger and {Hernquist}, Lars and {Torrey}, Paul and {Vogelsberger}, Mark and {Weinberger}, Rainer and {Marinacci}, Federico},
        title = "{Simulating galaxy formation with the IllustrisTNG model}",
      journal = {\mnras},
         year = 2018,
        month = jan,
       volume = {473},
       number = {3},
        pages = {4077-4106},
          doi = {10.1093/mnras/stx2656},
archivePrefix = {arXiv},
       eprint = {1703.02970},
 primaryClass = {astro-ph.GA},
       adsurl = {https://ui.adsabs.harvard.edu/abs/2018MNRAS.473.4077P}
}

@ARTICLE{2023ApJ...955...94F,
       author = {{Ferreira}, Leonardo and {Conselice}, Christopher J. and {Sazonova}, Elizaveta and {Ferrari}, Fabricio and {Caruana}, Joseph and {Tohill}, Cl{\'a}r-Br{\'\i}d and {Lucatelli}, Geferson and {Adams}, Nathan and {Irodotou}, Dimitrios and {Marshall}, Madeline A. and {Roper}, Will J. and {Lovell}, Christopher C. and {Verma}, Aprajita and {Austin}, Duncan and {Trussler}, James and {Wilkins}, Stephen M.},
        title = "{The JWST Hubble Sequence: The Rest-frame Optical Evolution of Galaxy Structure at 1.5 < z < 6.5}",
      journal = {\apj},
         year = 2023,
        month = oct,
       volume = {955},
       number = {2},
          eid = {94},
        pages = {94},
          doi = {10.3847/1538-4357/acec76},
archivePrefix = {arXiv},
       eprint = {2210.01110},
 primaryClass = {astro-ph.GA},
       adsurl = {https://ui.adsabs.harvard.edu/abs/2023ApJ...955...94F}
}

@ARTICLE{2014ApJ...788...28V,
       author = {{van der Wel}, A. and {Franx}, M. and {van Dokkum}, P.~G. and {Skelton}, R.~E. and {Momcheva}, I.~G. and {Whitaker}, K.~E. and {Brammer}, G.~B. and {Bell}, E.~F. and {Rix}, H. -W. and {Wuyts}, S. and {Ferguson}, H.~C. and {Holden}, B.~P. and {Barro}, G. and {Koekemoer}, A.~M. and {Chang}, Yu-Yen and {McGrath}, E.~J. and {H{\"a}ussler}, B. and {Dekel}, A. and {Behroozi}, P. and {Fumagalli}, M. and {Leja}, J. and {Lundgren}, B.~F. and {Maseda}, M.~V. and {Nelson}, E.~J. and {Wake}, D.~A. and {Patel}, S.~G. and {Labb{\'e}}, I. and {Faber}, S.~M. and {Grogin}, N.~A. and {Kocevski}, D.~D.},
        title = "{3D-HST+CANDELS: The Evolution of the Galaxy Size-Mass Distribution since z = 3}",
      journal = {\apj},
         year = 2014,
        month = jun,
       volume = {788},
       number = {1},
          eid = {28},
        pages = {28},
          doi = {10.1088/0004-637X/788/1/28},
archivePrefix = {arXiv},
       eprint = {1404.2844},
 primaryClass = {astro-ph.GA},
       adsurl = {https://ui.adsabs.harvard.edu/abs/2014ApJ...788...28V}
}

@ARTICLE{2023MNRAS.518.6011D,
       author = {{Donnan}, C.~T. and {McLeod}, D.~J. and {Dunlop}, J.~S. and {McLure}, R.~J. and {Carnall}, A.~C. and {Begley}, R. and {Cullen}, F. and {Hamadouche}, M.~L. and {Bowler}, R.~A.~A. and {Magee}, D. and {McCracken}, H.~J. and {Milvang-Jensen}, B. and {Moneti}, A. and {Targett}, T.},
        title = "{The evolution of the galaxy UV luminosity function at redshifts z ≃ 8 - 15 from deep JWST and ground-based near-infrared imaging}",
      journal = {\mnras},
         year = 2023,
        month = feb,
       volume = {518},
       number = {4},
        pages = {6011-6040},
          doi = {10.1093/mnras/stac3472},
archivePrefix = {arXiv},
       eprint = {2207.12356},
 primaryClass = {astro-ph.GA},
       adsurl = {https://ui.adsabs.harvard.edu/abs/2023MNRAS.518.6011D}
}

@ARTICLE{2022ApJ...938L..15C,
       author = {{Castellano}, Marco and {Fontana}, Adriano and {Treu}, Tommaso and {Santini}, Paola and {Merlin}, Emiliano and {Leethochawalit}, Nicha and {Trenti}, Michele and {Vanzella}, Eros and {Mestric}, Uros and {Bonchi}, Andrea and {Belfiori}, Davide and {Nonino}, Mario and {Paris}, Diego and {Polenta}, Gianluca and {Roberts-Borsani}, Guido and {Boyett}, Kristan and {Brada{\v{c}}}, Maru{\v{s}}a and {Calabr{\`o}}, Antonello and {Glazebrook}, Karl and {Grillo}, Claudio and {Mascia}, Sara and {Mason}, Charlotte and {Mercurio}, Amata and {Morishita}, Takahiro and {Nanayakkara}, Themiya and {Pentericci}, Laura and {Rosati}, Piero and {Vulcani}, Benedetta and {Wang}, Xin and {Yang}, Lilan},
        title = "{Early Results from GLASS-JWST. III. Galaxy Candidates at z  9-15}",
      journal = {\apjl},
         year = 2022,
        month = oct,
       volume = {938},
       number = {2},
          eid = {L15},
        pages = {L15},
          doi = {10.3847/2041-8213/ac94d0},
archivePrefix = {arXiv},
       eprint = {2207.09436},
 primaryClass = {astro-ph.GA},
       adsurl = {https://ui.adsabs.harvard.edu/abs/2022ApJ...938L..15C}
}

@ARTICLE{2024ApJ...972..143C,
       author = {{Castellano}, Marco and {Napolitano}, Lorenzo and {Fontana}, Adriano and {Roberts-Borsani}, Guido and {Treu}, Tommaso and {Vanzella}, Eros and {Zavala}, Jorge A. and {Arrabal Haro}, Pablo and {Calabr{\`o}}, Antonello and {Llerena}, Mario and {Mascia}, Sara and {Merlin}, Emiliano and {Paris}, Diego and {Pentericci}, Laura and {Santini}, Paola and {Bakx}, Tom J.~L.~C. and {Bergamini}, Pietro and {Cupani}, Guido and {Dickinson}, Mark and {Filippenko}, Alexei V. and {Glazebrook}, Karl and {Grillo}, Claudio and {Kelly}, Patrick L. and {Malkan}, Matthew A. and {Mason}, Charlotte A. and {Morishita}, Takahiro and {Nanayakkara}, Themiya and {Rosati}, Piero and {Sani}, Eleonora and {Wang}, Xin and {Yoon}, Ilsang},
        title = "{JWST NIRSpec Spectroscopy of the Remarkable Bright Galaxy GHZ2/GLASS-z12 at Redshift 12.34}",
      journal = {\apj},
         year = 2024,
        month = sep,
       volume = {972},
       number = {2},
          eid = {143},
        pages = {143},
          doi = {10.3847/1538-4357/ad5f88},
archivePrefix = {arXiv},
       eprint = {2403.10238},
 primaryClass = {astro-ph.GA},
       adsurl = {https://ui.adsabs.harvard.edu/abs/2024ApJ...972..143C}
}

@ARTICLE{2023ApJ...948L..14C,
       author = {{Castellano}, Marco and {Fontana}, Adriano and {Treu}, Tommaso and {Merlin}, Emiliano and {Santini}, Paola and {Bergamini}, Pietro and {Grillo}, Claudio and {Rosati}, Piero and {Acebron}, Ana and {Leethochawalit}, Nicha and {Paris}, Diego and {Bonchi}, Andrea and {Belfiori}, Davide and {Calabr{\`o}}, Antonello and {Correnti}, Matteo and {Nonino}, Mario and {Polenta}, Gianluca and {Trenti}, Michele and {Boyett}, Kristan and {Brammer}, G. and {Broadhurst}, Tom and {Caminha}, Gabriel B. and {Chen}, Wenlei and {Filippenko}, Alexei V. and {Fortuni}, Flaminia and {Glazebrook}, Karl and {Mascia}, Sara and {Mason}, Charlotte A. and {Menci}, Nicola and {Meneghetti}, Massimo and {Mercurio}, Amata and {Metha}, Benjamin and {Morishita}, Takahiro and {Nanayakkara}, Themiya and {Pentericci}, Laura and {Roberts-Borsani}, Guido and {Roy}, Namrata and {Vanzella}, Eros and {Vulcani}, Benedetta and {Yang}, Lilan and {Wang}, Xin},
        title = "{Early Results from GLASS-JWST. XIX. A High Density of Bright Galaxies at z {\ensuremath{\approx}} 10 in the A2744 Region}",
      journal = {\apjl},
         year = 2023,
        month = may,
       volume = {948},
       number = {2},
          eid = {L14},
        pages = {L14},
          doi = {10.3847/2041-8213/accea5},
archivePrefix = {arXiv},
       eprint = {2212.06666},
 primaryClass = {astro-ph.GA},
       adsurl = {https://ui.adsabs.harvard.edu/abs/2023ApJ...948L..14C}
}

@ARTICLE{2023MNRAS.520.4554D,
       author = {{Donnan}, C.~T. and {McLeod}, D.~J. and {McLure}, R.~J. and {Dunlop}, J.~S. and {Carnall}, A.~C. and {Cullen}, F. and {Magee}, D.},
        title = "{The abundance of z {\ensuremath{\gtrsim}} 10 galaxy candidates in the HUDF using deep JWST NIRCam medium-band imaging}",
      journal = {\mnras},
         year = 2023,
        month = apr,
       volume = {520},
       number = {3},
        pages = {4554-4561},
          doi = {10.1093/mnras/stad471},
archivePrefix = {arXiv},
       eprint = {2212.10126},
 primaryClass = {astro-ph.GA},
       adsurl = {https://ui.adsabs.harvard.edu/abs/2023MNRAS.520.4554D}
}

@ARTICLE{2025A&A...699A.360C,
       author = {{Costantin}, L. and {Gillman}, S. and {Boogaard}, L.~A. and {P{\'e}rez-Gonz{\'a}lez}, P.~G. and {Iani}, E. and {Rinaldi}, P. and {Melinder}, J. and {Crespo G{\'o}mez}, A. and {Colina}, L. and {Greve}, T.~R. and {{\"O}stlin}, G. and {Wright}, G. and {Alonso-Herrero}, A. and {{\'A}lvarez-M{\'a}rquez}, J. and {Annunziatella}, M. and {Bik}, A. and {Caputi}, K.~I. and {Dicken}, D. and {Eckart}, A. and {Hjorth}, J. and {Ilbert}, O. and {Jermann}, I. and {Labiano}, A. and {Langeroodi}, D. and {Pei{\ss}ker}, F. and {Pye}, J.~P. and {Tikkanen}, T.~V. and {van der Werf}, P.~P. and {Walter}, F. and {Ward}, M. and {G{\"u}del}, M. and {Henning}, T.~K.},
        title = "{MIDIS: Near-infrared rest-frame morphology of massive galaxies at 3 < z < 5 in the Hubble eXtreme Deep Field}",
      journal = {\aap},
         year = 2025,
        month = jul,
       volume = {699},
          eid = {A360},
        pages = {A360},
          doi = {10.1051/0004-6361/202451330},
archivePrefix = {arXiv},
       eprint = {2407.00153},
 primaryClass = {astro-ph.GA},
       adsurl = {https://ui.adsabs.harvard.edu/abs/2025A&A...699A.360C}
}

@ARTICLE{2020MNRAS.492.3073L,
       author = {{Lacerda}, Eduardo A.~D. and {S{\'a}nchez}, Sebasti{\'a}n F. and {Cid Fernandes}, R. and {L{\'o}pez-Cob{\'a}}, Carlos and {Espinosa-Ponce}, Carlos and {Galbany}, L.},
        title = "{Galaxies hosting an active galactic nucleus: a view from the CALIFA survey}",
      journal = {\mnras},
         year = 2020,
        month = mar,
       volume = {492},
       number = {3},
        pages = {3073-3090},
          doi = {10.1093/mnras/staa008},
archivePrefix = {arXiv},
       eprint = {2001.00099},
 primaryClass = {astro-ph.GA},
       adsurl = {https://ui.adsabs.harvard.edu/abs/2020MNRAS.492.3073L}
}

@ARTICLE{1926ApJ....64..321H,
       author = {{Hubble}, E.~P.},
        title = "{Extragalactic nebulae.}",
      journal = {\apj},
         year = 1926,
        month = dec,
       volume = {64},
        pages = {321-369},
          doi = {10.1086/143018},
       adsurl = {https://ui.adsabs.harvard.edu/abs/1926ApJ....64..321H}
}

@BOOK{1961hag..book.....S,
       author = {{Sandage}, Allan},
        title = "{The Hubble Atlas of Galaxies}",
         year = 1961,
       adsurl = {https://ui.adsabs.harvard.edu/abs/1961hag..book.....S}
}

@ARTICLE{1976ApJ...206..883V,
       author = {{van den Bergh}, S.},
        title = "{A new classification system for galaxies.}",
      journal = {\apj},
         year = 1976,
        month = jun,
       volume = {206},
        pages = {883-887},
          doi = {10.1086/154452},
       adsurl = {https://ui.adsabs.harvard.edu/abs/1976ApJ...206..883V}
}

@ARTICLE{2003ApJS..149..289B,
       author = {{Bell}, Eric F. and {McIntosh}, Daniel H. and {Katz}, Neal and {Weinberg}, Martin D.},
        title = "{The Optical and Near-Infrared Properties of Galaxies. I. Luminosity and Stellar Mass Functions}",
      journal = {\apjs},
         year = 2003,
        month = dec,
       volume = {149},
       number = {2},
        pages = {289-312},
          doi = {10.1086/378847},
archivePrefix = {arXiv},
       eprint = {astro-ph/0302543},
 primaryClass = {astro-ph},
       adsurl = {https://ui.adsabs.harvard.edu/abs/2003ApJS..149..289B}
}

@ARTICLE{2003MNRAS.341...54K,
       author = {{Kauffmann}, Guinevere and {Heckman}, Timothy M. and {White}, Simon D.~M. and {Charlot}, St{\'e}phane and {Tremonti}, Christy and {Peng}, Eric W. and {Seibert}, Mark and {Brinkmann}, Jon and {Nichol}, Robert C. and {SubbaRao}, Mark and {York}, Don},
        title = "{The dependence of star formation history and internal structure on stellar mass for {}10$^{5}$ low-redshift galaxies}",
      journal = {\mnras},
         year = 2003,
        month = may,
       volume = {341},
       number = {1},
        pages = {54-69},
          doi = {10.1046/j.1365-8711.2003.06292.x},
archivePrefix = {arXiv},
       eprint = {astro-ph/0205070},
 primaryClass = {astro-ph},
       adsurl = {https://ui.adsabs.harvard.edu/abs/2003MNRAS.341...54K}
}

@ARTICLE{2014ARA&A..52..291C,
       author = {{Conselice}, Christopher J.},
        title = "{The Evolution of Galaxy Structure Over Cosmic Time}",
      journal = {\araa},
         year = 2014,
        month = aug,
       volume = {52},
        pages = {291-337},
          doi = {10.1146/annurev-astro-081913-040037},
archivePrefix = {arXiv},
       eprint = {1403.2783},
 primaryClass = {astro-ph.GA},
       adsurl = {https://ui.adsabs.harvard.edu/abs/2014ARA&A..52..291C}
}

@ARTICLE{2015MNRAS.452.4347K,
       author = {{Kannan}, Rahul and {Macci{\`o}}, Andrea V. and {Fontanot}, Fabio and {Moster}, Benjamin P. and {Karman}, Wouter and {Somerville}, Rachel S.},
        title = "{From discs to bulges: effect of mergers on the morphology of galaxies}",
      journal = {\mnras},
         year = 2015,
        month = oct,
       volume = {452},
       number = {4},
        pages = {4347-4360},
          doi = {10.1093/mnras/stv1633},
archivePrefix = {arXiv},
       eprint = {1507.04746},
 primaryClass = {astro-ph.GA},
       adsurl = {https://ui.adsabs.harvard.edu/abs/2015MNRAS.452.4347K}
}

@ARTICLE{2022ApJ...940L..14N,
       author = {{Naidu}, Rohan P. and {Oesch}, Pascal A. and {van Dokkum}, Pieter and {Nelson}, Erica J. and {Suess}, Katherine A. and {Brammer}, Gabriel and {Whitaker}, Katherine E. and {Illingworth}, Garth and {Bouwens}, Rychard and {Tacchella}, Sandro and {Matthee}, Jorryt and {Allen}, Natalie and {Bezanson}, Rachel and {Conroy}, Charlie and {Labbe}, Ivo and {Leja}, Joel and {Leonova}, Ecaterina and {Magee}, Dan and {Price}, Sedona H. and {Setton}, David J. and {Strait}, Victoria and {Stefanon}, Mauro and {Toft}, Sune and {Weaver}, John R. and {Weibel}, Andrea},
        title = "{Two Remarkably Luminous Galaxy Candidates at z {\ensuremath{\approx}} 10-12 Revealed by JWST}",
      journal = {\apjl},
         year = 2022,
        month = nov,
       volume = {940},
       number = {1},
          eid = {L14},
        pages = {L14},
          doi = {10.3847/2041-8213/ac9b22},
archivePrefix = {arXiv},
       eprint = {2207.09434},
 primaryClass = {astro-ph.GA},
       adsurl = {https://ui.adsabs.harvard.edu/abs/2022ApJ...940L..14N}
}

@ARTICLE{2003PASP..115..763C,
       author = {{Chabrier}, Gilles},
        title = "{Galactic Stellar and Substellar Initial Mass Function}",
      journal = {\pasp},
         year = 2003,
        month = jul,
       volume = {115},
       number = {809},
        pages = {763-795},
          doi = {10.1086/376392},
archivePrefix = {arXiv},
       eprint = {astro-ph/0304382},
 primaryClass = {astro-ph},
       adsurl = {https://ui.adsabs.harvard.edu/abs/2003PASP..115..763C}
}

@ARTICLE{1998ApJ...498..541K,
       author = {{Kennicutt}, Jr., Robert C.},
        title = "{The Global Schmidt Law in Star-forming Galaxies}",
      journal = {\apj},
         year = 1998,
        month = may,
       volume = {498},
       number = {2},
        pages = {541-552},
          doi = {10.1086/305588},
archivePrefix = {arXiv},
       eprint = {astro-ph/9712213},
 primaryClass = {astro-ph},
       adsurl = {https://ui.adsabs.harvard.edu/abs/1998ApJ...498..541K}
}

@ARTICLE{2015MNRAS.454.2691M,
       author = {{Muratov}, Alexander L. and {Kere{\v{s}}}, Du{\v{s}}an and {Faucher-Gigu{\`e}re}, Claude-Andr{\'e} and {Hopkins}, Philip F. and {Quataert}, Eliot and {Murray}, Norman},
        title = "{Gusty, gaseous flows of FIRE: galactic winds in cosmological simulations with explicit stellar feedback}",
      journal = {\mnras},
         year = 2015,
        month = dec,
       volume = {454},
       number = {3},
        pages = {2691-2713},
          doi = {10.1093/mnras/stv2126},
archivePrefix = {arXiv},
       eprint = {1501.03155},
 primaryClass = {astro-ph.GA},
       adsurl = {https://ui.adsabs.harvard.edu/abs/2015MNRAS.454.2691M}
}

@ARTICLE{2015ARA&A..53...51S,
       author = {{Somerville}, Rachel S. and {Dav{\'e}}, Romeel},
        title = "{Physical Models of Galaxy Formation in a Cosmological Framework}",
      journal = {\araa},
         year = 2015,
        month = aug,
       volume = {53},
        pages = {51-113},
          doi = {10.1146/annurev-astro-082812-140951},
archivePrefix = {arXiv},
       eprint = {1412.2712},
 primaryClass = {astro-ph.GA},
       adsurl = {https://ui.adsabs.harvard.edu/abs/2015ARA&A..53...51S}
}

@ARTICLE{2016ApJ...824...57C,
       author = {{Christensen}, Charlotte R. and {Dav{\'e}}, Romeel and {Governato}, Fabio and {Pontzen}, Andrew and {Brooks}, Alyson and {Munshi}, Ferah and {Quinn}, Thomas and {Wadsley}, James},
        title = "{In-N-Out: The Gas Cycle from Dwarfs to Spiral Galaxies}",
      journal = {\apj},
         year = 2016,
        month = jun,
       volume = {824},
       number = {1},
          eid = {57},
        pages = {57},
          doi = {10.3847/0004-637X/824/1/57},
archivePrefix = {arXiv},
       eprint = {1508.00007},
 primaryClass = {astro-ph.GA},
       adsurl = {https://ui.adsabs.harvard.edu/abs/2016ApJ...824...57C}
}

@ARTICLE{2006MNRAS.365...11C,
       author = {{Croton}, Darren J. and {Springel}, Volker and {White}, Simon D.~M. and {De Lucia}, G. and {Frenk}, C.~S. and {Gao}, L. and {Jenkins}, A. and {Kauffmann}, G. and {Navarro}, J.~F. and {Yoshida}, N.},
        title = "{The many lives of active galactic nuclei: cooling flows, black holes and the luminosities and colours of galaxies}",
      journal = {\mnras},
         year = 2006,
        month = jan,
       volume = {365},
       number = {1},
        pages = {11-28},
          doi = {10.1111/j.1365-2966.2005.09675.x},
archivePrefix = {arXiv},
       eprint = {astro-ph/0508046},
 primaryClass = {astro-ph},
       adsurl = {https://ui.adsabs.harvard.edu/abs/2006MNRAS.365...11C}
}

@ARTICLE{2019MNRAS.488.3143B,
       author = {{Behroozi}, Peter and {Wechsler}, Risa H. and {Hearin}, Andrew P. and {Conroy}, Charlie},
        title = "{UNIVERSEMACHINE: The correlation between galaxy growth and dark matter halo assembly from z = 0-10}",
      journal = {\mnras},
         year = 2019,
        month = sep,
       volume = {488},
       number = {3},
        pages = {3143-3194},
          doi = {10.1093/mnras/stz1182},
archivePrefix = {arXiv},
       eprint = {1806.07893},
 primaryClass = {astro-ph.GA},
       adsurl = {https://ui.adsabs.harvard.edu/abs/2019MNRAS.488.3143B}
}

@ARTICLE{2006MNRAS.373.1389C,
       author = {{Conselice}, Christopher J.},
        title = "{The fundamental properties of galaxies and a new galaxy classification system}",
      journal = {\mnras},
         year = 2006,
        month = dec,
       volume = {373},
       number = {4},
        pages = {1389-1408},
          doi = {10.1111/j.1365-2966.2006.11114.x},
archivePrefix = {arXiv},
       eprint = {astro-ph/0610016},
 primaryClass = {astro-ph},
       adsurl = {https://ui.adsabs.harvard.edu/abs/2006MNRAS.373.1389C}
}

@ARTICLE{2006MNRAS.368....2D,
       author = {{Dekel}, Avishai and {Birnboim}, Yuval},
        title = "{Galaxy bimodality due to cold flows and shock heating}",
      journal = {\mnras},
         year = 2006,
        month = may,
       volume = {368},
       number = {1},
        pages = {2-20},
          doi = {10.1111/j.1365-2966.2006.10145.x},
archivePrefix = {arXiv},
       eprint = {astro-ph/0412300},
 primaryClass = {astro-ph},
       adsurl = {https://ui.adsabs.harvard.edu/abs/2006MNRAS.368....2D}
}

@ARTICLE{2009ApJ...691.1168H,
       author = {{Hopkins}, Philip F. and {Cox}, Thomas J. and {Younger}, Joshua D. and {Hernquist}, Lars},
        title = "{How do Disks Survive Mergers?}",
      journal = {\apj},
         year = 2009,
        month = feb,
       volume = {691},
       number = {2},
        pages = {1168-1201},
          doi = {10.1088/0004-637X/691/2/1168},
archivePrefix = {arXiv},
       eprint = {0806.1739},
 primaryClass = {astro-ph},
       adsurl = {https://ui.adsabs.harvard.edu/abs/2009ApJ...691.1168H}
}

@ARTICLE{2003MNRAS.343..978S,
       author = {{Shen}, Shiyin and {Mo}, H.~J. and {White}, Simon D.~M. and {Blanton}, Michael R. and {Kauffmann}, Guinevere and {Voges}, Wolfgang and {Brinkmann}, J. and {Csabai}, Istvan},
        title = "{The size distribution of galaxies in the Sloan Digital Sky Survey}",
      journal = {\mnras},
         year = 2003,
        month = aug,
       volume = {343},
       number = {3},
        pages = {978-994},
          doi = {10.1046/j.1365-8711.2003.06740.x},
archivePrefix = {arXiv},
       eprint = {astro-ph/0301527},
 primaryClass = {astro-ph},
       adsurl = {https://ui.adsabs.harvard.edu/abs/2003MNRAS.343..978S}
}

@ARTICLE{2009ApJ...699L.178N,
       author = {{Naab}, Thorsten and {Johansson}, Peter H. and {Ostriker}, Jeremiah P.},
        title = "{Minor Mergers and the Size Evolution of Elliptical Galaxies}",
      journal = {\apjl},
         year = 2009,
        month = jul,
       volume = {699},
       number = {2},
        pages = {L178-L182},
          doi = {10.1088/0004-637X/699/2/L178},
archivePrefix = {arXiv},
       eprint = {0903.1636},
 primaryClass = {astro-ph.CO},
       adsurl = {https://ui.adsabs.harvard.edu/abs/2009ApJ...699L.178N}
}

@ARTICLE{2021ApJ...915...71V,
       author = {{Villaescusa-Navarro}, Francisco and {Angl{\'e}s-Alc{\'a}zar}, Daniel and {Genel}, Shy and {Spergel}, David N. and {Somerville}, Rachel S. and {Dave}, Romeel and {Pillepich}, Annalisa and {Hernquist}, Lars and {Nelson}, Dylan and {Torrey}, Paul and {Narayanan}, Desika and {Li}, Yin and {Philcox}, Oliver and {La Torre}, Valentina and {Maria Delgado}, Ana and {Ho}, Shirley and {Hassan}, Sultan and {Burkhart}, Blakesley and {Wadekar}, Digvijay and {Battaglia}, Nicholas and {Contardo}, Gabriella and {Bryan}, Greg L.},
        title = "{The CAMELS Project: Cosmology and Astrophysics with Machine-learning Simulations}",
      journal = {\apj},
         year = 2021,
        month = jul,
       volume = {915},
       number = {1},
          eid = {71},
        pages = {71},
          doi = {10.3847/1538-4357/abf7ba},
archivePrefix = {arXiv},
       eprint = {2010.00619},
 primaryClass = {astro-ph.CO},
       adsurl = {https://ui.adsabs.harvard.edu/abs/2021ApJ...915...71V}
}

@ARTICLE{2017MNRAS.472L.109A,
       author = {{Angl{\'e}s-Alc{\'a}zar}, Daniel and {Faucher-Gigu{\`e}re}, Claude-Andr{\'e} and {Quataert}, Eliot and {Hopkins}, Philip F. and {Feldmann}, Robert and {Torrey}, Paul and {Wetzel}, Andrew and {Kere{\v{s}}}, Du{\v{s}}an},
        title = "{Black holes on FIRE: stellar feedback limits early feeding of galactic nuclei}",
      journal = {\mnras},
         year = 2017,
        month = nov,
       volume = {472},
       number = {1},
        pages = {L109-L114},
          doi = {10.1093/mnrasl/slx161},
archivePrefix = {arXiv},
       eprint = {1707.03832},
 primaryClass = {astro-ph.GA},
       adsurl = {https://ui.adsabs.harvard.edu/abs/2017MNRAS.472L.109A}
}

@ARTICLE{2020ApJS..248...32W,
       author = {{Weinberger}, Rainer and {Springel}, Volker and {Pakmor}, R{\"u}diger},
        title = "{The AREPO Public Code Release}",
      journal = {\apjs},
         year = 2020,
        month = jun,
       volume = {248},
       number = {2},
          eid = {32},
        pages = {32},
          doi = {10.3847/1538-4365/ab908c},
archivePrefix = {arXiv},
       eprint = {1909.04667},
 primaryClass = {astro-ph.IM},
       adsurl = {https://ui.adsabs.harvard.edu/abs/2020ApJS..248...32W}
}

@ARTICLE{2010MNRAS.401..791S,
       author = {{Springel}, Volker},
        title = "{E pur si muove: Galilean-invariant cosmological hydrodynamical simulations on a moving mesh}",
      journal = {\mnras},
         year = 2010,
        month = jan,
       volume = {401},
       number = {2},
        pages = {791-851},
          doi = {10.1111/j.1365-2966.2009.15715.x},
archivePrefix = {arXiv},
       eprint = {0901.4107},
 primaryClass = {astro-ph.CO},
       adsurl = {https://ui.adsabs.harvard.edu/abs/2010MNRAS.401..791S}
}

@ARTICLE{2020A&A...641A...6P,
       author = {{Planck Collaboration} and {Aghanim}, N. and {Akrami}, Y. and {Ashdown}, M. and {Aumont}, J. and {Baccigalupi}, C. and {Ballardini}, M. and {Banday}, A.~J. and {Barreiro}, R.~B. and {Bartolo}, N. and {Basak}, S. and {Battye}, R. and {Benabed}, K. and {Bernard}, J.-P. and {Bersanelli}, M. and {Bielewicz}, P. and {Bock}, J.~J. and {Bond}, J.~R. and {Borrill}, J. and {Bouchet}, F.~R. and {Boulanger}, F. and {Bucher}, M. and {Burigana}, C. and {Butler}, R.~C. and {Calabrese}, E. and {Cardoso}, J.-F. and {Carron}, J. and {Challinor}, A. and {Chiang}, H.~C. and {Chluba}, J. and {Colombo}, L.~P.~L. and {Combet}, C. and {Contreras}, D. and {Crill}, B.~P. and {Cuttaia}, F. and {de Bernardis}, P. and {de Zotti}, G. and {Delabrouille}, J. and {Delouis}, J.-M. and {Di Valentino}, E. and {Diego}, J.~M. and {Dor{\'e}}, O. and {Douspis}, M. and {Ducout}, A. and {Dupac}, X. and {Dusini}, S. and {Efstathiou}, G. and {Elsner}, F. and {En{\ss}lin}, T.~A. and {Eriksen}, H.~K. and {Fantaye}, Y. and {Farhang}, M. and {Fergusson}, J. and {Fernandez-Cobos}, R. and {Finelli}, F. and {Forastieri}, F. and {Frailis}, M. and {Fraisse}, A.~A. and {Franceschi}, E. and {Frolov}, A. and {Galeotta}, S. and {Galli}, S. and {Ganga}, K. and {G{\'e}nova-Santos}, R.~T. and {Gerbino}, M. and {Ghosh}, T. and {Gonz{\'a}lez-Nuevo}, J. and {G{\'o}rski}, K.~M. and {Gratton}, S. and {Gruppuso}, A. and {Gudmundsson}, J.~E. and {Hamann}, J. and {Handley}, W. and {Hansen}, F.~K. and {Herranz}, D. and {Hildebrandt}, S.~R. and {Hivon}, E. and {Huang}, Z. and {Jaffe}, A.~H. and {Jones}, W.~C. and {Karakci}, A. and {Keih{\"a}nen}, E. and {Keskitalo}, R. and {Kiiveri}, K. and {Kim}, J. and {Kisner}, T.~S. and {Knox}, L. and {Krachmalnicoff}, N. and {Kunz}, M. and {Kurki-Suonio}, H. and {Lagache}, G. and {Lamarre}, J.-M. and {Lasenby}, A. and {Lattanzi}, M. and {Lawrence}, C.~R. and {Le Jeune}, M. and {Lemos}, P. and {Lesgourgues}, J. and {Levrier}, F. and {Lewis}, A. and {Liguori}, M. and {Lilje}, P.~B. and {Lilley}, M. and {Lindholm}, V. and {L{\'o}pez-Caniego}, M. and {Lubin}, P.~M. and {Ma}, Y.-Z. and {Mac{\'\i}as-P{\'e}rez}, J.~F. and {Maggio}, G. and {Maino}, D. and {Mandolesi}, N. and {Mangilli}, A. and {Marcos-Caballero}, A. and {Maris}, M. and {Martin}, P.~G. and {Martinelli}, M. and {Mart{\'\i}nez-Gonz{\'a}lez}, E. and {Matarrese}, S. and {Mauri}, N. and {McEwen}, J.~D. and {Meinhold}, P.~R. and {Melchiorri}, A. and {Mennella}, A. and {Migliaccio}, M. and {Millea}, M. and {Mitra}, S. and {Miville-Desch{\^e}nes}, M.-A. and {Molinari}, D. and {Montier}, L. and {Morgante}, G. and {Moss}, A. and {Natoli}, P. and {N{\o}rgaard-Nielsen}, H.~U. and {Pagano}, L. and {Paoletti}, D. and {Partridge}, B. and {Patanchon}, G. and {Peiris}, H.~V. and {Perrotta}, F. and {Pettorino}, V. and {Piacentini}, F. and {Polastri}, L. and {Polenta}, G. and {Puget}, J.-L. and {Rachen}, J.~P. and {Reinecke}, M. and {Remazeilles}, M. and {Renzi}, A. and {Rocha}, G. and {Rosset}, C. and {Roudier}, G. and {Rubi{\~n}o-Mart{\'\i}n}, J.~A. and {Ruiz-Granados}, B. and {Salvati}, L. and {Sandri}, M. and {Savelainen}, M. and {Scott}, D. and {Shellard}, E.~P.~S. and {Sirignano}, C. and {Sirri}, G. and {Spencer}, L.~D. and {Sunyaev}, R. and {Suur-Uski}, A.-S. and {Tauber}, J.~A. and {Tavagnacco}, D. and {Tenti}, M. and {Toffolatti}, L. and {Tomasi}, M. and {Trombetti}, T. and {Valenziano}, L. and {Valiviita}, J. and {Van Tent}, B. and {Vibert}, L. and {Vielva}, P. and {Villa}, F. and {Vittorio}, N. and {Wandelt}, B.~D. and {Wehus}, I.~K. and {White}, M. and {White}, S.~D.~M. and {Zacchei}, A. and {Zonca}, A.},
        title = "{Planck 2018 results. VI. Cosmological parameters}",
      journal = {\aap},
         year = 2020,
        month = sep,
       volume = {641},
          eid = {A6},
        pages = {A6},
          doi = {10.1051/0004-6361/201833910},
archivePrefix = {arXiv},
       eprint = {1807.06209},
 primaryClass = {astro-ph.CO},
       adsurl = {https://ui.adsabs.harvard.edu/abs/2020A&A...641A...6P}
}

@ARTICLE{2019MNRAS.487L..61Z,
       author = {{Zaragoza-Cardiel}, Javier and {Fritz}, Jacopo and {Aretxaga}, Itziar and {Mayya}, Divakara and {Rosa-Gonz{\'a}lez}, Daniel and {Beckman}, John E. and {Bruzual}, Gustavo and {Charlot}, Stephane and {Lomel{\'\i}-N{\'u}{\~n}ez}, Luis},
        title = "{Detection of the self-regulation of star formation in galaxy discs}",
      journal = {\mnras},
         year = 2019,
        month = jul,
       volume = {487},
       number = {1},
        pages = {L61-L66},
          doi = {10.1093/mnrasl/slz093},
archivePrefix = {arXiv},
       eprint = {1906.01641},
 primaryClass = {astro-ph.GA},
       adsurl = {https://ui.adsabs.harvard.edu/abs/2019MNRAS.487L..61Z}
}

@ARTICLE{2022MNRAS.513.3768I,
       author = {{Irodotou}, Dimitrios and {Fragkoudi}, Francesca and {Pakmor}, Ruediger and {Grand}, Robert J.~J. and {Gadotti}, Dimitri A. and {Costa}, Tiago and {Springel}, Volker and {G{\'o}mez}, Facundo A. and {Marinacci}, Federico},
        title = "{The effects of AGN feedback on the structural and dynamical properties of Milky Way-mass galaxies in cosmological simulations}",
      journal = {\mnras},
         year = 2022,
        month = jul,
       volume = {513},
       number = {3},
        pages = {3768-3787},
          doi = {10.1093/mnras/stac1143},
archivePrefix = {arXiv},
       eprint = {2110.11368},
 primaryClass = {astro-ph.GA},
       adsurl = {https://ui.adsabs.harvard.edu/abs/2022MNRAS.513.3768I}
}

@ARTICLE{2019MNRAS.489.2702F,
       author = {{Frigo}, Matteo and {Naab}, Thorsten and {Hirschmann}, Michaela and {Choi}, Ena and {Somerville}, Rachel S. and {Krajnovic}, Davor and {Dav{\'e}}, Romeel and {Cappellari}, Michele},
        title = "{The impact of AGN on stellar kinematics and orbits in simulated massive galaxies}",
      journal = {\mnras},
         year = 2019,
        month = oct,
       volume = {489},
       number = {2},
        pages = {2702-2722},
          doi = {10.1093/mnras/stz2318},
archivePrefix = {arXiv},
       eprint = {1811.11059},
 primaryClass = {astro-ph.GA},
       adsurl = {https://ui.adsabs.harvard.edu/abs/2019MNRAS.489.2702F}
}

@ARTICLE{2024ApJ...960..104S,
       author = {{Sun}, Wen and {Ho}, Luis C. and {Zhuang}, Ming-Yang and {Ma}, Chao and {Chen}, Changhao and {Li}, Ruancun},
        title = "{The Structure and Morphology of Galaxies during the Epoch of Reionization Revealed by JWST}",
      journal = {\apj},
         year = 2024,
        month = jan,
       volume = {960},
       number = {2},
          eid = {104},
        pages = {104},
          doi = {10.3847/1538-4357/acf1f6},
archivePrefix = {arXiv},
       eprint = {2308.09076},
 primaryClass = {astro-ph.GA},
       adsurl = {https://ui.adsabs.harvard.edu/abs/2024ApJ...960..104S}
}

@ARTICLE{2020ApJ...889...32S,
       author = {{Suh}, Hyewon and {Civano}, Francesca and {Trakhtenbrot}, Benny and {Shankar}, Francesco and {Hasinger}, G{\"u}nther and {Sanders}, David B. and {Allevato}, Viola},
        title = "{No Significant Evolution of Relations between Black Hole Mass and Galaxy Total Stellar Mass Up to z {\ensuremath{\sim}} 2.5}",
      journal = {\apj},
         year = 2020,
        month = jan,
       volume = {889},
       number = {1},
          eid = {32},
        pages = {32},
          doi = {10.3847/1538-4357/ab5f5f},
archivePrefix = {arXiv},
       eprint = {1912.02824},
 primaryClass = {astro-ph.GA},
       adsurl = {https://ui.adsabs.harvard.edu/abs/2020ApJ...889...32S}
}

@ARTICLE{2025ApJ...979..215T,
       author = {{Tanaka}, Takumi S. and {Silverman}, John D. and {Ding}, Xuheng and {Jahnke}, Knud and {Trakhtenbrot}, Benny and {Lambrides}, Erini and {Onoue}, Masafusa and {Andika}, Irham Taufik and {Bongiorno}, Angela and {Faisst}, Andreas L. and {Gillman}, Steven and {Hayward}, Christopher C. and {Hirschmann}, Michaela and {Koekemoer}, Anton and {Kokorev}, Vasily and {Liu}, Zhaoxuan and {Magdis}, Georgios E. and {Renzini}, Alvio and {Casey}, Caitlin and {Drakos}, Nicole E. and {Franco}, Maximilien and {Gozaliasl}, Ghassem and {Kartaltepe}, Jeyhan and {Liu}, Daizhong and {McCracken}, Henry Joy and {Rhodes}, Jason and {Robertson}, Brant and {Toft}, Sune},
        title = "{The M$_{BH}$─M$_{{\ensuremath{*}}}$ Relation up to z {\ensuremath{\sim}} 2 through Decomposition of COSMOS-Web NIRCam Images}",
      journal = {\apj},
         year = 2025,
        month = feb,
       volume = {979},
       number = {2},
          eid = {215},
        pages = {215},
          doi = {10.3847/1538-4357/ad9d0a},
archivePrefix = {arXiv},
       eprint = {2401.13742},
 primaryClass = {astro-ph.GA},
       adsurl = {https://ui.adsabs.harvard.edu/abs/2025ApJ...979..215T}
}

@ARTICLE{2020ApJ...888...37D,
       author = {{Ding}, Xuheng and {Silverman}, John and {Treu}, Tommaso and {Schulze}, Andreas and {Schramm}, Malte and {Birrer}, Simon and {Park}, Daeseong and {Jahnke}, Knud and {Bennert}, Vardha N. and {Kartaltepe}, Jeyhan S. and {Koekemoer}, Anton M. and {Malkan}, Matthew A. and {Sanders}, David},
        title = "{The Mass Relations between Supermassive Black Holes and Their Host Galaxies at 1 < z < 2 HST-WFC3}",
      journal = {\apj},
         year = 2020,
        month = jan,
       volume = {888},
       number = {1},
          eid = {37},
        pages = {37},
          doi = {10.3847/1538-4357/ab5b90},
archivePrefix = {arXiv},
       eprint = {1910.11875},
 primaryClass = {astro-ph.GA},
       adsurl = {https://ui.adsabs.harvard.edu/abs/2020ApJ...888...37D}
}

@ARTICLE{2016ApJ...821..114H,
       author = {{Huang}, Song and {Ho}, Luis C. and {Peng}, Chien Y. and {Li}, Zhao-Yu and {Barth}, Aaron J.},
        title = "{The Carnegie-Irvine Galaxy Survey. IV. A Method to Determine the Average Mass Ratio of Mergers That Built Massive Elliptical Galaxies}",
      journal = {\apj},
         year = 2016,
        month = apr,
       volume = {821},
       number = {2},
          eid = {114},
        pages = {114},
          doi = {10.3847/0004-637X/821/2/114},
archivePrefix = {arXiv},
       eprint = {1603.00876},
 primaryClass = {astro-ph.GA},
       adsurl = {https://ui.adsabs.harvard.edu/abs/2016ApJ...821..114H}
}

@ARTICLE{2013ApJ...768L..28H,
       author = {{Huang}, Song and {Ho}, Luis C. and {Peng}, Chien Y. and {Li}, Zhao-Yu and {Barth}, Aaron J.},
        title = "{Fossil Evidence for the Two-phase Formation of Elliptical Galaxies}",
      journal = {\apjl},
         year = 2013,
        month = may,
       volume = {768},
       number = {2},
          eid = {L28},
        pages = {L28},
          doi = {10.1088/2041-8205/768/2/L28},
archivePrefix = {arXiv},
       eprint = {1304.2299},
 primaryClass = {astro-ph.CO},
       adsurl = {https://ui.adsabs.harvard.edu/abs/2013ApJ...768L..28H}
}

@ARTICLE{2013ApJ...766...47H,
       author = {{Huang}, Song and {Ho}, Luis C. and {Peng}, Chien Y. and {Li}, Zhao-Yu and {Barth}, Aaron J.},
        title = "{The Carnegie-Irvine Galaxy Survey. III. The Three-component Structure of Nearby Elliptical Galaxies}",
      journal = {\apj},
         year = 2013,
        month = mar,
       volume = {766},
       number = {1},
          eid = {47},
        pages = {47},
          doi = {10.1088/0004-637X/766/1/47},
archivePrefix = {arXiv},
       eprint = {1212.2639},
 primaryClass = {astro-ph.CO},
       adsurl = {https://ui.adsabs.harvard.edu/abs/2013ApJ...766...47H}
}

@ARTICLE{2018ApJ...852...36G,
       author = {{Greene}, J.~E. and {Leauthaud}, A. and {Emsellem}, E. and {Ge}, J. and {Arag{\'o}n-Salamanca}, A. and {Greco}, J. and {Lin}, Y.-T. and {Mao}, S. and {Masters}, K. and {Merrifield}, M. and {More}, S. and {Okabe}, N. and {Schneider}, D.~P. and {Thomas}, D. and {Wake}, D.~A. and {Pan}, K. and {Bizyaev}, D. and {Oravetz}, D. and {Simmons}, A. and {Yan}, R. and {van den Bosch}, F.},
        title = "{SDSS-IV MaNGA: Uncovering the Angular Momentum Content of Central and Satellite Early-type Galaxies}",
      journal = {\apj},
         year = 2018,
        month = jan,
       volume = {852},
       number = {1},
          eid = {36},
        pages = {36},
          doi = {10.3847/1538-4357/aa9bde},
archivePrefix = {arXiv},
       eprint = {1708.07845},
 primaryClass = {astro-ph.GA},
       adsurl = {https://ui.adsabs.harvard.edu/abs/2018ApJ...852...36G}
}

@ARTICLE{2020ApJ...889...37W,
       author = {{Wang}, Enci and {Wang}, Huiyuan and {Mo}, Houjun and {van den Bosch}, Frank C. and {Yang}, Xiaohu},
        title = "{The Dearth of Differences between Central and Satellite Galaxies. III. Environmental Dependencies of Mass-Size and Mass-Structure Relations}",
      journal = {\apj},
         year = 2020,
        month = jan,
       volume = {889},
       number = {1},
          eid = {37},
        pages = {37},
          doi = {10.3847/1538-4357/ab6217},
archivePrefix = {arXiv},
       eprint = {1912.05969},
 primaryClass = {astro-ph.GA},
       adsurl = {https://ui.adsabs.harvard.edu/abs/2020ApJ...889...37W}
}

\end{document}